\documentclass[reprint,nofootinbib,superscriptaddress,amsmath,amssymb,aps,prd]{revtex4-1}
\usepackage{amsmath,mathrsfs,amsbsy,color,graphicx,bm,amsthm,amsfonts}
\usepackage{bbm}
\usepackage{times}
\usepackage{units}
\usepackage{dcolumn}
\usepackage{graphicx}
\usepackage{epsfig}
\usepackage{epstopdf}
\usepackage[colorlinks,linkcolor=red,anchorcolor=green,citecolor=blue,CJKbookmarks=True]{hyperref}
\DeclareMathSymbol{\shortminus}{\mathbin}{AMSa}{"39}
\usepackage{mathrsfs}
\usepackage{braket}
\usepackage{amssymb}
\usepackage{txfonts}
\usepackage{float}
\usepackage{enumitem}
\usepackage{multirow}

\newcommand{\meq}[1]{(\ref{#1})}

\newcommand{\pp}{\partial}

\begin{document}

\title{Quasinormal Modes of a Massive Scalar Field in Slowly Rotating Einstein–Bumblebee Black Holes}

\author{Weike Deng}
\email[]{wkdeng@hnit.edu.cn} 
\affiliation{School of Science, Hunan Institute of Technology, Hengyang 421002, P. R. China}
\affiliation{Department of Physics, Key Laboratory of Low Dimensional Quantum Structures and Quantum Control of Ministry of Education, and Synergetic Innovation Center for Quantum Effects and Applications, Hunan Normal
University, Changsha, Hunan 410081, P. R. China}

\author{Wentao Liu}
\affiliation{Department of Physics, Key Laboratory of Low Dimensional Quantum Structures and Quantum Control of Ministry of Education, and Synergetic Innovation Center for Quantum Effects and Applications, Hunan Normal
	University, Changsha, Hunan 410081, P. R. China}
	
\author{Fen Long}
\affiliation{School of Mathematics and Physics, University of South China, Hengyang 421001, P. R. China}

\author{Kui Xiao}
\affiliation{School of Science, Hunan Institute of Technology, Hengyang 421002, P. R. China}

\author{Jiliang Jing}
\email[]{jljing@hunnu.edu.cn (Corresponding authors)} 
\affiliation{Department of Physics, Key Laboratory of Low Dimensional Quantum Structures and Quantum Control of Ministry of Education, and Synergetic Innovation Center for Quantum Effects and Applications, Hunan Normal
University, Changsha, Hunan 410081, P. R. China}

\begin{abstract}

In this study, we examine the impacts of black hole spin, Lorentz-violating parameter, and the scalar field's mass on quasinormal modes (QNMs) of rotating Einstein-Bumblebee black holes, including computations up to the second-order expansion in rotation parameters. 
We investigate two classes of Lorentz-violating rotating black holes: one constructed via the Newman–Janis algorithm and the other obtained by solving the field equations through a series expansion.
Within the slow-rotation approximation framework, we derive the master equations governing a massive scalar field and compute the corresponding QNM frequencies numerically using both the continued fraction method and the matrix method.
The numerical results indicate that the QNM frequencies exhibit increased sensitivity to negative $\ell$ variations, which reduces the influence of the field mass parameter $\tilde{\mu}$. 
Meanwhile, the spectral "cube" of NJA black holes shows slight compression for $m>0$ with $ \ell>0 $ and expansion for $m<0$ with $ \ell>0 $ compared to another black holes, where $m$ is approximately proportional to the spin parameter at first order, while richer structures and lifted degeneracy emerge at second order.

\end{abstract}

\maketitle
\section{Introduction}

As a fundamental symmetry, Lorentz invariance is pivotal in quantum field theory and general relativity. 
Nevertheless, evidence from unified canonical theories and high-energy cosmic ray observations \cite{Takeda:1998ps} indicates that spontaneous  Lorentz symmetry breaking (LSB) might emerge at elevated energy scales. 
Typically, effects of Lorentz violation are empirically detectable only at low energies, and these effects can be described through effective field theory \cite{Kalb:1974yc,Kostelecky:1994rn,Kostelecky1998,Wu:2022xwy,Devecioglu:2024uyi,Liu:2025lwj}.
Lately, models that modify gravitational characteristics through non-minimal coupling to a spontaneous Lorentz-violating field have attracted significant research interest, exemplified by the Einstein-Bumblebee gravity model with a vector field and the Einstein-Kalb-Ramond gravity model with an antisymmetric tensor field.
In these models, various solutions have been obtained, including but not limited to those incorporating the effects of a cosmological constant, electric charge, Tab-NUT, and global monopole, among others \cite{Casana2018,Ovgun2019,Gullu2020,Poulis:2021nqh,Maluf2021,Ding:2021iwv,Xu:2022frb,Filho:2022yrk,AraujoFilho:2024ykw,Chen:2025ypx,Yang:2023wtu,Duan:2023gng,Ding:2022qcy,Liu:2024oas,Liu:2024lve,Liu:2025fxj,Belchior:2025xam,Ding2022,Ding:2020kfr,Islam:2024sph,Afrin:2024khy}.
Our analysis primarily centers on the bumblebee model, in which the bumblebee vector field $B_{\mu}$, possessing a nonzero vacuum expectation value (VEV), induces spontaneous Lorentz symmetry breaking, resulting in a background spacetime geometry that is not fully symmetric. 
As a result, the bumblebee gravity theory reveals novel physical phenomena, rendering it essential to advancements in modern physics \cite{Liu:2022dcn,Xu:2023xqh,Zhang:2023wwk,Chen:2020qyp,Chen2020,Chen:2023cjd,Ge:2025xuy,EslamPanah:2025zcm,Liu:2024wpa,AraujoFilho:2024ctw,Liu:2025bpp,AraujoFilho:2025hkm,AraujoFilho:2024ctw,AraujoFilho:2025fwd,Shi:2025ywa,Shi:2025plr,Guo:2023nkd,Mai:2023ggs,Mai:2024lgk}.

Black hole perturbation theory is an important method for studying the physical evolution of black holes \cite{Regge:1957td,Zerilli:1970se,Zerilli:1974ai,Teukolsky:1972my,Jing:2022vks,Jing:2023okh,Long:2023vph,Long:2024axi,Wu:2023sye,Lutfuoglu:2025bsf,Lutfuoglu:2025qkt,Konoplya:2004wg,Konoplya:2013rxa,Konoplya:2004wg,Konoplya:2011qq}.
Numerous astrophysical processes of black holes can be interpreted as minor deviations from the background spacetime. 
For instance, quasinormal modes (QNMs) serve as effective characterizations of the late stages of binary black hole mergers or gravitational collapse \cite{Vishveshwara:1970zz,Berti:2009kk,Berti:2025hly}. 
To obtain the QNMs of the field in spacetime, we need derive the decoupled second-order differential equation in the frequency domain.
In spherically symmetric spacetimes, the perturbed fields can be consistently decomposed into spherical harmonics, ensuring automatic decoupling of the angular and radial components \cite{Zerilli:1970wzz,Thompson:2016fxe,Jing:2021ahx,Lenzi:2021wpc,Liu2023,Tan:2024aym,Tan:2024qij,Wu:2023spa,Liu:2025reu}. 
In contrast, such decomposition becomes invalid for rotating spacetimes. 
Historically, the Teukolsky-like equation, derived within the Newman-Penrose formalism, has been used to characterize perturbations of rotating spacetimes \cite{Teukolsky:1973ha,Teukolsky:1974yv,Jing:2023vzq,Deng:2024ayh}.
Nevertheless, decoupling perturbation equations for massive fields or modified gravity rotating spacetimes remains a significant challenge.
Pani et al. recently introduced a novel approach, suggesting that when the slowly rotating background spacetime closely resembles a spherically symmetric spacetime, the perturbation equations can be separated \cite{Pani2011,Pani2012,Pani2012prl,Pani2013IJMPA,Pani2013prd,Pani2013prl}. 
This approach builds upon Kojima's studies of perturbations in slowly rotating neutron stars \cite{Kojima}. 
In the slowly rotating limit, assuming the dimensionless rotation parameter $ \tilde{a}=a/M $ of the background spacetime is sufficiently small, the perturbation equations can be transformed into coupled ordinary differential equations by expanding the perturbation fields---whether scalar, vector, or tensor---in a harmonic basis.
This framework has inspired extensive studies of black hole perturbations, particularly investigations of QNMs in rotating spacetimes considering the effects of the cosmological constant, charge, dilaton and Chern-Simons term, etc \cite{Tattersall2018,Blazquez-Salcedo:2022eik,Brito:2018hjh,Wagle:2021tam,Feng:2024ygo,Wu:2025euf}.

This work seeks to employ the slow-rotating approximation to compute the QNM frequencies of massive scalar field perturbations associated with two types of rotating Einstein-Bumblebee black holes.
Employing numerical methods, we compute the QNM frequencies of these perturbations to examine the influence of spin, Lorentz violation, and field mass parameters on their QNM frequencies.
The manuscript is organized as follows.
In Sec. \ref{sec2}, we provide a concise overview of the Einstein-Bumblebee theory and the NJA-generated rotating spacetime. 
In Sec. \ref{sec3}, we formulate the radial Schr\"odinger-like equation governing the massive scalar field perturbations associated with second-order slowly rotating Einstein--Bumblebee black holes.
In Sec. \ref{sec4}, we outline the application of the continued fraction method and the matrix method to this spacetime, and we present the corresponding numerical findings.
Throughout our paper, we adopt the conventions $ c=G=1 $ and use them in the description of the metric whose signature is $ (-,+,+,+) $.

\section{Lorentz-violating black holes}\label{sec2}

We provide a brief introduction to the Einstein-bumblebee gravity model, an extension of General Relativity. 
The action describing the coupling of the bumblebee field $B_\mu$ to spacetime curvature is given by \cite{Casana2018}:
\begin{equation}
\begin{aligned}\label{Action}
\mathcal{S}_B=&\int d^4x \sqrt{-g}\left[\frac{1}{2\kappaup}\left(R-2\Lambda\right)+\frac{\varrho}{2\kappa} B^\mu B^\nu R_{\mu\nu}\right.\\ 
&\left.-\frac{1}{4}B^{\mu\nu}B_{\mu\nu}-V\left(B^\mu B_\mu\pm b^2\right)\right],
\end{aligned}
\end{equation}
Here, $\kappa = 8\pi G_N$ represents the gravitational coupling constant (with $G_N = 1$ in natural units), $\Lambda$ is the cosmological constant, and $\varrho$ characterizes the non-minimal curvature coupling between the bumblebee field $B_\mu$ and gravity. 
The antisymmetric field strength tensor, defined as $B_{\mu\nu} \equiv \partial_\mu B_\nu - \partial_\nu B_\mu$, governs the dynamics of the bumblebee field. 
The potential $V$ induces spontaneous LSB through a non-zero vacuum expectation value $\langle B_\mu \rangle = b_\mu$, reaching its minimum at $B_\mu B^\mu = \pm b^2$. 
Here, $b \in \mathbb{R}^+$ specifies the symmetry-breaking scale, with the $\pm$ sign indicating timelike ($+$) or spacelike ($-$) configurations of $B_\mu$.

Based on the action given in Eq. \meq{Action}, the field equations for the gravity sector are derived as follows:
\begin{align}
R_{\mu\nu}-\frac{1}{2}g_{\mu\nu}\left(R-2\Lambda\right)=\kappa T^{B}_{\mu\nu}.
\end{align}
The bumblebee energy-momentum tensor $T^{B}_{\mu\nu}$ is defined by the following expression:
\begin{equation}\label{TBab}
T^{B}_{\mu\nu}=B_{\mu\alpha}B^\alpha_\nu-\frac{1}{4}g_{\mu\nu}B^{\alpha\beta}B_{\alpha\beta}-g_{\mu\nu}V+2B_\mu B_\nu V'
+\frac{\varrho}{2\kappa}\mathcal{B}_{\mu\nu},
\end{equation}
with
\begin{equation}
\begin{aligned}
\mathcal{B}_{\mu\nu}=&
g_{\mu\nu}B^{\alpha}B^{\beta}R_{\alpha\beta}-4B_{(\mu} B^\alpha R_{\nu) \alpha}+\nabla_\alpha\nabla_\mu\left(B^\alpha B_\nu\right)
\\&+\nabla_\alpha\nabla_\nu\left(B^\alpha B_\mu\right)
-\nabla^2\left(B_\mu B_\nu\right)-g_{\mu\nu}\nabla_\alpha\nabla_\beta\left(B^\alpha B^\beta\right),
\end{aligned}\
\end{equation}
The potential derivative $ V' $ is given by:
\begin{align}
V'=\frac{\partial V(x)}{\partial x}\Big|_{x=B^\mu B_\mu\pm b^2}.
\end{align}
The field equations governing the bumblebee field are expressed as:
\begin{align} 
\nabla^\mu B_{\mu\nu}=2V'B_\nu-\frac{\varrho}{\kappa}B^\mu R_{\mu\nu},
\end{align}
Henceforth, we assume the bumblebee field is set to its vacuum expectation value, given by $ B^\mu=b^\mu $.

When the bumblebee field $ B_\mu $ is fixed at its vacuum expectation value $ b_\mu $, the spontaneous breaking of Lorentz symmetry generates a vacuum solution. 
The bumblebee field thus takes the form $ B_\mu=b_\mu $, resulting in $ V=V'=0 $. 
In these conditions, and without considering the influence of the cosmological constant, a spherically symmetric solution that violates Lorentz symmetry is obtained by Casana et al \cite{Casana2018}, as
\begin{equation}\label{Casana}
ds^2=-F(r)dt^2+\frac{(1+\ell)}{F(r)}dr^2+r^2d\theta^2+r^2\sin^2\theta d\varphi^2,
\end{equation}
where $ F(r)=1-2M/r $, consistent with the Schwarzschild case, and the LSB parameter is defined by $ \ell=\varrho b^2 $.

The metric describes a purely radial Lorentz-violating solution outside a spherical body, corresponding to a modified Schwarzschild solution.
Numerous efforts have been made to identify a Kerr-like Lorentz-violating black hole, and recently, S. G. Ghosh et al. \cite{Islam:2024sph} successfully derived such a solution using a modified Newman-Janis algorithm \cite{Azreg-Ainou:2014pra}. 
The original Newman-Janis algorith \cite{Newman:1965tw} offers an effective approach for generating rotating spacetimes from a static, spherically symmetric metric without solving field equations.
Reviewing the study of S. G. Ghosh \cite{Islam:2024sph}, by implementing a transformation, namely $ T \rightarrow t\sqrt{1+\ell} $, it find that the metric \meq{Casana} is transformed into a Schwarzschild-like solution, given by:
\begin{equation}
ds^2=-\frac{F(r)}{(1+\ell)}dT^2+\frac{(1+\ell)}{F(r)}dr^2+r^2d\theta^2+r^2\sin^2\theta d\varphi^2.
\end{equation}
By employing NJA, S. G. Ghosh et al. derived a rotating spacetime, expressed as:
\begin{equation}\label{Ghosh}
\begin{aligned}
ds^2=&-\left(1-\frac{2\mathcal{M}}{\Sigma}\right)dT^2+\frac{\Sigma}{\bar{\Delta}}dr^2+\Sigma d\theta^2\\
&-\frac{4a\mathcal{M}r}{\Sigma}\sin^2\theta dtd\varphi+\frac{\mathcal{A}\sin^2\theta}{\Sigma}d\varphi^2,
\end{aligned}
\end{equation}
where
\begin{equation}
\begin{aligned}
&\bar{\Delta}=r^2+a^2-2\mathcal{M}r,&& \Sigma=r^2+a^2\cos^2\theta,\\
&\mathcal{M}=\frac{M(1+\frac{r\ell}{2M})}{1+\ell},&& \mathcal{A}=\left(r^2+a^2\right)^2-a^2\Delta \sin^2\theta.
\end{aligned}
\end{equation}
The black hole's mass is represented by $ M $, the LSB parameter by $ \ell $, and the spin parameter by $ a $. 
A non-zero value of $ \ell $ causes a deviation from the Kerr spacetime, indicating a violation of Lorentz symmetry.

When $ a=0 $, the spacetime fails to revert to Casana's initial Lorentz-violating black hole in this gravitational model; hence, a coordinate transformation is applied to the metric \meq{Ghosh} to facilitate comparison of its properties with the static model. 
To achieve this, we perform the coordinate transformation $ t = T/\sqrt{1+\ell} $, resulting in a Kerr-like Lorentz-violating black hole spacetime, given by:
\begin{equation}\label{NJA2}
\begin{aligned}
\! ds^2\!=&ds^2_\text{Kerr}\!+\!\frac{\ell r^2}{\Sigma}\left(dt+\frac{a\sin^2\theta}{\sqrt{1+\ell}}d\varphi\right)^2\!
+\frac{\ell\Sigma(r^2-2Mr)}{\Delta^2+\ell a^2\Delta}dr^2\\
&-\ell dt^2+\frac{\left(\lambda_\ell dt-a\sin^2\theta d\varphi\right)}{(1+\ell)\Sigma}2\ell Mr a\sin^2\theta d\varphi,
\end{aligned}
\end{equation}
where $ ds^2_\text{Kerr} $ represents the Kerr metric, and
\begin{equation}
\begin{aligned}
&\Delta=r^2+a^2-2Mr,\\
&\lambda_\ell=2+2/\ell-2\sqrt{1+\ell}(1/\ell+r/M).
\end{aligned}
\end{equation}
Since the coordinate transformation we adopt does not affect spatial coordinates, the black hole horizon of this spacetime remains consistent with the Ref. \cite{Islam:2024sph}, as given by:
\begin{equation}
r_{\pm}=M\pm\sqrt{M^2-a^2(1+\ell)}.
\end{equation}

\section{MASSIVE SCALAR PERTURBATIONS OF SLOWLY ROTATING Lorentz-violating BLACK HOLES}\label{sec3}

By leveraging the dynamical characteristics of scalar fields, we can assess the plausibility of the Lorentz-violating rotating spacetime produced by NJA, comparing it to the established second-order slow-rotation approximation solution to approximately ascertain whether its properties fall within acceptable limits \cite{Liu:2022dcn}.
For this we consider the dimensionless rotational parameter $ \tilde{a}=a/M $.
In what follows we will expand on the metric and all other relevant quantities to second order in $ \tilde{a} $.
At this order, the event horizon $ r_+ $ and the Cauchy horizon $ r_- $ can be expressed as follows:
\begin{equation}
r_+=2M-(1+\ell)\frac{\tilde{a}^2M}{2},\quad\quad r_-=(1+\ell)\frac{\tilde{a}^2M}{2},
\end{equation}
which are consistent with the results for second-order slow-rotation Lorentz-violating black holes\footnote{In the following text, we will simply refer to it as LV-Ding BHs and the metric \meq{NJA2} obtained by NJA as LV-NJA BHs.} reported by Ding et al. \cite{Ding2020,Liu:2022dcn}.

Considering that the scalar field coupling to the bumblebee field is neglected, then the massive Klein-Gordon equation reads
\begin{equation}
\frac{1}{\sqrt{-g}}\pp_\mu \left( \sqrt{-g}g^{\mu\nu}\pp_\nu \phi \right)=\mu^2\phi,
\end{equation}
where $ m_s=\mu \hbar $ represents the scalar field's mass.
We express the field as a sum of spherical harmonics:
\begin{equation}
\phi=\sum_{lm}\frac{\psi_l(r)}{\sqrt{r^2+\tilde{a}^2M^2}}e^{-i\omega t}Y^{lm}(\theta,\varphi),
\end{equation}
and perform a second-order expansion of $ \tilde{a} $.
Subsequently, by utilizing the eigenvalue equation for the angular part:
\begin{equation}
\left[\frac{1}{\sin\theta}\frac{\pp}{\pp\theta}\left(\sin\theta \frac{\pp}{\pp \theta} \right)
+\frac{1}{\sin^2\theta}\frac{\pp^2}{\pp \varphi^2}\right]Y^{lm}=l(l+1)Y^{lm},
\end{equation}
we can obtain the following expression:
\begin{equation}\label{Meq}
A_lY^l+D_l\cos^2\theta Y^l=0,
\end{equation}
where a sum over $(l,m)$ is implicit.
For an axially symmetric background, perturbations corresponding to different values of $m$ are decoupled; we therefore ignore the index $m$ in the subsequent discussions, where the detailed expressions for $A_l$ and $D_l$ are provided below:
\begin{widetext}
\begin{equation}\label{EQAAA}
\begin{aligned}
A_l=&\bigg[-\frac{l(l+1)}{r^3}-\frac{2M}{r^2(1+\ell)}+\frac{2M\mu^2-r(\mu^2-\omega^2)}{r(r-2M)}
-\frac{2m M \tilde{a}\omega(2M+r\ell) }{r^3(r-2M)\sqrt{1+\ell}}\bigg]\psi_l(r)
+M^2\tilde{a}^2\bigg[\frac{l(l+1)}{2r^5}\\
&+\frac{(1+\ell)m^2}{r^4(r-2M)}-\frac{(1-2\ell)r-9M}{r^6(1+\ell)}+\frac{\mu^2}{2r^3}
-\frac{8M^2-2Mr(1-3\ell)+r^2(1+\ell+2\ell^2)}{2(1+\ell)r^3(r-2M)^2}\omega^2 \bigg]\psi_l(r)\\
&+\bigg[\frac{2Mr^2-(5M+2r\ell)M^2\tilde{a}^2}{r^5(1+\ell)} \bigg]\psi'_{l}(r)
+\bigg[\frac{r-2M}{r^2(1+\ell)}+\frac{M^2\tilde{a}^2(2M+r+r\ell)}{2r^4(1+\ell)} \bigg]\psi''_{l}(r),
\end{aligned}
\end{equation}
\begin{align}\label{EQDDD}
D_l=&M^2\tilde{a}^2\left[\frac{l(l+1)}{r^5}+\frac{2M}{r^6(1+\ell)}-\frac{(2M+r\ell)\omega^2}{r^3(r-2M)(1+\ell)} \right]\psi_l(r)
-\frac{2M^3\tilde{a}^2}{r^5(1+\ell)}\psi'_{l}(r)-\frac{(r-2M)M^2\tilde{a}^2}{r^4(1+\ell)}\psi''_l(r).
\end{align}
\end{widetext}

It is significant that $ D_l $ scales with $ \tilde{a}^2 $, rendering the second term in the previous equation zero to the first order in $ \tilde{a} $.

First, we consider a first-order expansion with respect to the rotation, where the scalar equation is already decoupled, and it can be cast in the form:
\begin{equation}
\left[\frac{d^2}{dr^2_*}+(1+\ell)\omega^2-V^{(0)}_l-V^{(1)}_l\right]\psi_l=0,
\end{equation}
where $ dr/dr_*=F $ and $ F=1-2M/r $, and $ V^{(0)}_l $ and $ v^{(1)}_l $ are respectively the zeroth-order and first-order terms of the effective potential with respect to the rotation parameter $ \tilde{a} $:
\begin{align}\label{V00}
V^{(0)}_l=&F\left[\frac{2M}{r^3}+\frac{(1+\ell)l(l+1)}{r^2}+(1+\ell)\mu^2\right],\\ \label{V11}
V^{(1)}_l=&\frac{2mM^2\sqrt{1+\ell}(2M+r\ell)\tilde{a}\omega}{r^3}.
\end{align}
Here, $ V^{(0)}_l $ ensures that when the rotation parameter is zero, the scalar field's dynamical properties in spacetime are consistent with the static spacetime given by Casana \cite{Casana2018}. 
However, $ V^{(1)}_l $ differs from the first-order term in equation (31) of Ref. \cite{Liu:2022dcn}, which reflects that the corrections due to rotation differ between LV-NJA BHs and LV-Ding BHs.
The coupling between perturbations with indices $ l\pm1 $ disappears due to a straightforward reason. 
Klein-Gordon perturbations, being polar quantities, are expected to couple with axial perturbations of indices $ l\pm1 $ at first order, as dictated by a selection rule analogous to the Laporte rule \cite{Pani2013IJMPA}. 
However, such axial perturbations are absent in the scalar case.

After that, let's shift our focus to the second-order slow-rotation approximation.
For this, perturbations with harmonic index $ l $ couple to those of the same parity with indices $ l\pm2 $. 
However, this coupling does not affect the eigenfrequencies, as discussed in \cite{Pani2012}.
Therefore, for given values of $ l $ and $ m $, we must solve a single scalar equation, expressed schematically as:
\begin{equation}
\mathcal{P}_l+\tilde{a}m\bar{\mathcal{P}}_l+\tilde{a}^2\hat{\mathcal{P}}_l=0.
\end{equation}
To express the perturbation equation in the form above, we separate the angular part of equation \meq{Meq}.
The angular part is separated by applying the Kojima identities \cite{Kojima}, expressed as:
\begin{equation}\label{Kojimaid}
\begin{aligned}
\cos\theta Y^{l}=&\mathcal{Q}_{l+1}Y^{l+1}+\mathcal{Q}_{l}Y^{l-1}, \\
\sin\theta \pp_\theta Y^{l}=&\mathcal{Q}_{l+1}lY^{l+1}-\mathcal{Q}_{l}(l+1)Y^{l-1}, \\
\cos^2\theta Y^{l}=&\left(\mathcal{Q}^2_{l+1}+\mathcal{Q}^2_{l}\right)Y^{l} \\
& +\mathcal{Q}_{l+1}\mathcal{Q}_{l+2}Y^{l+2} +\mathcal{Q}_{l}\mathcal{Q}_{l-1}Y^{l-2},\\
\cos\theta \sin\theta \pp_\theta Y^{l}=&\left[l\mathcal{Q}^2_{l+1}-(l+1)\mathcal{Q}^2_{l}\right]Y^{l}\\
&+\mathcal{Q}_{l+1}\mathcal{Q}_{l+2}lY^{l+2}\\
&-\mathcal{Q}_{l}\mathcal{Q}_{l-1}(l+1)Y^{l-2}.
\end{aligned}
\end{equation}
The coefficients $\mathcal{Q}_{l}$, defined as
\begin{align}
\mathcal{Q}_{l} = \sqrt{\frac{l^2-m^2}{4l^2-1}},
\end{align}
are associated with the Clebsch-Gordan coefficients. 
Additionally, they relate to the orthogonality property of scalar spherical harmonics, given by:
\begin{equation}
\int Y^l Y^{*l'} d\Omega = \delta^{ll'}.
\end{equation}
The resulting equation is given schematically as:
\begin{equation}\label{EQAQD}
\begin{aligned}
A_l+\left(\mathcal{Q}^2_{l+1}+\mathcal{Q}^2_l \right)D_l+\mathcal{Q}_{l-1}\mathcal{Q}_lD_{l-2}
+\mathcal{Q}_{l+2}\mathcal{Q}_{l+1}D_{l+2}=0.
\end{aligned}
\end{equation}
Indeed, by iteratively applying the identity in Eq. \ref{Kojimaid}, the perturbation equations can be separated for any order of $ \tilde{a} $.

By employing the explicit coefficients from Eqs. \meq{EQAAA}-\meq{EQDDD}, the scalar field equation \meq{EQAQD} are expressed schematically as:
\begin{equation}\label{EQUWV}
\begin{aligned}
&\!\!\!\frac{d^2\psi_l}{dx^2}+\left[(1+\ell)\omega^2-\left(V^{(0)}_l+V^{(1)}_l+V^{(2)}_l\right) \right]\psi_l+\tilde{a}^2\times\\
&\!\!\!\bigg[ U_{l+2}\psi_{l+2}+U_{l-2}\psi_{l-2}
+W_{l+2}\frac{d^2\psi_{l+2}}{dx^2}+W_{l-2}\frac{d^2\psi_{l-2}}{dx^2} \bigg]=0,
\end{aligned}
\end{equation}
where the tortoise coordinate is defined by $ dr/dx \equiv f $, with
\begin{equation}
f = \frac{(1+\ell)}{r^2+\tilde{a}^2M^2}\left(\frac{r^2-2Mr}{1+\ell}+\tilde{a}^2M^2\right),
\end{equation}
expanded to second order of spin $ \tilde{a} $. 
The explicit expressions for $ U_{l\pm2} $ and $ W_{l\pm2} $ are not needed here, while $ V^{(2)}_l $, representing the second-order correction to the effective potential, has its explicit form presented below.
Note that the coupling to the $ l\pm2 $ terms is proportional to $ \tilde{a}^2 $.

For a calculation accurate to second order in $ \tilde{a} $, the terms in parenthesis can be evaluated at zeroth order, and therefore the functions $ \psi^{(0)}_{l\pm 2} $ must be solutions of
\begin{equation}
\frac{d^2\psi^{(0)}_{l\pm2}}{dx^2}+\left[(1+\ell)\omega^2-V^{(0)}_{l\pm2} \right]\psi^{(0)}_{l\pm2}=0.
\end{equation}
After substituting these relations into equation \meq{EQUWV} and expanding to second order in spin $ \tilde{a} $, the field equations are simplified to:
\begin{equation}\label{MEQQQ}
\begin{aligned}
&\frac{d^2\psi_l}{dx^2} + \left[(1+\ell)\omega^2 - \mathcal{V}_l \right]\psi_l 
=\frac{\tilde{a}^2M^2F}{r^2}  \\
&\times\left[(1+\ell)\mu^2-\omega^2\right] \left[\mathcal{Q}_{l+1}\mathcal{Q}_{l+2}\psi^{(0)}_{l+2}+\mathcal{Q}_{l-1}\mathcal{Q}_l\psi^{(0)}_{l-2} \right].
\end{aligned}
\end{equation}
The potential $ \mathcal{V}_l $ is given by
\begin{equation}
\mathcal{V}_l=V^{(0)}_l + V^{(1)}_l + V^{(2)}_l,
\end{equation}
where $V^{(0)}_l$ and $V^{(1)}_l$ are specified in Eqs. \meq{V00} and \meq{V11}, respectively, and the explicit form of $V^{(2)}_l$ is provided below as
\begin{equation}
\begin{aligned}
\!\!\! V^{(2)}_l=
&\frac{\tilde{a}^2M^2}{r^2(1+\ell)^{-2}}\bigg[\frac{l(l+1)}{r^2}+\mu^2\!-\!\frac{m^2}{r^2}\!-\!\frac{l(l+1)+r^2\mu^2}{r^2(1+\ell)F^{-1}/2}\bigg]\\
&+\!\frac{\tilde{a}^2M^2}{r^2F^{-1}}\bigg[\Lambda_\ell+\omega^2+\left((1+\ell)\mu^2\!-\!\omega^2\right)\left(\mathcal{Q}^2_l+\mathcal{Q}^2_{l+1}\right) \bigg],\\
\end{aligned}
\end{equation}
with
\begin{equation}
\Lambda_\ell=\frac{24M^2+r^2(1-2\ell)+6rM(\ell-2)}{r^3(r-2M)}.
\end{equation}
This potential coincides with Pani's result when $ \ell\rightarrow 0 $. 

As discussed by Pani et al.\cite{Pani2012}, the couplings to terms with indices $ l\pm2 $ can be neglected in the calculation of the modes. 
In the Bumblebee model case, to obtain the single variable equation for $ \psi_l $, this can be shown explicitly by defining the master variables:
\begin{equation}\label{mastervar}
Z_l=\psi_l+\tilde{a}^2M^2\left[c_l\psi_{l-2}-c_{l+2}\psi_{l+2}\right],
\end{equation}
where $ c_l $ read as
\begin{equation}
c_l=\frac{(1+\ell)\mu^2-\omega^2}{2(1+\ell)(2l-1)}\mathcal{Q}_{l-1}\mathcal{Q}_l.
\end{equation}
Then, the master equation for scalar perturbation is given by
\begin{equation}\label{mastereq}
\frac{d^2}{dx^2}Z_l+\left[(1+\ell)\omega^2-\mathcal{V}_l \right]Z_l=0.
\end{equation}
The structure of the above equation is consistent with that of slow-rotating LV-Ding black holes, differing only in the rotational correction of the effective potential. 
Although both are based on the Bumblebee model, it is worth noting that the master variable \meq{mastervar} applies only to LV-NJA black holes, whose structure differs from that described by equation (39) in Ref. \cite{Liu:2022dcn}, which pertains to LV-Ding black holes.

For the second-order term in the wave equation \meq{mastereq}, its essence is
\begin{equation}
\frac{d^2}{dx^2}Z_l=\left(f^2\frac{d^2}{dr^2}+f'f\frac{d}{dr}\right)Z_l,
\end{equation}
where
\begin{equation}
f=1-\frac{2M}{r}+\frac{\ell\tilde{a}^2M^2}{r^2}+\frac{2\tilde{a}^2M^3}{r^3}+\mathcal{O}(\tilde{a}^3),
\end{equation}
For the purpose of computing the eigenfrequencies, it is desirable to represent $f$ in terms of $r_{\pm}$. 
However, as the $r^{-3}$ term in $f$ leads to three singularities, it is not feasible to represent $f$ in terms of $r_{\pm}$.
In this regard, we can define
\begin{equation}
\bar{f}=\left(1+\tilde{a}^2M^2/r^2\right)f=1-\frac{2M}{r}+\frac{(1+\ell)\tilde{a}^2M^2}{r^2}+\mathcal{O}(\tilde{a}^3).
\end{equation}
The above equation can be rewritten as
\begin{equation}
\bar{f}=\left(1-\frac{r_+}{r}\right)\left(1-\frac{r_-}{r}\right).
\end{equation}
It is desirable to substitute $\bar{f}$ for $f$ in the master equation. 
For this, a functional transformation is applied to the variable $\Psi_l(r)$,
\begin{equation}\label{Eqnn}
\Psi_l(r)=\frac{1}{n(r)}Z_l(r),
\end{equation}
allowing the second-order term in the master equation to be expressed as:
\begin{equation}\label{EQnz}
\frac{\bar{f}^2}{nf^2}\frac{d^2}{dx^2}Z_l=
\bar{f}^2\frac{d^2}{dr^2}\Psi_l+\frac{\bar{f}^2\left(nf'+2fn' \right)}{nf}\frac{d}{dr}\Psi_l+C_0\Psi_l.
\end{equation}
Here, the specific form of $C_0$ is unnecessary. 
The goal is to eliminate the first-order term with the new tortoise coordinate, requiring the following equation to be satisfied:
\begin{equation}
\bar{f}\left(nf'+2fn' \right)-n\bar{f}'f=0.
\end{equation}
By expanding the ordinary differential equation for $n$ to second order in spin $a$, the equation becomes solvable, yielding the solution:
\begin{equation}\label{Eqnr}
n(r)=\exp\left[A_1(r)+A_2(r)+A_3(r)+A_4(r) \right],
\end{equation}
where
\begin{equation}
\begin{aligned}
\! A_1&=-\frac{M}{r(1+\ell)},\\
\! A_2&=\frac{\ln(r)}{(1+\ell)^2\tilde{a}^2}-\frac{\ln(r)}{2(1+\ell)},\\
\! A_3&=\frac{[(1+\ell)\tilde{a}^2-2]}{4(1+\ell)^2\tilde{a}^2}\ln\left[\Delta+(1+2\ell)\tilde{a}^2M^2 \right],\\
\! A_4&=\frac{2-3(1+\ell)\tilde{a}^2}{2(1+\ell)^2\tilde{a}^2\sqrt{2(1+\ell)\tilde{a}^2-1}}
\arctan[\tfrac{r-M}{M\sqrt{2(1+\ell)\tilde{a}^2-1}} ].
\end{aligned}
\end{equation}

Substituting Eqs. \meq{Eqnn}, \meq{EQnz}, and \meq{Eqnr} into \meq{mastereq} yields the final master equation:
\begin{equation}\label{eqNJA}
\frac{d^2}{dx_*^2}\Psi_l+\left[(1+\ell)\omega^2-\mathcal{V}^\text{NJA}_l \right]\Psi_l=0,
\end{equation}
where the tortoise coordinate is given by
\begin{equation}
 x_*=\int\bar{f}^{-1}dr=r+\frac{r_+^2\ln\left(r-r_+\right)}{r_+-r_-}-\frac{r_-^2\ln\left(r-r_-\right)}{r_+-r_-},
\end{equation}
and the effective potential is give by
\begin{equation}\label{VVVNJA}
\begin{aligned}
\!\!\!\mathcal{V}_l^\text{NJA}\!\!=&V^{(0)}_l\!
+\!\frac{2mM\tilde{a}\omega\sqrt{1+\ell}(2M+r\ell)}{r^3}+\frac{(1+\ell)^2\tilde{a}^2M^2}{r^4}\\
&\times\bigg[l^2+l-m^2+r^2\mu^2-\frac{2(r-3M+r^3\omega^2)}{r(1+\ell)} \bigg]\\
&+\frac{\tilde{a}^2M^2F}{r^2}\left[\omega^2+\left((1+\ell)\mu^2-\omega^2\right)\left(\mathcal{Q}^2_l+Q^2_{l+1}\right) \right],
\end{aligned}
\end{equation}
\begin{equation}\label{VVVDing}
\begin{aligned}
\!\!\!\mathcal{V}_l^\text{Ding}=&V^{(0)}_l
+\frac{4mM^2\tilde{a}\omega(1+\ell)^{3/2}}{r^3}+\frac{(1+\ell)^2\tilde{a}^2M^2}{r^4}\\
&\times\bigg[l^2+l-m^2+r^2\mu^2-\frac{2(r-3M+r^3\omega^2)}{r(1+\ell)} \bigg]\\
&+\frac{\tilde{a}^2M^2F}{r^2}\left[\omega^2+(1+\ell)^2(\mu^2-\omega^2)\left(\mathcal{Q}^2_l+Q^2_{l+1}\right) \right]\\
&-\frac{\tilde{a}^2M^2\omega^2\ell\left[r\ell+2M(2+\ell)\right]}{r^3}.
\end{aligned}
\end{equation}
The function transformation affects solely the second-order rotation term in the effective potential. 
Thus, we can utilize a conventional numerical approach to address equation \meq{eqNJA}. 
Notably, equation \meq{VVVDing} describes the effective potential of LV-Ding black holes within the slow-rotation approximation. 
The primary equation differs from equation (31) in Ref. \cite{Liu:2022dcn}, as we applied the same function transformation to align its form with \meq{eqNJA} in this paper, facilitating the calculation of eigenfrequencies and comparison with LV-NJA BHs.

\section{The Eigenvalue Problem For Quasinormal Modes}\label{sec4}
\subsection{Boundary conditions}

In this study, we investigate how the Lorentz-violating parameter $\ell$ influences the QNMs. 
In the study by Liu et al. \cite{Liu:2022dcn}, the QNMs were examined in the massless limit. 
Therefore, we focus on evaluating the impact of field mass and comparing the effects of the rotation parameter on two distinct spacetimes.
The perturbation equations \meq{eqNJA}, together with appropriate boundary conditions at the horizon and at infinity, form an eigenvalue problem for the frequency spectrum.
Thus, for our asymptotically flat spacetime, the quasi-normal frequencies (QNFs), denoted by $\omega$, are derived by solving the boundary condition
\begin{equation}
\Psi_l\sim
\begin{cases}
e^{-i\sqrt{1+\ell}(\omega-m\Omega_H)}x_* &\text{for}~~x_*\rightarrow~-\infty ,\\
e^{i\sqrt{\omega^2-\mu^2} x_*} &\text{for}~~x_*\rightarrow~+\infty,
\end{cases}
\end{equation}
which means that the perturbation field is ingoing at the event horizon and outgoing at infinity.
Here, the horizon angular frequency $ \Omega_H=\sqrt{1+\ell}\tilde{a}/(4M) $ and expanded it to second order.
We specify an ansatz for the radial function $ \Psi_l $ that respects the physical boundary conditions.
The ansatz takes the form
\begin{align}\label{psia2}
\Psi_l(r)= & e^{-\sqrt{1+\ell}qr}(r-r_-)^{\sqrt{1+\ell}\chi}\left(\frac{r-r_+}{r-r_-}\right)^{-i\sqrt{1+\ell}\rho}R(r),
\end{align}
where 
\begin{align}
q&=\pm\sqrt{\mu^2-\omega^2},\\
\chi&=\frac{M(2\omega^2-\mu^2)}{q},\\
\rho&=\frac{2Mr_+\left(\omega-m\Omega_H\right)}{r_+-r_-}.
\end{align}
The parameter $q$ in the exponential term depends on the boundary condition applied at infinity, with $Re(q)>0$ for QNMs and $Re(q)<0$ for quasibound states.
In subsequent subsections, we will analyze QNMs via the continued fraction method and the matrix method, respectively.

\subsection{Continued fraction method}
Following Leaver's pioneering work \cite{Leaver:1985ax}, the continued fraction method (CFM) has emerged as a highly accurate approach for determining the QNMs.
Within this framework, the eigenfunction is represented as a series with coefficients satisfying a finite-term recurrence relation.
By substituting the ansatz \meq{psia2} into equations \meq{eqNJA}, we obtain the solutions $ R_l(r) $, expanded at the event horizon, as:
\begin{align}
R_l=\sum_{n=0}^{\infty}d_n\left(\frac{r-r_+}{r-r_-} \right)^n.
\end{align}
Subsequently, equation \eqref{eqNJA}, which governs the scalar fields of Lv-NJA and Lv-Ding black holes, enables the definition of expansion coefficients through a six-term recurrence relation in both spacetimes.
These relations are presented in schematic form as
\begin{equation}\label{grav0-4}
\begin{aligned}
d_1=&{\mathcal{C}}_{1,0}~d_0,\\
d_2=&{\mathcal{C}}_{2,0}~d_0+{\mathcal{C}}_{2,1}~d_1,\\
d_3=&{\mathcal{C}}_{3,0}~d_0+{\mathcal{C}}_{3,1}~d_1+{\mathcal{C}}_{3,2}~d_2,\\
d_4=&{\mathcal{C}}_{4,0}~d_0+{\mathcal{C}}_{4,1}~d_1+{\mathcal{C}}_{4,2}~d_2+{\mathcal{C}}_{4,3}~d_3,
\end{aligned}
\end{equation}
with
\begin{equation}
\begin{aligned}\label{6oder}
&d_{n+1}\alpha_n+d_{n}\beta_n+d_{n-1}\gamma_n \\
&+d_{n-2}\sigma_n+d_{n-3}\tau_n+d_{n-4}\delta_n=0,&&n=4,5,6\cdots
\end{aligned}
\end{equation}
where ${\mathcal{C}}_{i,j}$, as well as $ \alpha_n $, $ \beta_n $, $ \gamma_n $, $ \sigma_n $, $ \tau_n $ and $ \delta_n $ are functions consisting of $ l,m, \ell, \tilde{a}, M\mu,M\omega $ and $ n $, whose explicit form we do not present here for brevity.
By applying Gaussian elimination \cite{Leaver:1990zz,Percival:2020skc,Guo:2022rms}, the n-term recurrence relation is reduced to a three-term recurrence relation, expressed as:  
\begin{align}
&\tilde{d}_1 \tilde{\alpha}_0+\tilde{d}_0\tilde{\beta}_0=0,\\
&\tilde{\alpha}_n\tilde{d}_{n+1}+\tilde{\beta}_n\tilde{d}_n+\tilde{\gamma}_n\tilde{d}_{n-1}=0.
\end{align}
With this three-term recurrence relation, the ratio of successive $ \tilde{d}_n $ can be formulated in two ways, one is given by infinite continued fraction as
\begin{equation}
\frac{\tilde{d}_{n+1}}{\tilde{d}_n}=\frac{-\tilde{\gamma}_{n+1}}{\tilde{\beta}_{n+1}-\frac{\tilde{\alpha}_{n+1}\tilde{\gamma}_{n+2}}
{\tilde{\beta}_{n+2}-\frac{\tilde{\alpha}_{n+2}\tilde{\gamma}_{n+3}}{\tilde{\beta}_{n+3}-...}}},
\end{equation}
Such a continued fraction is usually denoted by the following notation:
\begin{equation}\label{Leaver222}
\frac{\tilde{d}_{n+1}}{\tilde{d}_n}=\frac{-\tilde{\gamma}_{n+1}}{\tilde{\beta}_{n+1}-}
\frac{\tilde{\alpha}_{n+1}\tilde{\gamma}_{n+2}}{\tilde{\beta}_{n+2}-}
\frac{\tilde{\alpha}_{n+2}\tilde{\gamma}_{n+3}}{\tilde{\beta}_{n+3}-}\dots
\end{equation}
Equation \eqref{Leaver222} can be regarded as the boundary condition for $\tilde{d}_n$ when $n \to \infty$. 
If we set $n=0$, another boundary condition is obtained:
\begin{equation}
\frac{\tilde{d}_1}{\tilde{d}_0}=\frac{-\tilde{\gamma}_1}{\tilde{\beta}_1-}
\frac{\tilde{\alpha}_1\tilde{\gamma}_2}{\tilde{\beta}_2-}\frac{\tilde{\alpha}_2\tilde{\gamma}_3}{\tilde{\beta}_3-}\dots
\end{equation}
Based on $\tilde{d}_1 \tilde{\alpha}_0 + \tilde{d}_0 \tilde{\beta}_0 = 0$, we have:
\begin{equation}
\frac{\tilde{d}_1}{\tilde{d}_0} = -\frac{\tilde{\beta}_0}{\tilde{\alpha}_0}.
\end{equation}
Clearly, the two expressions are equivalent, leading to the characteristic equation for the QNM frequency:
\begin{equation}
0 = \tilde{\beta}_0 - \frac{\tilde{\alpha}_0 \tilde{\gamma}_1}{\tilde{\beta}_1 -}
\frac{\tilde{\alpha}_1 \tilde{\gamma}_2}{\tilde{\beta}_2 -} \frac{\tilde{\alpha}_2 \tilde{\gamma}_3}{\tilde{\beta}_3 -} \dots
\end{equation}
Therefore, the quasinormal frequencies are determined by solving the algebraic equation \eqref{6oder} for sufficiently large $ n $.

\subsection{Matrix method}

To verify the accuracy of the continued fraction method in computing the QNMs of a massive scalar field in the Lv BHs spacetime, the matrix method (MM) is employed for comparison. 
This numerical method was developed by Lin and his collaborators \cite{Lin:2016sch,Lin:2017oag,Lin:2019mmf,Lin:2022ynv,Shen:2022xdp,Lei2021,Liu:2023MOG,Liu:2024oeq}. 
Compared to the continued fraction method, it imposes less stringent conditions on the trial solution, as it only requires the boundary conditions to be satisfied in order to obtain sufficiently accurate results.
Incorporating the boundary conditions specified in the prior subsection, we apply a transformation of coordinates:
\begin{align}\label{transyr}
y=\frac{r-r_+}{r-r_-},
\end{align}
thereby defining the QNMs computation domain as $ y \in [0,1] $. 
With the boundary conditions in mind, we propose that the wave function takes the form:
\begin{align}\label{transchipsi}
\chi(y)=y(1-y)\psi_l(y),
\end{align}
resulting in the conditions at the event horizon and spatial infinity:
\begin{align}\label{chi01}
\chi(0)=\chi(1)=0.
\end{align}
These conditions ensure the homogeneity of the subsequent matrix equation.
It can be shown that all equations governing perturbation fields may be expressed as:
\begin{align}\label{qicieq}
\tilde{\mathcal{C}}_2(y,\omega)\chi''(y)+\tilde{\mathcal{C}}_1(y,\omega)\chi'(y)+\tilde{\mathcal{C}}_0(y,\omega)\chi(y)=0,
\end{align}
with the coefficients $\tilde{\mathcal{C}}_j(y,\omega)$ determined by plugging the wave function's properties into the relevant field equations. 
For example, by combining equations \eqref{transyr} and \eqref{transchipsi}, one may insert \eqref{psia2} into \eqref{eqNJA} to yield an equation consistent with the structure of \eqref{qicieq}. 
It should be noted that each $\tilde{\mathcal{C}}_j$ (where $j=0,1,2$) depends linearly on $\omega$, permitting the breakdown: $\tilde{\mathcal{C}}_j(y,\omega)=\tilde{\mathcal{C}}_{j,0}(y)+\omega \tilde{\mathcal{C}}_{j,1}(y)$. 
Employing the matrix technique to discretize \eqref{qicieq}, we define evenly distributed grid points across the interval $[0,1]$.
By performing a Taylor series expansion of the function $\chi(y)$ about each grid point, we construct the associated differential matrices. 
As a result, equation \eqref{qicieq} is recast into a matrix algebraic form:
\begin{align}
\left(\mathcal{M}_0+\omega\mathcal{M}_1\right)\chi(y)=0,
\end{align}
where $\mathcal{M}_0$ and $\mathcal{M}_1$ are $N\times N$ matrices constructed from the functions $\tilde{\mathcal{C}}_j$ and their derivatives, with their dimensions determined by the number of grid points adopted.
Once these matrices are evaluated as outlined in \cite{Lin:2016sch,Lin:2017oag,Lin:2019mmf}, finding the QNMs reduces to a straightforward algebraic task.

\subsection{Numerical results}
Using the continued fraction method and the matrix method, we numerically calculated QNMs and show the result in this subsection.
Without loss of generality, we set $ M=1 $.
For massive scalar perturbations in slowly rotating Lv black hole spacetimes, the continued fraction method yields results with a convergence of $10^{-5}$ at the 12th order ($ n=12 $), while the matrix method achieves the same accuracy by setting $N=15$.

\subsubsection{Error analysis between the two methods}
Prior to presenting the QNM frequencies, we evaluate the discrepancy between the two methods by analyzing the $ l=m=2 $ and $ n=0 $ mode of Lv-Ding black holes, the primary mode in astrophysical contexts.
Here, the percentage error of the real part, for example, is defined as \cite{Liu:2023uft}
\begin{equation}
\Delta\text{Re}(\omega)=100\times\frac{\text{Re}\left(\omega_\text{CFM}\right)-\text{Re}\left(\omega_\text{MM}\right)}
{\text{Re}\left(\omega_\text{CFM}\right)}.
\end{equation}
\begin{figure}[h]
\centering
\includegraphics[width=0.9\linewidth]{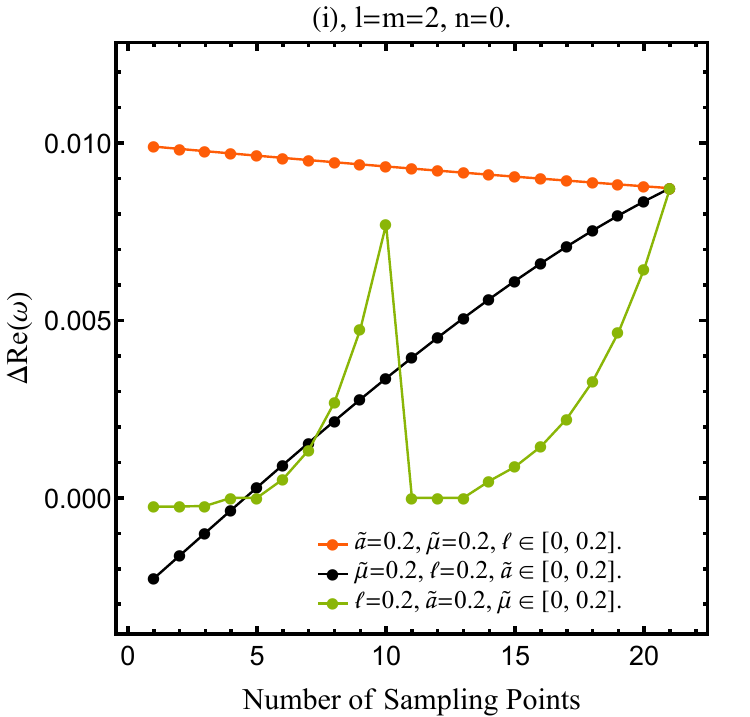}
\includegraphics[width=0.9\linewidth]{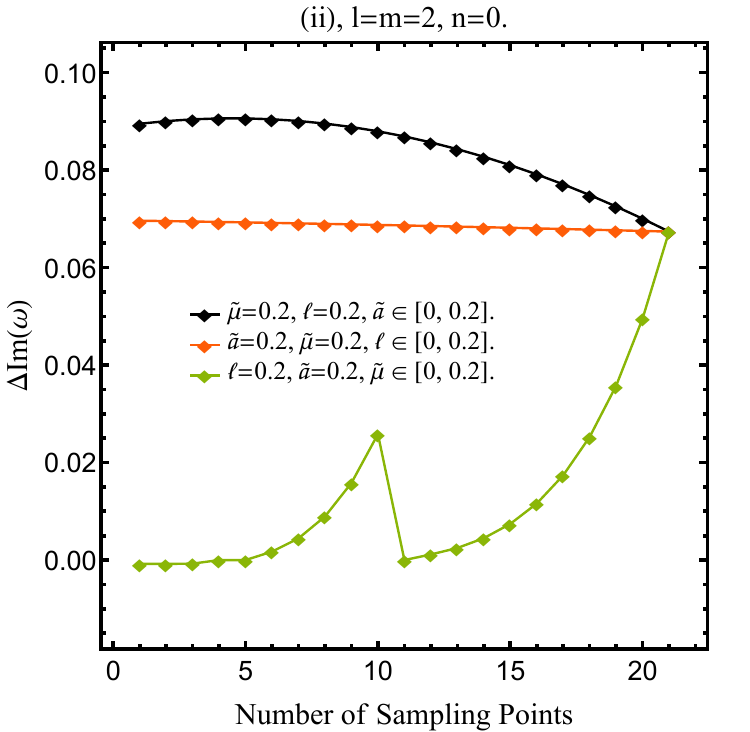}
\caption{The percentage error between CFM and MM results is analyzed. 
(i) and (ii) represent real and imaginary components, respectively. 
Data points are sampled at intervals of 0.01 for the dimensionless parameters $ \tilde{a} $, $ \tilde{\mu} $ and $ \ell $.}
\label{fig1}
\end{figure}
Figs. \ref{fig1}(i)-\ref{fig1}(ii) shows the percentage errors in the real and imaginary parts of $ \omega=\omega_\text{R}+i\omega_\text{I} $, as computed by the two methods.
All parameters depicted are dimensionless, expressed as $ \tilde{a}=a/M $ and $ \tilde{\mu}=\mu M $.
For each line segment, we computed the error values of 21 points over the range from 0 to 0.2, with a step size of $0.01$. 
For the yellow line segment, the black hole parameters are set as $\tilde{a}=0.2$ and $\tilde{\mu}=0.2$, with the Lv parameter $\ell$ varying from $0$ to $0.2$. 
The black line segment tracks the variation of spin $\tilde{a}$, whereas the green line segment tracks the variation of field mass $\tilde{\mu}$.
The figures clearly illustrates the influence of various parameters on computational errors. 
The error in the imaginary part primarily arises from the field mass, as $\textit{Wolfram Mathematica}$ requires significant computational effort to process the $\sqrt{\mu^2-\omega^2}$ term, prompting a Eighth-order Taylor series expansion of the field mass $\tilde{\mu}$,
\begin{equation}
q=i\omega-\frac{i\tilde{\mu}^2}{2\omega}-\frac{i\tilde{\mu}^4}{8\omega^3}-\frac{i\tilde{\mu}^6}{16\omega^5}-\frac{5i\tilde{\mu}^8}{128\omega^7}
+\mathcal{O}\left(\tilde{\mu}^{10}\right).
\end{equation}
For specific values of $\tilde{\mu}$, the $\omega$-related terms were truncated to a precision of $10^{-8}$. 
As a result, the green line segment depicts the error's variation with increasing field mass, typically rising but reaching a minimum at specific values (e.g., $\tilde{\mu}=0.1$) before rising again.
We achieved a balance between precision and computational efficiency, with an accuracy of $10^{-4}$, which corresponds to a percentage error of $0.1\%$.
The error in the real part primarily results from the coupling effect between the rotation parameter and the mass of the scalar field.
When either of $ \tilde{a} $ or $ \tilde{\mu} $ is sufficiently small, the error becomes negligible. 
Even when both parameters are sufficiently large, the error remains within $0.01\%$.
Thus, we have sufficient evidence to believe that our results are accurate. 
Next, we will demonstrate the QNMs of two Lv black holes under the influence of three parameters.

\subsubsection{The monopole modes}

Here, we first examine the monopole fundamental modes within a massive scalar field, characterized by $ l=m=0, n=0 $, which represent the most significant modes of the scalar field. At present, research in the field of relativistic quantum information mainly revolves around this model \cite{Fuentes-Schuller:2004iaz,Wu:2025ncd,Wu:2023vis,Huang:2024vyc,Wu:2024qhd,Li:2025bzd}.
Under these conditions, the influence of the black hole's spin is suppressed. 
The equation governing the dynamical properties simplifies to
\begin{equation}
\frac{d^2}{dr^2_*}\Psi_0+\left[(1+\ell)(\omega^2-F\mu^2)-2FM/r^3 \right]\Psi_0=0,
\end{equation}
It is clear that, in this case, the eigenfrequencies of the equation strongly depend on the field mass. 
Consequently, when addressing the monopole modes, we avoid any approximations for the field mass $\tilde{\mu}$. 
The simplicity of the equation further enables direct calculations with minimal computational resources.

\begin{figure}[h]
\centering
\includegraphics[width=1\linewidth]{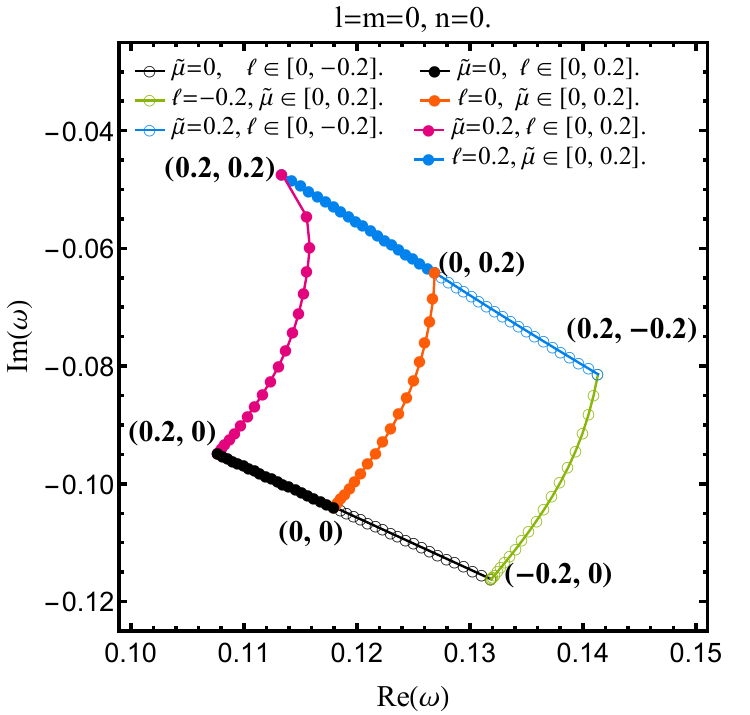}
\caption{Complex scalar frequencies corresponding to the $ l=m=n=0 $ mode, shown for various values of the Lorentz-violation parameter and field mass.}
\label{fig2}
\end{figure}
In Fig. \ref{fig2}, we illustrate the fundamental modes across various parameters, with a uniform parameter interval of $0.01$ between all points. 
The ``coordinates" indicate the parameter values, as exemplified by the point $(0, 0.2)$ where the blue and orange line segments intersect, denoted by $(\ell,\tilde{\mu})=(0,0.2)$.
The Lorentz violation parameter $\ell$ amplifies both the real and imaginary components of the frequency, whereas the field mass parameter $ \tilde{\mu} $, influenced by $\ell$, causes the real part to rise initially and then fall, with the imaginary part consistently decreasing.
Irrespective of the field mass parameter, our analysis demonstrates the linear nature of the Lorentz violation effect, yet the line segments illustrating variations in the Lorentz violation parameter across different field mass parameters are non-parallel. 
This arises from the influence of Lorentz violation on the effect of the field mass parameter on QNM frequencies. 
For a positive $ \ell $, the impact of a given field mass parameter increment increases with the violation magnitude, whereas for a negative $ \ell $, the impact decreases as the violation magnitude grows. 
With negative variation of the parameter $\ell$, the spacing between data points progressively widens. 
These observations yield the first conclusion of this paper: 
(i) QNM frequencies demonstrate increased sensitivity to negative variation of the parameter $\ell$, consequently suppressing the sensitivity to the field mass parameter $\tilde{\mu}$.

\subsubsection{The dipole modes}
We now focus on the dipole fundamental modes with $l=1, m=\pm1, n=0$. 
This part examines the differences between the Lv-NJA and Lv-Ding spacetimes by analyzing the influence of the three parameters ($\tilde{a},\ell,\tilde{\mu}$) on their QNM frequencies.

\begin{figure}[h]
\centering
\includegraphics[width=1\linewidth]{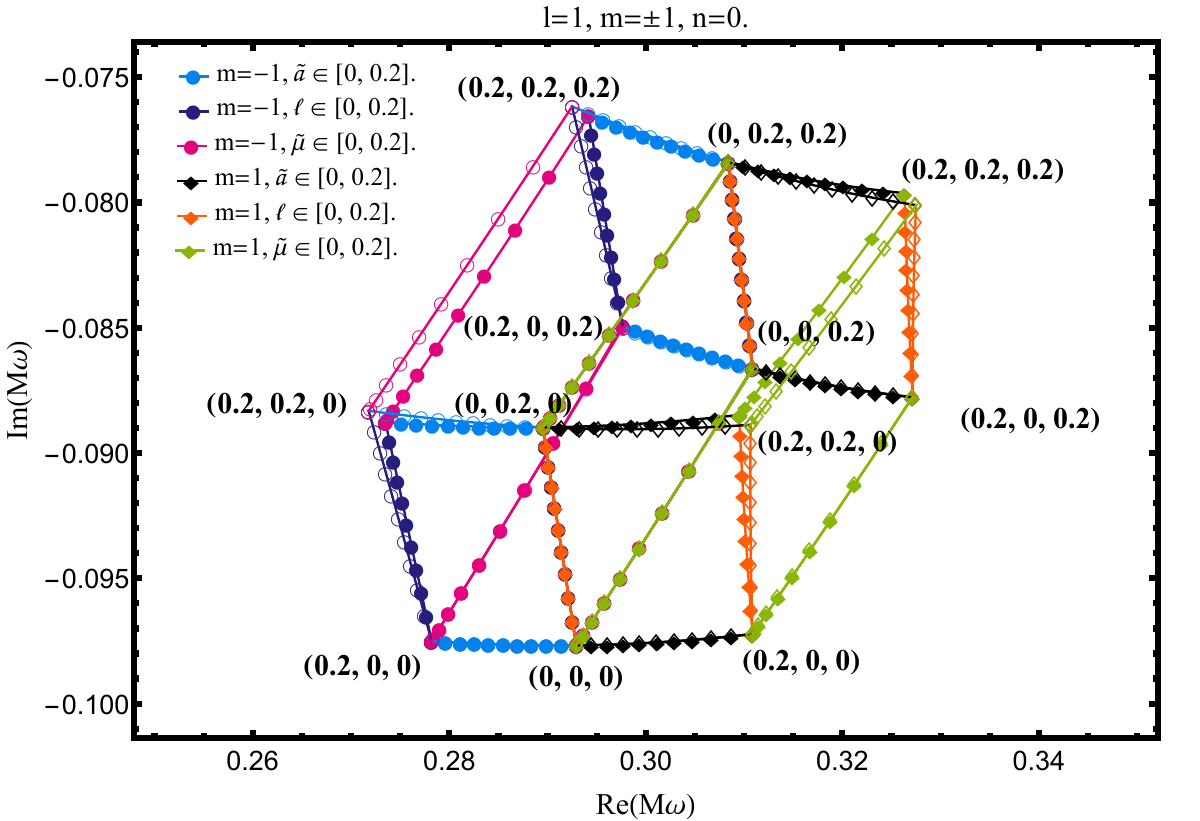}
\caption{Complex scalar frequencies corresponding to the $ l=1, m=\pm1, n=0 $ modes, shown for various values of the spin, Lorentz-violation parameter and field mass.}
\label{fig3}
\end{figure}
Fig. \ref{fig3} presents four cases of $m=1$ and $m=-1$ for both the Lv-NJA and Lv-Ding spacetimes.
Fig. \ref{fig3} depicts filled points for Lv-Ding spacetimes and hollow points for Lv-NJA spacetimes, with circular markers indicating $m=-1$ and square markers indicating $m=1$, with a parameter interval of 0.02 between points.
Just like Fig. \ref{fig2}, the ``coordinates'' represent the parameter values, for example, the point $(0, 0, 0)$ represented by $(\tilde{a},\ell,\tilde{\mu})=(0,0,0)$.
Each case presents a ``axonometric projection'' in the parameter space spanned by $\tilde{a},\ell,\tilde{\mu}$.
As an example, for Lv-Ding with $m=1$ in Fig. \ref{fig3}, four black line segments depict combinations of $\ell, \tilde{\mu}$, under which QNM frequencies change with the increasing spin parameter $\tilde{a}$.
As the spin parameter $\tilde{a}$ increases, the imaginary part of QNM frequencies rises for small $\tilde{\mu}$ but falls for large $\tilde{\mu}$, while the real part consistently rises.
Yellow line segments indicate the parameter $\ell$ as the independent variable, leading to an increased imaginary part and a decreased real part of QNM frequencies, while green line segments indicate the field mass parameter $\tilde{\mu}$, leading to increases in both parts, resembling the monopole  modes, thereby maintaining conclusion (i).
The other three cases follow similar trends to the case described above, with $m=-1$ showing a trend in contrast to $m=1$ regarding the spin parameter $\tilde{a}$, and Lv-NJA exhibiting a greater QNM frequency variation than Lv-Ding under the same $\tilde{a}$, induced by the Lorentz violation parameter $\ell$.
Additionally, Fig. \ref{fig3} visually demonstrates the differences in dynamical properties between Lv-NJA and Lv-Ding, which can be summarized as: (ii) Compared to the rotating spacetime obtained through the slow-rotation approximation, the spectral ``cube'' of NJA displays a slight compression in the upper-right region ($m>0$) and a slight expansion in the upper-left region ($m<0$).

\subsubsection{The quadrupole modes}
\begin{figure}[h]
\centering
\includegraphics[width=1\linewidth]{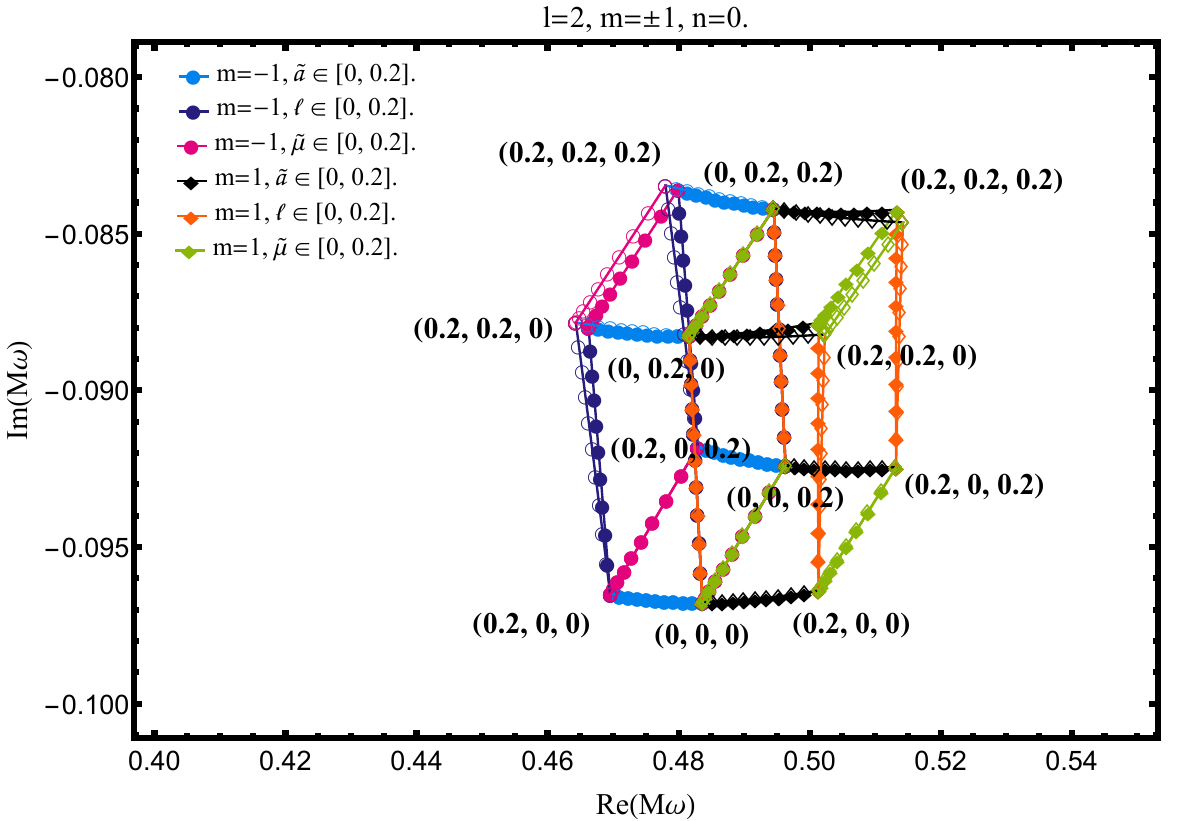}
\caption{Complex scalar frequencies corresponding to the $ l=2, m=\pm1, n=0 $ modes, shown for various values of the spin, Lorentz-violation parameter and field mass.}
\label{fig4}
\end{figure}

\begin{figure}[h]
\centering
\includegraphics[width=1\linewidth]{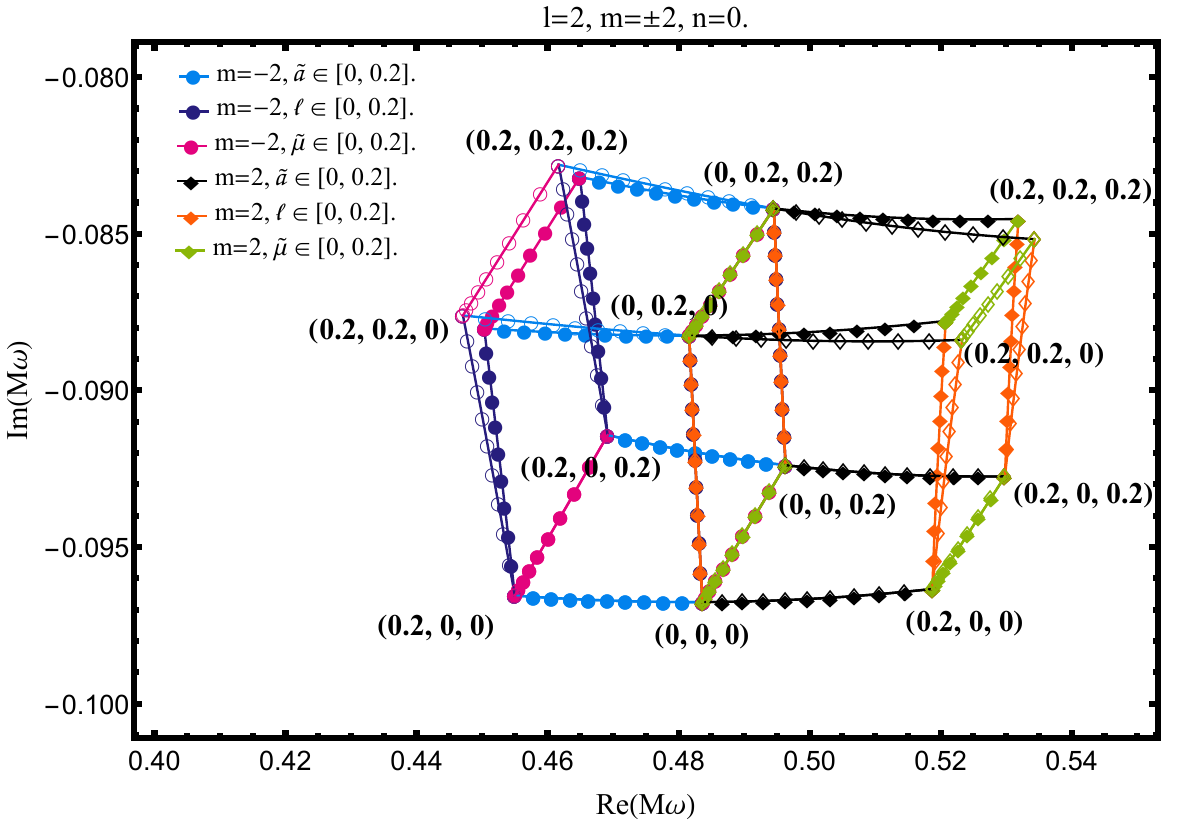}
\caption{Complex scalar frequencies corresponding to the $ l=1, m=\pm2, n=0 $ modes, shown for various values of the spin, Lorentz-violation parameter and field mass.}
\label{fig5}
\end{figure}

Finally, we examine the quadrupole modes by $l=2$ and $n=0$, with possible $m$ values given by $m=0, \pm1, \pm2$. 
The outcomes for $m=\pm1$ and $m=\pm2$ are illustrated in Figs. \ref{fig4} and \ref{fig5}, respectively.

Fig. \ref{fig4} illustrates that this mode differs significantly from the dipole modes, particularly noticeable in the axonometric projection's angle, transitioning from an overhead to a lateral perspective. 
A large angular quantum number enhances the Lorentz violation effect, resulting in a taller overall structure.
While Lorentz violation and field mass impacts on frequency align with those for $l=1$, the spin parameter's effect undergoes a change for $m=1$ and $\tilde{\mu}=0.2$ due to the $l$ parameter's dominance over the frequency, which suppresses the field mass influence, with the frequency range caused by the field mass reducing as $l$ increases.
Fig. \ref{fig5} shows that the plot for $m=\pm2$ appears similar to the $m=\pm1$ plot but does not perfectly coincide, with the $m=1, \tilde{a}=0.2$ position being slightly higher than the $m=2, \tilde{a}=0.1$ position. 
It is incorrect to presume that the $m$ parameter is solely a multiple of the spin parameter, a feature evident only under the second-order slow-rotation approximation, whereas in the first-order approximation, $m$ and $\tilde{a}$ are degenerate, as reflected in the effective potential.
For this, we arrive at the conclusion: (iii) In the first-order slow-rotation approximation, $ m $ is approximately a multiple of the spin parameter due to the effective potential, whereas richer structures and lifted degeneracy arise in the second-order approximation.
Additionally, the conclusions (i) and (ii) obtained for $l=0,1$ remain valid for $l=2$.

\begin{table*}[t]
\renewcommand{\arraystretch}{1.25}
\centering
\setlength\tabcolsep{1.3mm}{
\begin{tabular}{ccccccccc}
\hline\hline
 \multirow{2}{*}{Variables}
&\multirow{2}{*}{Fixed parameters} & \multirow{2}{*}{ $(f_1,q_1)$}&\multirow{2}{*}{ $(f_2,q_2)$} & \multirow{2}{*}{ $(f_3,q_3)$ }
\\ \\
\hline
$x=\tilde{a}$&  $(\ell=0,~~~~\tilde{\mu}=0) $  &  $ (0.483742, -0.096753)  $   & $(0.143912, -0.000229)  $   & $ (0.152141, 0.011275)   $   \\
$x=\tilde{a}$&  $(\ell=0.2,~\tilde{\mu}=0) $  &  $ (0.481763, -0.088245)  $   & $(0.168661, -0.002688)  $   & $ (0.190170, 0.009624)   $   \\
$x=\tilde{a}$&  $(~\ell=0,~~~\tilde{\mu}=0.2) $  &  $ (0.496420, -0.092384)  $   & $(0.136827, -0.003765)  $   & $ (0.147920, 0.009604)   $   \\
$x=\tilde{a}$&  $(\ell=0.2,~\tilde{\mu}=0.2) $  &  $ (0.494675, -0.084188)  $   & $(0.160708, -0.006343)  $   & $ (0.184384, 0.006980)   $   \\
\\
$x=\ell$&  $(\tilde{a}=0,~~~~\tilde{\mu}=0) $  &  $ (0.483639, -0.096751)  $   & $(-0.011898, 0.048320)  $   & $ (0.009095, -0.029278)   $   \\
$x=\ell$&  $(\tilde{a}=0.2,~\tilde{\mu}=0) $  &  $ (0.518722, -0.096334)  $   & $(~~0.022362, 0.045261)  $   & $ (0.001751, -0.027919)   $   \\
$x=\ell$&  $(~\tilde{a}=0,~~~\tilde{\mu}=0.2) $  &  $ (0.496320, -0.092380)  $   & $(-0.010525, 0.046591)  $   & $ (0.008113, -0.028433)   $   \\
$x=\ell$&  $(\tilde{a}=0.2,~\tilde{\mu}=0.2) $  &  $ (0.529810, -0.092740)  $   & $(~~0.022500, 0.043201)  $   & $ (0.000974, -0.026998)   $   \\
\\
$x=\tilde{\mu}$&  $(\tilde{a}=0,~~~~\ell=0) $  &  $ (0.483646, -0.096755)  $   & $(-0.000257, -0.000201)  $   & $ (0.318073, 0.110808)   $   \\
$x=\tilde{\mu}$&  $(\tilde{a}=0.2,~\ell=0) $  &  $ (0.518727, -0.096338)  $   & $(-0.000228, -0.000196)  $   & $ (0.278130, 0.090633)   $   \\
$x=\tilde{\mu}$&  $(~\tilde{a}=0,~~~\ell=0.2) $  &  $ (0.481625, -0.088248)  $   & $(-0.000271, -0.000186)  $   & $ (0.324026, 0.102214)   $   \\
$x=\tilde{\mu}$&  $(\tilde{a}=0.2,~\ell=0.2) $  &  $ (0.523267, -0.088388)  $   & $(-0.000238, -0.000168)  $   & $ (0.278094, 0.081131)   $   \\
\hline\hline
\end{tabular}}
\caption{ The fitting coefficients are obtained for the massive scalar field in the LV--NJA black-hole spacetime, focusing on the fundamental mode with $l=m=2$. }
\label{tab1}
\end{table*}
To facilitate straightforward access and comparison, we present numerical fits for the fundamental scalar $l=m=2$ modes in the Lv-NJA black holes, with explicit coefficients supplied in the Table \ref{tab1}.
A common choice is a fractional-power polynomial ansatz. 
For the gravitational perturbation with $l=m=2$, Berti et al. tabulated the well-known coefficients \cite{Berti:2009kk}.
Here, for massvie scalar perturbations we adopt the second-order polynomial ansatz. 
The following fitting functions:
\begin{equation}
\begin{aligned}
M\omega_R=f_1+f_2 x+f_3x^2,\\
 M\omega_I=q_1+q_2x+q_3x^2,
\end{aligned}
\end{equation}
where $f_i$ and $q_i$ are coefficients obtained by fitting to numerical QNM data.

\section{CONCLUSIONS AND OUTLOOKS}\label{sec5}

In this work, we investigate perturbations of massive scalar fields in rotating spacetimes under the Einstein-Bumblebee gravity framework, employing the second-order slow-rotation approximation. 
Within this theoretical framework, the initial rotating spacetime, presented by Ding et al., adheres to the corresponding field equations under the second-order slow-rotation expansion, while a subsequent rotating black hole spacetime was developed by Ghosh et al. using NJA. 
As these two spacetimes differ, the dynamics of scalar fields within them can be analyzed to validate NJA and assess its deviation from the standard theoretical framework. 
Utilizing the Kojima identity and the slow-rotation decoupling framework introduced by Pani et al., we derived the equations governing massive scalar perturbations in both spacetimes under the second-order slow-rotation approximation. 
By employing both the continued fraction method and the matrix method, we computed the QNMs for massive scalar fields.
Our results show that, the QNMs errors obtained by the two methods are negligible. 
We emphasize not the individual impacts of parameters like spin $ \tilde{a} $, Lorentz violation parameter $ \ell $, and field mass $ \tilde{\mu} $ on QNM frequencies, but rather the collective effects stemming from their interactions. 
To examine the combined influence of these three parameters, we depict the spectrum in a ``cubic'' representation, which clearly illustrates the effects of their coupling. 
In this analysis, we identified the three most significant characteristics of QNMs within the Einstein--Bumblebee gravity framework, as (i), (ii), and (iii) presented in the previous section.

An interesting direction for future research is the investigation of the QNM frequencies of the massive vector field, specifically the Proca field, in the context of the Bumblebee gravity model.
The separation of variables for the Proca field in rotating black holes was achieved by Frolov et al. in 2018 \cite{Frolov:2018ezx}; however, their approach is not directly applicable to theories beyond general relativity. 
Within the slow-rotating approximation framework, we aim to approximately derive the Proca field's QNMs by solving coupled equations, providing a significant extension to this research.
On the other hand, in addition to employing dynamical methods to analyze the properties of two rotating Einstein-Bumblebee black holes, we can further evaluate whether the NJA-generated solution fundamentally deviates from the standard framework by examining their thermodynamic topological properties \cite{Wei:2022dzw,Wei:2024gfz,Wu:2024asq,Liu:2025iyl,Zhu:2024zcl,Wu:2024rmv,Wu:2023fcw,Wu:2023xpq,Wu:2023sue,Wu:2022whe,Liu:2024soc,Liu:2024iec,Cunha:2024ajc} to determine if they belong to the same topological class.
Typically, the slow-rotation approximation does not alter the thermodynamic classification of black holes.

\acknowledgments

This work was supported by the National Natural Science Foundation of China (Grant Nos. 12035005, 12205140), and the Natural Science Foundation of Hunan Province (Grant Nos. 2023JJ40523, 2022JJ30220).
\appendix


\begin{thebibliography}{125}%
\makeatletter
\providecommand \@ifxundefined [1]{%
 \@ifx{#1\undefined}
}%
\providecommand \@ifnum [1]{%
 \ifnum #1\expandafter \@firstoftwo
 \else \expandafter \@secondoftwo
 \fi
}%
\providecommand \@ifx [1]{%
 \ifx #1\expandafter \@firstoftwo
 \else \expandafter \@secondoftwo
 \fi
}%
\providecommand \natexlab [1]{#1}%
\providecommand \enquote  [1]{``#1''}%
\providecommand \bibnamefont  [1]{#1}%
\providecommand \bibfnamefont [1]{#1}%
\providecommand \citenamefont [1]{#1}%
\providecommand \href@noop [0]{\@secondoftwo}%
\providecommand \href [0]{\begingroup \@sanitize@url \@href}%
\providecommand \@href[1]{\@@startlink{#1}\@@href}%
\providecommand \@@href[1]{\endgroup#1\@@endlink}%
\providecommand \@sanitize@url [0]{\catcode `\\12\catcode `\$12\catcode
  `\&12\catcode `\#12\catcode `\^12\catcode `\_12\catcode `\%12\relax}%
\providecommand \@@startlink[1]{}%
\providecommand \@@endlink[0]{}%
\providecommand \url  [0]{\begingroup\@sanitize@url \@url }%
\providecommand \@url [1]{\endgroup\@href {#1}{\urlprefix }}%
\providecommand \urlprefix  [0]{URL }%
\providecommand \Eprint [0]{\href }%
\providecommand \doibase [0]{https://doi.org/}%
\providecommand \selectlanguage [0]{\@gobble}%
\providecommand \bibinfo  [0]{\@secondoftwo}%
\providecommand \bibfield  [0]{\@secondoftwo}%
\providecommand \translation [1]{[#1]}%
\providecommand \BibitemOpen [0]{}%
\providecommand \bibitemStop [0]{}%
\providecommand \bibitemNoStop [0]{.\EOS\space}%
\providecommand \EOS [0]{\spacefactor3000\relax}%
\providecommand \BibitemShut  [1]{\csname bibitem#1\endcsname}%
\let\auto@bib@innerbib\@empty
\bibitem [{\citenamefont {Takeda}\ \emph {et~al.}(1998)\citenamefont {Takeda}
  \emph {et~al.}}]{Takeda:1998ps}%
  \BibitemOpen
  \bibfield  {author} {\bibinfo {author} {\bibfnamefont {M.}~\bibnamefont
  {Takeda}} \emph {et~al.},\ }\bibfield  {title} {\bibinfo {title} {{Extension
  of the cosmic ray energy spectrum beyond the predicted
  Greisen-Zatsepin-Kuz'min cutoff}},\ }\href
  {https://doi.org/10.1103/PhysRevLett.81.1163} {\bibfield  {journal} {\bibinfo
   {journal} {Phys. Rev. Lett.}\ }\textbf {\bibinfo {volume} {81}},\ \bibinfo
  {pages} {1163} (\bibinfo {year} {1998})},\ \Eprint
  {https://arxiv.org/abs/astro-ph/9807193} {arXiv:astro-ph/9807193}
  \BibitemShut {NoStop}%
\bibitem [{\citenamefont {Kalb}\ and\ \citenamefont
  {Ramond}(1974)}]{Kalb:1974yc}%
  \BibitemOpen
  \bibfield  {author} {\bibinfo {author} {\bibfnamefont {M.}~\bibnamefont
  {Kalb}}\ and\ \bibinfo {author} {\bibfnamefont {P.}~\bibnamefont {Ramond}},\
  }\bibfield  {title} {\bibinfo {title} {{Classical direct interstring
  action}},\ }\href {https://doi.org/10.1103/PhysRevD.9.2273} {\bibfield
  {journal} {\bibinfo  {journal} {Phys. Rev. D}\ }\textbf {\bibinfo {volume}
  {9}},\ \bibinfo {pages} {2273} (\bibinfo {year} {1974})}\BibitemShut
  {NoStop}%
\bibitem [{\citenamefont {Kostelecky}\ and\ \citenamefont
  {Potting}(1995)}]{Kostelecky:1994rn}%
  \BibitemOpen
  \bibfield  {author} {\bibinfo {author} {\bibfnamefont {V.~A.}\ \bibnamefont
  {Kostelecky}}\ and\ \bibinfo {author} {\bibfnamefont {R.}~\bibnamefont
  {Potting}},\ }\bibfield  {title} {\bibinfo {title} {{CPT, strings, and meson
  factories}},\ }\href {https://doi.org/10.1103/PhysRevD.51.3923} {\bibfield
  {journal} {\bibinfo  {journal} {Phys. Rev. D}\ }\textbf {\bibinfo {volume}
  {51}},\ \bibinfo {pages} {3923} (\bibinfo {year} {1995})},\ \Eprint
  {https://arxiv.org/abs/hep-ph/9501341} {arXiv:hep-ph/9501341} \BibitemShut
  {NoStop}%
\bibitem [{\citenamefont {Colladay}\ and\ \citenamefont
  {Kostelecky}(1998)}]{Kostelecky1998}%
  \BibitemOpen
  \bibfield  {author} {\bibinfo {author} {\bibfnamefont {D.}~\bibnamefont
  {Colladay}}\ and\ \bibinfo {author} {\bibfnamefont {V.~A.}\ \bibnamefont
  {Kostelecky}},\ }\bibfield  {title} {\bibinfo {title} {{Lorentz violating
  extension of the standard model}},\ }\href
  {https://doi.org/10.1103/PhysRevD.58.116002} {\bibfield  {journal} {\bibinfo
  {journal} {Phys. Rev. D}\ }\textbf {\bibinfo {volume} {58}},\ \bibinfo
  {pages} {116002} (\bibinfo {year} {1998})},\ \Eprint
  {https://arxiv.org/abs/9809521} {arXiv:9809521 [hep-ph]} \BibitemShut
  {NoStop}%
\bibitem [{\citenamefont {Wu}\ and\ \citenamefont {Zeng}(2022)}]{Wu:2022xwy}%
  \BibitemOpen
  \bibfield  {author} {\bibinfo {author} {\bibfnamefont {S.~M.}\ \bibnamefont
  {Wu}}\ and\ \bibinfo {author} {\bibfnamefont {H.~S.}\ \bibnamefont {Zeng}},\
  }\bibfield  {title} {\bibinfo {title} {{Genuine tripartite nonlocality and
  entanglement in curved spacetime}},\ }\href
  {https://doi.org/10.1140/epjc/s10052-021-09954-4} {\bibfield  {journal}
  {\bibinfo  {journal} {Eur. Phys. J. C}\ }\textbf {\bibinfo {volume} {82}},\
  \bibinfo {pages} {4} (\bibinfo {year} {2022})},\ \Eprint
  {https://arxiv.org/abs/2201.02333} {arXiv:2201.02333 [quant-ph]} \BibitemShut
  {NoStop}%
\bibitem [{\citenamefont {Devecioglu}\ and\ \citenamefont
  {Park}(2024)}]{Devecioglu:2024uyi}%
  \BibitemOpen
  \bibfield  {author} {\bibinfo {author} {\bibfnamefont {D.~O.}\ \bibnamefont
  {Devecioglu}}\ and\ \bibinfo {author} {\bibfnamefont {M.-I.}\ \bibnamefont
  {Park}},\ }\bibfield  {title} {\bibinfo {title} {{Rotating black holes in a
  viable Lorentz-violating gravity: finding exact solutions without tears}},\
  }\href {https://doi.org/10.1140/epjc/s10052-024-13209-3} {\bibfield
  {journal} {\bibinfo  {journal} {Eur. Phys. J. C}\ }\textbf {\bibinfo {volume}
  {84}},\ \bibinfo {pages} {852} (\bibinfo {year} {2024})},\ \Eprint
  {https://arxiv.org/abs/2402.02253} {arXiv:2402.02253 [hep-th]} \BibitemShut
  {NoStop}%
\bibitem [{\citenamefont {Liu}\ \emph {et~al.}(2025{\natexlab{a}})\citenamefont
  {Liu}, \citenamefont {Huang}, \citenamefont {Wu},\ and\ \citenamefont
  {Wang}}]{Liu:2025lwj}%
  \BibitemOpen
  \bibfield  {author} {\bibinfo {author} {\bibfnamefont {W.}~\bibnamefont
  {Liu}}, \bibinfo {author} {\bibfnamefont {H.}~\bibnamefont {Huang}}, \bibinfo
  {author} {\bibfnamefont {D.}~\bibnamefont {Wu}},\ and\ \bibinfo {author}
  {\bibfnamefont {J.}~\bibnamefont {Wang}},\ }\bibfield  {title} {\bibinfo
  {title} {{Lorentz violation signatures in the low-energy sector of
  Ho{\v{r}}ava gravity from black hole shadow observations}},\ }\href@noop {}
  {\  (\bibinfo {year} {2025}{\natexlab{a}})},\ \Eprint
  {https://arxiv.org/abs/2506.13504} {arXiv:2506.13504 [gr-qc]} \BibitemShut
  {NoStop}%
\bibitem [{\citenamefont {Casana}\ \emph {et~al.}(2018)\citenamefont {Casana},
  \citenamefont {Cavalcante}, \citenamefont {Poulis},\ and\ \citenamefont
  {Santos}}]{Casana2018}%
  \BibitemOpen
  \bibfield  {author} {\bibinfo {author} {\bibfnamefont {R.}~\bibnamefont
  {Casana}}, \bibinfo {author} {\bibfnamefont {A.}~\bibnamefont {Cavalcante}},
  \bibinfo {author} {\bibfnamefont {F.~P.}\ \bibnamefont {Poulis}},\ and\
  \bibinfo {author} {\bibfnamefont {E.~B.}\ \bibnamefont {Santos}},\ }\bibfield
   {title} {\bibinfo {title} {{Exact Schwarzschild-like solution in a bumblebee
  gravity model}},\ }\href {https://doi.org/10.1103/PhysRevD.97.104001}
  {\bibfield  {journal} {\bibinfo  {journal} {Phys. Rev. D}\ }\textbf {\bibinfo
  {volume} {97}},\ \bibinfo {pages} {104001} (\bibinfo {year} {2018})},\
  \Eprint {https://arxiv.org/abs/1711.02273} {arXiv:1711.02273 [gr-qc]}
  \BibitemShut {NoStop}%
\bibitem [{\citenamefont {\"Ovg\"un}\ \emph {et~al.}(2019)\citenamefont
  {\"Ovg\"un}, \citenamefont {Jusufi},\ and\ \citenamefont
  {Sakall\i{}}}]{Ovgun2019}%
  \BibitemOpen
  \bibfield  {author} {\bibinfo {author} {\bibfnamefont {A.}~\bibnamefont
  {\"Ovg\"un}}, \bibinfo {author} {\bibfnamefont {K.}~\bibnamefont {Jusufi}},\
  and\ \bibinfo {author} {\bibfnamefont {I.}~\bibnamefont {Sakall\i{}}},\
  }\bibfield  {title} {\bibinfo {title} {{Exact traversable wormhole solution
  in bumblebee gravity}},\ }\href {https://doi.org/10.1103/PhysRevD.99.024042}
  {\bibfield  {journal} {\bibinfo  {journal} {Phys. Rev. D}\ }\textbf {\bibinfo
  {volume} {99}},\ \bibinfo {pages} {024042} (\bibinfo {year} {2019})},\
  \Eprint {https://arxiv.org/abs/1804.09911} {arXiv:1804.09911 [gr-qc]}
  \BibitemShut {NoStop}%
\bibitem [{\citenamefont {G\"ull\"u}\ and\ \citenamefont
  {\"Ovg\"un}(2022)}]{Gullu2020}%
  \BibitemOpen
  \bibfield  {author} {\bibinfo {author} {\bibfnamefont {I.}~\bibnamefont
  {G\"ull\"u}}\ and\ \bibinfo {author} {\bibfnamefont {A.}~\bibnamefont
  {\"Ovg\"un}},\ }\bibfield  {title} {\bibinfo {title} {{Schwarzschild-like
  black hole with a topological defect in bumblebee gravity}},\ }\href
  {https://doi.org/10.1016/j.aop.2021.168721} {\bibfield  {journal} {\bibinfo
  {journal} {Annals Phys.}\ }\textbf {\bibinfo {volume} {436}},\ \bibinfo
  {pages} {168721} (\bibinfo {year} {2022})},\ \Eprint
  {https://arxiv.org/abs/2012.02611} {arXiv:2012.02611 [gr-qc]} \BibitemShut
  {NoStop}%
\bibitem [{\citenamefont {Poulis}\ and\ \citenamefont
  {Soares}(2022)}]{Poulis:2021nqh}%
  \BibitemOpen
  \bibfield  {author} {\bibinfo {author} {\bibfnamefont {F.~P.}\ \bibnamefont
  {Poulis}}\ and\ \bibinfo {author} {\bibfnamefont {M.~A.~C.}\ \bibnamefont
  {Soares}},\ }\bibfield  {title} {\bibinfo {title} {{Exact modifications on a
  vacuum spacetime due to a gradient bumblebee field at its vacuum expectation
  value}},\ }\href {https://doi.org/10.1140/epjc/s10052-022-10547-y} {\bibfield
   {journal} {\bibinfo  {journal} {Eur. Phys. J. C}\ }\textbf {\bibinfo
  {volume} {82}},\ \bibinfo {pages} {613} (\bibinfo {year} {2022})},\ \Eprint
  {https://arxiv.org/abs/2112.04040} {arXiv:2112.04040 [gr-qc]} \BibitemShut
  {NoStop}%
\bibitem [{\citenamefont {Maluf}\ and\ \citenamefont
  {Neves}(2021)}]{Maluf2021}%
  \BibitemOpen
  \bibfield  {author} {\bibinfo {author} {\bibfnamefont {R.~V.}\ \bibnamefont
  {Maluf}}\ and\ \bibinfo {author} {\bibfnamefont {J.~C.~S.}\ \bibnamefont
  {Neves}},\ }\bibfield  {title} {\bibinfo {title} {{Black holes with a
  cosmological constant in bumblebee gravity}},\ }\href
  {https://doi.org/10.1103/PhysRevD.103.044002} {\bibfield  {journal} {\bibinfo
   {journal} {Phys. Rev. D}\ }\textbf {\bibinfo {volume} {103}},\ \bibinfo
  {pages} {044002} (\bibinfo {year} {2021})},\ \Eprint
  {https://arxiv.org/abs/2011.12841} {arXiv:2011.12841 [gr-qc]} \BibitemShut
  {NoStop}%
\bibitem [{\citenamefont {Ding}\ \emph
  {et~al.}(2022{\natexlab{a}})\citenamefont {Ding}, \citenamefont {Chen},\ and\
  \citenamefont {Fu}}]{Ding:2021iwv}%
  \BibitemOpen
  \bibfield  {author} {\bibinfo {author} {\bibfnamefont {C.}~\bibnamefont
  {Ding}}, \bibinfo {author} {\bibfnamefont {X.}~\bibnamefont {Chen}},\ and\
  \bibinfo {author} {\bibfnamefont {X.}~\bibnamefont {Fu}},\ }\bibfield
  {title} {\bibinfo {title} {{Einstein-Gauss-Bonnet gravity coupled to
  bumblebee field in four dimensional spacetime}},\ }\href
  {https://doi.org/10.1016/j.nuclphysb.2022.115688} {\bibfield  {journal}
  {\bibinfo  {journal} {Nucl. Phys. B}\ }\textbf {\bibinfo {volume} {975}},\
  \bibinfo {pages} {115688} (\bibinfo {year} {2022}{\natexlab{a}})},\ \Eprint
  {https://arxiv.org/abs/2102.13335} {arXiv:2102.13335 [gr-qc]} \BibitemShut
  {NoStop}%
\bibitem [{\citenamefont {Xu}\ \emph {et~al.}(2023{\natexlab{a}})\citenamefont
  {Xu}, \citenamefont {Liang},\ and\ \citenamefont {Shao}}]{Xu:2022frb}%
  \BibitemOpen
  \bibfield  {author} {\bibinfo {author} {\bibfnamefont {R.}~\bibnamefont
  {Xu}}, \bibinfo {author} {\bibfnamefont {D.}~\bibnamefont {Liang}},\ and\
  \bibinfo {author} {\bibfnamefont {L.}~\bibnamefont {Shao}},\ }\bibfield
  {title} {\bibinfo {title} {{Static spherical vacuum solutions in the
  bumblebee gravity model}},\ }\href
  {https://doi.org/10.1103/PhysRevD.107.024011} {\bibfield  {journal} {\bibinfo
   {journal} {Phys. Rev. D}\ }\textbf {\bibinfo {volume} {107}},\ \bibinfo
  {pages} {024011} (\bibinfo {year} {2023}{\natexlab{a}})},\ \Eprint
  {https://arxiv.org/abs/2209.02209} {arXiv:2209.02209 [gr-qc]} \BibitemShut
  {NoStop}%
\bibitem [{\citenamefont {Filho}\ \emph {et~al.}(2023)\citenamefont {Filho},
  \citenamefont {Nascimento}, \citenamefont {Petrov},\ and\ \citenamefont
  {Porf\'\i{}rio}}]{Filho:2022yrk}%
  \BibitemOpen
  \bibfield  {author} {\bibinfo {author} {\bibfnamefont {A.~A.~A.}\
  \bibnamefont {Filho}}, \bibinfo {author} {\bibfnamefont {J.~R.}\ \bibnamefont
  {Nascimento}}, \bibinfo {author} {\bibfnamefont {A.~Y.}\ \bibnamefont
  {Petrov}},\ and\ \bibinfo {author} {\bibfnamefont {P.~J.}\ \bibnamefont
  {Porf\'\i{}rio}},\ }\bibfield  {title} {\bibinfo {title} {{Vacuum solution
  within a metric-affine bumblebee gravity}},\ }\href
  {https://doi.org/10.1103/PhysRevD.108.085010} {\bibfield  {journal} {\bibinfo
   {journal} {Phys. Rev. D}\ }\textbf {\bibinfo {volume} {108}},\ \bibinfo
  {pages} {085010} (\bibinfo {year} {2023})},\ \Eprint
  {https://arxiv.org/abs/2211.11821} {arXiv:2211.11821 [gr-qc]} \BibitemShut
  {NoStop}%
\bibitem [{\citenamefont {Ara{\'u}jo~Filho}\ \emph {et~al.}(2024)\citenamefont
  {Ara{\'u}jo~Filho}, \citenamefont {Nascimento}, \citenamefont {Petrov},\ and\
  \citenamefont {Porf{\'\i}rio}}]{AraujoFilho:2024ykw}%
  \BibitemOpen
  \bibfield  {author} {\bibinfo {author} {\bibfnamefont {A.~A.}\ \bibnamefont
  {Ara{\'u}jo~Filho}}, \bibinfo {author} {\bibfnamefont {J.~R.}\ \bibnamefont
  {Nascimento}}, \bibinfo {author} {\bibfnamefont {A.~Y.}\ \bibnamefont
  {Petrov}},\ and\ \bibinfo {author} {\bibfnamefont {P.~J.}\ \bibnamefont
  {Porf{\'\i}rio}},\ }\bibfield  {title} {\bibinfo {title} {{An exact
  stationary axisymmetric vacuum solution within a metric-affine bumblebee
  gravity}},\ }\href {https://doi.org/10.1088/1475-7516/2024/07/004} {\bibfield
   {journal} {\bibinfo  {journal} {JCAP}\ }\textbf {\bibinfo {volume} {07}},\
  \bibinfo {pages} {004}},\ \Eprint {https://arxiv.org/abs/2402.13014}
  {arXiv:2402.13014 [gr-qc]} \BibitemShut {NoStop}%
\bibitem [{\citenamefont {Chen}\ and\ \citenamefont
  {Liu}(2025)}]{Chen:2025ypx}%
  \BibitemOpen
  \bibfield  {author} {\bibinfo {author} {\bibfnamefont {Y.-Q.}\ \bibnamefont
  {Chen}}\ and\ \bibinfo {author} {\bibfnamefont {H.-S.}\ \bibnamefont {Liu}},\
  }\bibfield  {title} {\bibinfo {title} {{Taub-NUT-like Black Holes in
  Einstein-Bumblebee Gravity}},\ }\href@noop {} {\  (\bibinfo {year} {2025})},\
  \Eprint {https://arxiv.org/abs/2505.23104} {arXiv:2505.23104 [gr-qc]}
  \BibitemShut {NoStop}%
\bibitem [{\citenamefont {Yang}\ \emph {et~al.}(2023)\citenamefont {Yang},
  \citenamefont {Chen}, \citenamefont {Duan},\ and\ \citenamefont
  {Zhao}}]{Yang:2023wtu}%
  \BibitemOpen
  \bibfield  {author} {\bibinfo {author} {\bibfnamefont {K.}~\bibnamefont
  {Yang}}, \bibinfo {author} {\bibfnamefont {Y.-Z.}\ \bibnamefont {Chen}},
  \bibinfo {author} {\bibfnamefont {Z.-Q.}\ \bibnamefont {Duan}},\ and\
  \bibinfo {author} {\bibfnamefont {J.-Y.}\ \bibnamefont {Zhao}},\ }\bibfield
  {title} {\bibinfo {title} {{Static and spherically symmetric black holes in
  gravity with a background Kalb-Ramond field}},\ }\href
  {https://doi.org/10.1103/PhysRevD.108.124004} {\bibfield  {journal} {\bibinfo
   {journal} {Phys. Rev. D}\ }\textbf {\bibinfo {volume} {108}},\ \bibinfo
  {pages} {124004} (\bibinfo {year} {2023})},\ \Eprint
  {https://arxiv.org/abs/2308.06613} {arXiv:2308.06613 [gr-qc]} \BibitemShut
  {NoStop}%
\bibitem [{\citenamefont {Duan}\ \emph {et~al.}(2024)\citenamefont {Duan},
  \citenamefont {Zhao},\ and\ \citenamefont {Yang}}]{Duan:2023gng}%
  \BibitemOpen
  \bibfield  {author} {\bibinfo {author} {\bibfnamefont {Z.-Q.}\ \bibnamefont
  {Duan}}, \bibinfo {author} {\bibfnamefont {J.-Y.}\ \bibnamefont {Zhao}},\
  and\ \bibinfo {author} {\bibfnamefont {K.}~\bibnamefont {Yang}},\ }\bibfield
  {title} {\bibinfo {title} {{Electrically charged black holes in gravity with
  a background Kalb\textendash{}Ramond field}},\ }\href
  {https://doi.org/10.1140/epjc/s10052-024-13188-5} {\bibfield  {journal}
  {\bibinfo  {journal} {Eur. Phys. J. C}\ }\textbf {\bibinfo {volume} {84}},\
  \bibinfo {pages} {798} (\bibinfo {year} {2024})},\ \Eprint
  {https://arxiv.org/abs/2310.13555} {arXiv:2310.13555 [gr-qc]} \BibitemShut
  {NoStop}%
\bibitem [{\citenamefont {Ding}\ \emph {et~al.}(2023)\citenamefont {Ding},
  \citenamefont {Shi}, \citenamefont {Chen}, \citenamefont {Zhou},\ and\
  \citenamefont {Liu}}]{Ding:2022qcy}%
  \BibitemOpen
  \bibfield  {author} {\bibinfo {author} {\bibfnamefont {C.}~\bibnamefont
  {Ding}}, \bibinfo {author} {\bibfnamefont {Y.}~\bibnamefont {Shi}}, \bibinfo
  {author} {\bibfnamefont {J.}~\bibnamefont {Chen}}, \bibinfo {author}
  {\bibfnamefont {Y.}~\bibnamefont {Zhou}},\ and\ \bibinfo {author}
  {\bibfnamefont {C.}~\bibnamefont {Liu}},\ }\bibfield  {title} {\bibinfo
  {title} {{High dimensional AdS-like black hole and phase transition in
  Einstein-bumblebee gravity*}},\ }\href
  {https://doi.org/10.1088/1674-1137/aca8f4} {\bibfield  {journal} {\bibinfo
  {journal} {Chin. Phys. C}\ }\textbf {\bibinfo {volume} {47}},\ \bibinfo
  {pages} {045102} (\bibinfo {year} {2023})},\ \Eprint
  {https://arxiv.org/abs/2201.06683} {arXiv:2201.06683 [gr-qc]} \BibitemShut
  {NoStop}%
\bibitem [{\citenamefont {Liu}\ \emph {et~al.}(2024{\natexlab{a}})\citenamefont
  {Liu}, \citenamefont {Wu},\ and\ \citenamefont {Wang}}]{Liu:2024oas}%
  \BibitemOpen
  \bibfield  {author} {\bibinfo {author} {\bibfnamefont {W.}~\bibnamefont
  {Liu}}, \bibinfo {author} {\bibfnamefont {D.}~\bibnamefont {Wu}},\ and\
  \bibinfo {author} {\bibfnamefont {J.}~\bibnamefont {Wang}},\ }\bibfield
  {title} {\bibinfo {title} {{Static neutral black holes in Kalb-Ramond
  gravity}},\ }\href {https://doi.org/10.1088/1475-7516/2024/09/017} {\bibfield
   {journal} {\bibinfo  {journal} {JCAP}\ }\textbf {\bibinfo {volume} {09}},\
  \bibinfo {pages} {017}},\ \Eprint {https://arxiv.org/abs/2406.13461}
  {arXiv:2406.13461 [hep-th]} \BibitemShut {NoStop}%
\bibitem [{\citenamefont {Liu}\ \emph {et~al.}(2024{\natexlab{b}})\citenamefont
  {Liu}, \citenamefont {Wu},\ and\ \citenamefont {Wang}}]{Liu:2024lve}%
  \BibitemOpen
  \bibfield  {author} {\bibinfo {author} {\bibfnamefont {W.}~\bibnamefont
  {Liu}}, \bibinfo {author} {\bibfnamefont {D.}~\bibnamefont {Wu}},\ and\
  \bibinfo {author} {\bibfnamefont {J.}~\bibnamefont {Wang}},\ }\bibfield
  {title} {\bibinfo {title} {{Shadow of slowly rotating Kalb-Ramond black
  holes}},\ }\href@noop {} {\  (\bibinfo {year} {2024}{\natexlab{b}})},\
  \Eprint {https://arxiv.org/abs/2407.07416} {arXiv:2407.07416 [gr-qc]}
  \BibitemShut {NoStop}%
\bibitem [{\citenamefont {Liu}\ \emph {et~al.}(2025{\natexlab{b}})\citenamefont
  {Liu}, \citenamefont {Wu}, \citenamefont {Wei},\ and\ \citenamefont
  {Liu}}]{Liu:2025fxj}%
  \BibitemOpen
  \bibfield  {author} {\bibinfo {author} {\bibfnamefont {J.-Z.}\ \bibnamefont
  {Liu}}, \bibinfo {author} {\bibfnamefont {S.-P.}\ \bibnamefont {Wu}},
  \bibinfo {author} {\bibfnamefont {S.-W.}\ \bibnamefont {Wei}},\ and\ \bibinfo
  {author} {\bibfnamefont {Y.-X.}\ \bibnamefont {Liu}},\ }\bibfield  {title}
  {\bibinfo {title} {{Exact black hole solutions in gravity with a background
  Kalb-Ramond field}},\ }\href@noop {} {\  (\bibinfo {year}
  {2025}{\natexlab{b}})},\ \Eprint {https://arxiv.org/abs/2505.07404}
  {arXiv:2505.07404 [gr-qc]} \BibitemShut {NoStop}%
\bibitem [{\citenamefont {Belchior}\ \emph {et~al.}(2025)\citenamefont
  {Belchior}, \citenamefont {Maluf}, \citenamefont {Petrov},\ and\
  \citenamefont {Porf{\'\i}rio}}]{Belchior:2025xam}%
  \BibitemOpen
  \bibfield  {author} {\bibinfo {author} {\bibfnamefont {F.~M.}\ \bibnamefont
  {Belchior}}, \bibinfo {author} {\bibfnamefont {R.~V.}\ \bibnamefont {Maluf}},
  \bibinfo {author} {\bibfnamefont {A.~Y.}\ \bibnamefont {Petrov}},\ and\
  \bibinfo {author} {\bibfnamefont {P.~J.}\ \bibnamefont {Porf{\'\i}rio}},\
  }\bibfield  {title} {\bibinfo {title} {{Global monopole in a Ricci-coupled
  Kalb{\textendash}Ramond bumblebee gravity}},\ }\href
  {https://doi.org/10.1140/epjc/s10052-025-14390-9} {\bibfield  {journal}
  {\bibinfo  {journal} {Eur. Phys. J. C}\ }\textbf {\bibinfo {volume} {85}},\
  \bibinfo {pages} {658} (\bibinfo {year} {2025})},\ \Eprint
  {https://arxiv.org/abs/2502.17267} {arXiv:2502.17267 [gr-qc]} \BibitemShut
  {NoStop}%
\bibitem [{\citenamefont {Ding}\ and\ \citenamefont
  {Chen}(2021)}]{Ding:2020kfr}%
  \BibitemOpen
  \bibfield  {author} {\bibinfo {author} {\bibfnamefont {C.}~\bibnamefont
  {Ding}}\ and\ \bibinfo {author} {\bibfnamefont {X.}~\bibnamefont {Chen}},\
  }\bibfield  {title} {\bibinfo {title} {{Slowly rotating Einstein-bumblebee
  black hole solution and its greybody factor in a Lorentz violation model}},\
  }\href {https://doi.org/10.1088/1674-1137/abce51} {\bibfield  {journal}
  {\bibinfo  {journal} {Chin. Phys. C}\ }\textbf {\bibinfo {volume} {45}},\
  \bibinfo {pages} {025106} (\bibinfo {year} {2021})},\ \Eprint
  {https://arxiv.org/abs/2008.10474} {arXiv:2008.10474 [gr-qc]} \BibitemShut
  {NoStop}%
\bibitem [{\citenamefont {Islam}\ \emph {et~al.}(2024)\citenamefont {Islam},
  \citenamefont {Ghosh},\ and\ \citenamefont {Maharaj}}]{Islam:2024sph}%
  \BibitemOpen
  \bibfield  {author} {\bibinfo {author} {\bibfnamefont {S.~U.}\ \bibnamefont
  {Islam}}, \bibinfo {author} {\bibfnamefont {S.~G.}\ \bibnamefont {Ghosh}},\
  and\ \bibinfo {author} {\bibfnamefont {S.~D.}\ \bibnamefont {Maharaj}},\
  }\bibfield  {title} {\bibinfo {title} {{Investigating rotating black holes in
  bumblebee gravity: insights from EHT observations}},\ }\href
  {https://doi.org/10.1088/1475-7516/2024/12/047} {\bibfield  {journal}
  {\bibinfo  {journal} {JCAP}\ }\textbf {\bibinfo {volume} {12}},\ \bibinfo
  {pages} {047}},\ \Eprint {https://arxiv.org/abs/2410.05395} {arXiv:2410.05395
  [gr-qc]} \BibitemShut {NoStop}%
\bibitem [{\citenamefont {Afrin}\ \emph {et~al.}(2024)\citenamefont {Afrin},
  \citenamefont {Ghosh},\ and\ \citenamefont {Wang}}]{Afrin:2024khy}%
  \BibitemOpen
  \bibfield  {author} {\bibinfo {author} {\bibfnamefont {M.}~\bibnamefont
  {Afrin}}, \bibinfo {author} {\bibfnamefont {S.~G.}\ \bibnamefont {Ghosh}},\
  and\ \bibinfo {author} {\bibfnamefont {A.}~\bibnamefont {Wang}},\ }\bibfield
  {title} {\bibinfo {title} {{Testing EGB gravity coupled to bumblebee field
  and black hole parameter estimation with EHT observations}},\ }\href
  {https://doi.org/10.1016/j.dark.2024.101642} {\bibfield  {journal} {\bibinfo
  {journal} {Phys. Dark Univ.}\ }\textbf {\bibinfo {volume} {46}},\ \bibinfo
  {pages} {101642} (\bibinfo {year} {2024})},\ \Eprint
  {https://arxiv.org/abs/2409.06218} {arXiv:2409.06218 [gr-qc]} \BibitemShut
  {NoStop}%
\bibitem [{\citenamefont {Liu}\ \emph {et~al.}(2023{\natexlab{a}})\citenamefont
  {Liu}, \citenamefont {Fang}, \citenamefont {Jing},\ and\ \citenamefont
  {Wang}}]{Liu:2022dcn}%
  \BibitemOpen
  \bibfield  {author} {\bibinfo {author} {\bibfnamefont {W.}~\bibnamefont
  {Liu}}, \bibinfo {author} {\bibfnamefont {X.}~\bibnamefont {Fang}}, \bibinfo
  {author} {\bibfnamefont {J.}~\bibnamefont {Jing}},\ and\ \bibinfo {author}
  {\bibfnamefont {J.}~\bibnamefont {Wang}},\ }\bibfield  {title} {\bibinfo
  {title} {{QNMs of slowly rotating Einstein\textendash{}Bumblebee black
  hole}},\ }\href {https://doi.org/10.1140/epjc/s10052-023-11231-5} {\bibfield
  {journal} {\bibinfo  {journal} {Eur. Phys. J. C}\ }\textbf {\bibinfo {volume}
  {83}},\ \bibinfo {pages} {83} (\bibinfo {year} {2023}{\natexlab{a}})},\
  \Eprint {https://arxiv.org/abs/2211.03156} {arXiv:2211.03156 [gr-qc]}
  \BibitemShut {NoStop}%
\bibitem [{\citenamefont {Xu}\ \emph {et~al.}(2023{\natexlab{b}})\citenamefont
  {Xu}, \citenamefont {Liang},\ and\ \citenamefont {Shao}}]{Xu:2023xqh}%
  \BibitemOpen
  \bibfield  {author} {\bibinfo {author} {\bibfnamefont {R.}~\bibnamefont
  {Xu}}, \bibinfo {author} {\bibfnamefont {D.}~\bibnamefont {Liang}},\ and\
  \bibinfo {author} {\bibfnamefont {L.}~\bibnamefont {Shao}},\ }\bibfield
  {title} {\bibinfo {title} {{Bumblebee Black Holes in Light of Event Horizon
  Telescope Observations}},\ }\href {https://doi.org/10.3847/1538-4357/acbdfb}
  {\bibfield  {journal} {\bibinfo  {journal} {Astrophys. J.}\ }\textbf
  {\bibinfo {volume} {945}},\ \bibinfo {pages} {148} (\bibinfo {year}
  {2023}{\natexlab{b}})},\ \Eprint {https://arxiv.org/abs/2302.05671}
  {arXiv:2302.05671 [gr-qc]} \BibitemShut {NoStop}%
\bibitem [{\citenamefont {Zhang}\ \emph {et~al.}(2023)\citenamefont {Zhang},
  \citenamefont {Wang},\ and\ \citenamefont {Jing}}]{Zhang:2023wwk}%
  \BibitemOpen
  \bibfield  {author} {\bibinfo {author} {\bibfnamefont {X.}~\bibnamefont
  {Zhang}}, \bibinfo {author} {\bibfnamefont {M.}~\bibnamefont {Wang}},\ and\
  \bibinfo {author} {\bibfnamefont {J.}~\bibnamefont {Jing}},\ }\bibfield
  {title} {\bibinfo {title} {{Quasinormal modes and late time tails of
  perturbation fields on a Schwarzschild-like black hole with a global monopole
  in the Einstein-bumblebee theory}},\ }\href
  {https://doi.org/10.1007/s11433-023-2153-6} {\bibfield  {journal} {\bibinfo
  {journal} {Sci. China Phys. Mech. Astron.}\ }\textbf {\bibinfo {volume}
  {66}},\ \bibinfo {pages} {100411} (\bibinfo {year} {2023})},\ \Eprint
  {https://arxiv.org/abs/2307.10856} {arXiv:2307.10856 [gr-qc]} \BibitemShut
  {NoStop}%
\bibitem [{\citenamefont {Chen}\ \emph
  {et~al.}(2020{\natexlab{a}})\citenamefont {Chen}, \citenamefont {Wang},\ and\
  \citenamefont {Jing}}]{Chen:2020qyp}%
  \BibitemOpen
  \bibfield  {author} {\bibinfo {author} {\bibfnamefont {S.}~\bibnamefont
  {Chen}}, \bibinfo {author} {\bibfnamefont {M.}~\bibnamefont {Wang}},\ and\
  \bibinfo {author} {\bibfnamefont {J.}~\bibnamefont {Jing}},\ }\bibfield
  {title} {\bibinfo {title} {{Polarization effects in Kerr black hole shadow
  due to the coupling between photon and bumblebee field}},\ }\href
  {https://doi.org/10.1007/JHEP07(2020)054} {\bibfield  {journal} {\bibinfo
  {journal} {JHEP}\ }\textbf {\bibinfo {volume} {07}},\ \bibinfo {pages}
  {054}},\ \Eprint {https://arxiv.org/abs/2004.08857} {arXiv:2004.08857
  [gr-qc]} \BibitemShut {NoStop}%
\bibitem [{\citenamefont {Chen}\ \emph {et~al.}(2023)\citenamefont {Chen},
  \citenamefont {Pan},\ and\ \citenamefont {Jing}}]{Chen:2023cjd}%
  \BibitemOpen
  \bibfield  {author} {\bibinfo {author} {\bibfnamefont {C.}~\bibnamefont
  {Chen}}, \bibinfo {author} {\bibfnamefont {Q.}~\bibnamefont {Pan}},\ and\
  \bibinfo {author} {\bibfnamefont {J.}~\bibnamefont {Jing}},\ }\bibfield
  {title} {\bibinfo {title} {{Quasinormal modes of a scalar perturbation around
  a rotating BTZ-like black hole in Einstein-bumblebee gravity}},\ }\href
  {https://doi.org/10.1016/j.physletb.2023.138186} {\bibfield  {journal}
  {\bibinfo  {journal} {Phys. Lett. B}\ }\textbf {\bibinfo {volume} {846}},\
  \bibinfo {pages} {138186} (\bibinfo {year} {2023})},\ \Eprint
  {https://arxiv.org/abs/2302.05861} {arXiv:2302.05861 [gr-qc]} \BibitemShut
  {NoStop}%
\bibitem [{\citenamefont {Ge}\ \emph {et~al.}(2025)\citenamefont {Ge},
  \citenamefont {Pan}, \citenamefont {Chen},\ and\ \citenamefont
  {Jing}}]{Ge:2025xuy}%
  \BibitemOpen
  \bibfield  {author} {\bibinfo {author} {\bibfnamefont {F.}~\bibnamefont
  {Ge}}, \bibinfo {author} {\bibfnamefont {Q.}~\bibnamefont {Pan}}, \bibinfo
  {author} {\bibfnamefont {S.}~\bibnamefont {Chen}},\ and\ \bibinfo {author}
  {\bibfnamefont {J.}~\bibnamefont {Jing}},\ }\bibfield  {title} {\bibinfo
  {title} {{Mass ladder operators and quasinormal modes of the static BTZ-like
  black hole in Einstein-bumblebee gravity}},\ }\href
  {https://doi.org/10.1088/1674-1137/ad83a9} {\bibfield  {journal} {\bibinfo
  {journal} {Chin. Phys. C}\ }\textbf {\bibinfo {volume} {49}},\ \bibinfo
  {pages} {015105} (\bibinfo {year} {2025})}\BibitemShut {NoStop}%
\bibitem [{\citenamefont {Eslam~Panah}(2025)}]{EslamPanah:2025zcm}%
  \BibitemOpen
  \bibfield  {author} {\bibinfo {author} {\bibfnamefont {B.}~\bibnamefont
  {Eslam~Panah}},\ }\bibfield  {title} {\bibinfo {title} {{Super-entropy
  bumblebee AdS black holes}},\ }\href
  {https://doi.org/10.1016/j.physletb.2025.139273} {\bibfield  {journal}
  {\bibinfo  {journal} {Phys. Lett. B}\ }\textbf {\bibinfo {volume} {861}},\
  \bibinfo {pages} {139273} (\bibinfo {year} {2025})},\ \Eprint
  {https://arxiv.org/abs/2501.09317} {arXiv:2501.09317 [gr-qc]} \BibitemShut
  {NoStop}%
\bibitem [{\citenamefont {Liu}\ \emph {et~al.}(2025{\natexlab{c}})\citenamefont
  {Liu}, \citenamefont {Wen},\ and\ \citenamefont {Wang}}]{Liu:2024wpa}%
  \BibitemOpen
  \bibfield  {author} {\bibinfo {author} {\bibfnamefont {W.}~\bibnamefont
  {Liu}}, \bibinfo {author} {\bibfnamefont {C.}~\bibnamefont {Wen}},\ and\
  \bibinfo {author} {\bibfnamefont {J.}~\bibnamefont {Wang}},\ }\bibfield
  {title} {\bibinfo {title} {{Lorentz violation alleviates gravitationally
  induced entanglement degradation}},\ }\href
  {https://doi.org/10.1007/JHEP01(2025)184} {\bibfield  {journal} {\bibinfo
  {journal} {JHEP}\ }\textbf {\bibinfo {volume} {01}},\ \bibinfo {pages}
  {184}},\ \Eprint {https://arxiv.org/abs/2410.21681} {arXiv:2410.21681
  [gr-qc]} \BibitemShut {NoStop}%
\bibitem [{\citenamefont {Ara\'ujo~Filho}(2024)}]{AraujoFilho:2024ctw}%
  \BibitemOpen
  \bibfield  {author} {\bibinfo {author} {\bibfnamefont {A.~A.}\ \bibnamefont
  {Ara\'ujo~Filho}},\ }\bibfield  {title} {\bibinfo {title} {{Particle creation
  and evaporation in Kalb-Ramond gravity}},\ }\href@noop {} {\  (\bibinfo
  {year} {2024})},\ \Eprint {https://arxiv.org/abs/2411.06841}
  {arXiv:2411.06841 [gr-qc]} \BibitemShut {NoStop}%
\bibitem [{\citenamefont {Liu}\ \emph {et~al.}(2025{\natexlab{d}})\citenamefont
  {Liu}, \citenamefont {Liu}, \citenamefont {Liu},\ and\ \citenamefont
  {Wang}}]{Liu:2025bpp}%
  \BibitemOpen
  \bibfield  {author} {\bibinfo {author} {\bibfnamefont {X.}~\bibnamefont
  {Liu}}, \bibinfo {author} {\bibfnamefont {W.}~\bibnamefont {Liu}}, \bibinfo
  {author} {\bibfnamefont {Z.}~\bibnamefont {Liu}},\ and\ \bibinfo {author}
  {\bibfnamefont {J.}~\bibnamefont {Wang}},\ }\bibfield  {title} {\bibinfo
  {title} {{Harvesting correlations from BTZ black hole coupled to a
  Lorentz-violating vector field}},\ }\href@noop {} {\  (\bibinfo {year}
  {2025}{\natexlab{d}})},\ \Eprint {https://arxiv.org/abs/2503.06404}
  {arXiv:2503.06404 [gr-qc]} \BibitemShut {NoStop}%
\bibitem [{\citenamefont {Ara\'ujo~Filho}(2025)}]{AraujoFilho:2025hkm}%
  \BibitemOpen
  \bibfield  {author} {\bibinfo {author} {\bibfnamefont {A.~A.}\ \bibnamefont
  {Ara\'ujo~Filho}},\ }\bibfield  {title} {\bibinfo {title} {{How does
  non-metricity affect particle creation and evaporation in bumblebee
  gravity?}},\ }\href@noop {} {\  (\bibinfo {year} {2025})},\ \Eprint
  {https://arxiv.org/abs/2501.00927} {arXiv:2501.00927 [gr-qc]} \BibitemShut
  {NoStop}%
\bibitem [{\citenamefont {Ara{\'u}jo~Filho}(2025)}]{AraujoFilho:2025fwd}%
  \BibitemOpen
  \bibfield  {author} {\bibinfo {author} {\bibfnamefont {A.~A.}\ \bibnamefont
  {Ara{\'u}jo~Filho}},\ }\bibfield  {title} {\bibinfo {title} {{Particle motion
  and thermal effects around a Kalb-Ramond black hole}},\ }\href@noop {} {\
  (\bibinfo {year} {2025})},\ \Eprint {https://arxiv.org/abs/2504.19246}
  {arXiv:2504.19246 [gr-qc]} \BibitemShut {NoStop}%
\bibitem [{\citenamefont {Shi}\ and\ \citenamefont
  {Ara{\'u}jo~Filho}(2025{\natexlab{a}})}]{Shi:2025ywa}%
  \BibitemOpen
  \bibfield  {author} {\bibinfo {author} {\bibfnamefont {Y.}~\bibnamefont
  {Shi}}\ and\ \bibinfo {author} {\bibfnamefont {A.~A.}\ \bibnamefont
  {Ara{\'u}jo~Filho}},\ }\bibfield  {title} {\bibinfo {title} {{The role of
  non-metricity on neutrino behavior in bumblebee gravity}},\ }\href@noop {} {\
   (\bibinfo {year} {2025}{\natexlab{a}})},\ \Eprint
  {https://arxiv.org/abs/2505.12551} {arXiv:2505.12551 [gr-qc]} \BibitemShut
  {NoStop}%
\bibitem [{\citenamefont {Shi}\ and\ \citenamefont
  {Ara{\'u}jo~Filho}(2025{\natexlab{b}})}]{Shi:2025plr}%
  \BibitemOpen
  \bibfield  {author} {\bibinfo {author} {\bibfnamefont {Y.}~\bibnamefont
  {Shi}}\ and\ \bibinfo {author} {\bibfnamefont {A.~A.}\ \bibnamefont
  {Ara{\'u}jo~Filho}},\ }\bibfield  {title} {\bibinfo {title} {{Effects of
  bumblebee gravity on neutrino motion}},\ }\href@noop {} {\  (\bibinfo {year}
  {2025}{\natexlab{b}})},\ \Eprint {https://arxiv.org/abs/2505.02290}
  {arXiv:2505.02290 [gr-qc]} \BibitemShut {NoStop}%
\bibitem [{\citenamefont {Guo}\ \emph {et~al.}(2023{\natexlab{a}})\citenamefont
  {Guo}, \citenamefont {Tan},\ and\ \citenamefont {Liu}}]{Guo:2023nkd}%
  \BibitemOpen
  \bibfield  {author} {\bibinfo {author} {\bibfnamefont {W.-D.}\ \bibnamefont
  {Guo}}, \bibinfo {author} {\bibfnamefont {Q.}~\bibnamefont {Tan}},\ and\
  \bibinfo {author} {\bibfnamefont {Y.-X.}\ \bibnamefont {Liu}},\ }\bibfield
  {title} {\bibinfo {title} {{Quasinormal modes and greybody factor of a
  Lorentz-violating black hole}},\ }\href@noop {} {\  (\bibinfo {year}
  {2023}{\natexlab{a}})},\ \Eprint {https://arxiv.org/abs/2312.16605}
  {arXiv:2312.16605 [gr-qc]} \BibitemShut {NoStop}%
\bibitem [{\citenamefont {Mai}\ \emph {et~al.}(2023)\citenamefont {Mai},
  \citenamefont {Xu}, \citenamefont {Liang},\ and\ \citenamefont
  {Shao}}]{Mai:2023ggs}%
  \BibitemOpen
  \bibfield  {author} {\bibinfo {author} {\bibfnamefont {Z.-F.}\ \bibnamefont
  {Mai}}, \bibinfo {author} {\bibfnamefont {R.}~\bibnamefont {Xu}}, \bibinfo
  {author} {\bibfnamefont {D.}~\bibnamefont {Liang}},\ and\ \bibinfo {author}
  {\bibfnamefont {L.}~\bibnamefont {Shao}},\ }\bibfield  {title} {\bibinfo
  {title} {{Extended thermodynamics of the bumblebee black holes}},\ }\href
  {https://doi.org/10.1103/PhysRevD.108.024004} {\bibfield  {journal} {\bibinfo
   {journal} {Phys. Rev. D}\ }\textbf {\bibinfo {volume} {108}},\ \bibinfo
  {pages} {024004} (\bibinfo {year} {2023})},\ \Eprint
  {https://arxiv.org/abs/2304.08030} {arXiv:2304.08030 [gr-qc]} \BibitemShut
  {NoStop}%
\bibitem [{\citenamefont {Mai}\ \emph {et~al.}(2024)\citenamefont {Mai},
  \citenamefont {Xu}, \citenamefont {Liang},\ and\ \citenamefont
  {Shao}}]{Mai:2024lgk}%
  \BibitemOpen
  \bibfield  {author} {\bibinfo {author} {\bibfnamefont {Z.-F.}\ \bibnamefont
  {Mai}}, \bibinfo {author} {\bibfnamefont {R.}~\bibnamefont {Xu}}, \bibinfo
  {author} {\bibfnamefont {D.}~\bibnamefont {Liang}},\ and\ \bibinfo {author}
  {\bibfnamefont {L.}~\bibnamefont {Shao}},\ }\bibfield  {title} {\bibinfo
  {title} {{Dynamic instability analysis for bumblebee black holes: The odd
  parity}},\ }\href {https://doi.org/10.1103/PhysRevD.109.084076} {\bibfield
  {journal} {\bibinfo  {journal} {Phys. Rev. D}\ }\textbf {\bibinfo {volume}
  {109}},\ \bibinfo {pages} {084076} (\bibinfo {year} {2024})},\ \Eprint
  {https://arxiv.org/abs/2401.07757} {arXiv:2401.07757 [gr-qc]} \BibitemShut
  {NoStop}%
\bibitem [{\citenamefont {Regge}\ and\ \citenamefont
  {Wheeler}(1957)}]{Regge:1957td}%
  \BibitemOpen
  \bibfield  {author} {\bibinfo {author} {\bibfnamefont {T.}~\bibnamefont
  {Regge}}\ and\ \bibinfo {author} {\bibfnamefont {J.~A.}\ \bibnamefont
  {Wheeler}},\ }\bibfield  {title} {\bibinfo {title} {{Stability of a
  Schwarzschild singularity}},\ }\href
  {https://doi.org/10.1103/PhysRev.108.1063} {\bibfield  {journal} {\bibinfo
  {journal} {Phys. Rev.}\ }\textbf {\bibinfo {volume} {108}},\ \bibinfo {pages}
  {1063} (\bibinfo {year} {1957})}\BibitemShut {NoStop}%
\bibitem [{\citenamefont {Zerilli}(1970{\natexlab{a}})}]{Zerilli:1970se}%
  \BibitemOpen
  \bibfield  {author} {\bibinfo {author} {\bibfnamefont {F.~J.}\ \bibnamefont
  {Zerilli}},\ }\bibfield  {title} {\bibinfo {title} {{Effective potential for
  even parity Regge-Wheeler gravitational perturbation equations}},\ }\href
  {https://doi.org/10.1103/PhysRevLett.24.737} {\bibfield  {journal} {\bibinfo
  {journal} {Phys. Rev. Lett.}\ }\textbf {\bibinfo {volume} {24}},\ \bibinfo
  {pages} {737} (\bibinfo {year} {1970}{\natexlab{a}})}\BibitemShut {NoStop}%
\bibitem [{\citenamefont {Zerilli}(1974)}]{Zerilli:1974ai}%
  \BibitemOpen
  \bibfield  {author} {\bibinfo {author} {\bibfnamefont {F.~J.}\ \bibnamefont
  {Zerilli}},\ }\bibfield  {title} {\bibinfo {title} {{Perturbation analysis
  for gravitational and electromagnetic radiation in a reissner-nordstroem
  geometry}},\ }\href {https://doi.org/10.1103/PhysRevD.9.860} {\bibfield
  {journal} {\bibinfo  {journal} {Phys. Rev. D}\ }\textbf {\bibinfo {volume}
  {9}},\ \bibinfo {pages} {860} (\bibinfo {year} {1974})}\BibitemShut {NoStop}%
\bibitem [{\citenamefont {Teukolsky}(1972)}]{Teukolsky:1972my}%
  \BibitemOpen
  \bibfield  {author} {\bibinfo {author} {\bibfnamefont {S.~A.}\ \bibnamefont
  {Teukolsky}},\ }\bibfield  {title} {\bibinfo {title} {{Rotating black holes -
  separable wave equations for gravitational and electromagnetic
  perturbations}},\ }\href {https://doi.org/10.1103/PhysRevLett.29.1114}
  {\bibfield  {journal} {\bibinfo  {journal} {Phys. Rev. Lett.}\ }\textbf
  {\bibinfo {volume} {29}},\ \bibinfo {pages} {1114} (\bibinfo {year}
  {1972})}\BibitemShut {NoStop}%
\bibitem [{\citenamefont {Jing}\ \emph
  {et~al.}(2022{\natexlab{a}})\citenamefont {Jing}, \citenamefont {Long},
  \citenamefont {Deng}, \citenamefont {Wang},\ and\ \citenamefont
  {Wang}}]{Jing:2022vks}%
  \BibitemOpen
  \bibfield  {author} {\bibinfo {author} {\bibfnamefont {J.}~\bibnamefont
  {Jing}}, \bibinfo {author} {\bibfnamefont {S.}~\bibnamefont {Long}}, \bibinfo
  {author} {\bibfnamefont {W.}~\bibnamefont {Deng}}, \bibinfo {author}
  {\bibfnamefont {M.}~\bibnamefont {Wang}},\ and\ \bibinfo {author}
  {\bibfnamefont {J.}~\bibnamefont {Wang}},\ }\bibfield  {title} {\bibinfo
  {title} {{New self-consistent effective one-body theory for spinless binaries
  based on the post-Minkowskian approximation}},\ }\href
  {https://doi.org/10.1007/s11433-022-1951-1} {\bibfield  {journal} {\bibinfo
  {journal} {Sci. China Phys. Mech. Astron.}\ }\textbf {\bibinfo {volume}
  {65}},\ \bibinfo {pages} {100411} (\bibinfo {year} {2022}{\natexlab{a}})},\
  \Eprint {https://arxiv.org/abs/2208.02420} {arXiv:2208.02420 [gr-qc]}
  \BibitemShut {NoStop}%
\bibitem [{\citenamefont {Jing}\ \emph
  {et~al.}(2023{\natexlab{a}})\citenamefont {Jing}, \citenamefont {Deng},
  \citenamefont {Long},\ and\ \citenamefont {Wang}}]{Jing:2023okh}%
  \BibitemOpen
  \bibfield  {author} {\bibinfo {author} {\bibfnamefont {J.}~\bibnamefont
  {Jing}}, \bibinfo {author} {\bibfnamefont {W.}~\bibnamefont {Deng}}, \bibinfo
  {author} {\bibfnamefont {S.}~\bibnamefont {Long}},\ and\ \bibinfo {author}
  {\bibfnamefont {J.}~\bibnamefont {Wang}},\ }\bibfield  {title} {\bibinfo
  {title} {{Effective metric of spinless binaries with radiation-reaction
  effect up to fourth post-Minkowskian order in effective-one-body theory}},\
  }\href {https://doi.org/10.1140/epjc/s10052-023-11705-6} {\bibfield
  {journal} {\bibinfo  {journal} {Eur. Phys. J. C}\ }\textbf {\bibinfo {volume}
  {83}},\ \bibinfo {pages} {608} (\bibinfo {year} {2023}{\natexlab{a}})},\
  \bibinfo {note} {[Erratum: Eur.Phys.J.C 83, 712 (2023)]},\ \Eprint
  {https://arxiv.org/abs/2307.05971} {arXiv:2307.05971 [gr-qc]} \BibitemShut
  {NoStop}%
\bibitem [{\citenamefont {Long}\ \emph {et~al.}(2023)\citenamefont {Long},
  \citenamefont {Zou},\ and\ \citenamefont {Jing}}]{Long:2023vph}%
  \BibitemOpen
  \bibfield  {author} {\bibinfo {author} {\bibfnamefont {S.}~\bibnamefont
  {Long}}, \bibinfo {author} {\bibfnamefont {Y.}~\bibnamefont {Zou}},\ and\
  \bibinfo {author} {\bibfnamefont {J.}~\bibnamefont {Jing}},\ }\bibfield
  {title} {\bibinfo {title} {{Reconstruction of gravitational waveforms of
  coalescing spinless binaries in EOB theory based on PM approximation}},\
  }\href {https://doi.org/10.1088/1361-6382/acfdee} {\bibfield  {journal}
  {\bibinfo  {journal} {Class. Quant. Grav.}\ }\textbf {\bibinfo {volume}
  {40}},\ \bibinfo {pages} {225006} (\bibinfo {year} {2023})}\BibitemShut
  {NoStop}%
\bibitem [{\citenamefont {Long}\ \emph {et~al.}(2024)\citenamefont {Long},
  \citenamefont {Deng},\ and\ \citenamefont {Jing}}]{Long:2024axi}%
  \BibitemOpen
  \bibfield  {author} {\bibinfo {author} {\bibfnamefont {S.}~\bibnamefont
  {Long}}, \bibinfo {author} {\bibfnamefont {W.}~\bibnamefont {Deng}},\ and\
  \bibinfo {author} {\bibfnamefont {J.}~\bibnamefont {Jing}},\ }\bibfield
  {title} {\bibinfo {title} {{Energy flux and waveforms by coalescing spinless
  binary system in effective one-body theory}},\ }\href
  {https://doi.org/10.1007/s11433-023-2354-1} {\bibfield  {journal} {\bibinfo
  {journal} {Sci. China Phys. Mech. Astron.}\ }\textbf {\bibinfo {volume}
  {67}},\ \bibinfo {pages} {260412} (\bibinfo {year} {2024})},\ \Eprint
  {https://arxiv.org/abs/2403.09211} {arXiv:2403.09211 [gr-qc]} \BibitemShut
  {NoStop}%
\bibitem [{\citenamefont {Wu}\ \emph {et~al.}(2023)\citenamefont {Wu},
  \citenamefont {Fan}, \citenamefont {Wang}, \citenamefont {Wu}, \citenamefont
  {Huang},\ and\ \citenamefont {Zeng}}]{Wu:2023sye}%
  \BibitemOpen
  \bibfield  {author} {\bibinfo {author} {\bibfnamefont {S.~M.}\ \bibnamefont
  {Wu}}, \bibinfo {author} {\bibfnamefont {X.~W.}\ \bibnamefont {Fan}},
  \bibinfo {author} {\bibfnamefont {R.~D.}\ \bibnamefont {Wang}}, \bibinfo
  {author} {\bibfnamefont {H.~Y.}\ \bibnamefont {Wu}}, \bibinfo {author}
  {\bibfnamefont {X.~L.}\ \bibnamefont {Huang}},\ and\ \bibinfo {author}
  {\bibfnamefont {H.~S.}\ \bibnamefont {Zeng}},\ }\bibfield  {title} {\bibinfo
  {title} {{Does Hawking effect always degrade fidelity of quantum
  teleportation in Schwarzschild spacetime?}},\ }\href
  {https://doi.org/10.1007/JHEP11(2023)232} {\bibfield  {journal} {\bibinfo
  {journal} {JHEP}\ }\textbf {\bibinfo {volume} {11}},\ \bibinfo {pages}
  {232}},\ \Eprint {https://arxiv.org/abs/2304.00984} {arXiv:2304.00984
  [gr-qc]} \BibitemShut {NoStop}%
\bibitem [{\citenamefont
  {L{\"u}tf{\"u}o{\u{g}}lu}(2025{\natexlab{a}})}]{Lutfuoglu:2025bsf}%
  \BibitemOpen
  \bibfield  {author} {\bibinfo {author} {\bibfnamefont {B.~C.}\ \bibnamefont
  {L{\"u}tf{\"u}o{\u{g}}lu}},\ }\bibfield  {title} {\bibinfo {title}
  {{Long-lived quasinormal modes in the Euler-Heisenberg electrodynamics}},\
  }\href@noop {} {\  (\bibinfo {year} {2025}{\natexlab{a}})},\ \Eprint
  {https://arxiv.org/abs/2508.13361} {arXiv:2508.13361 [gr-qc]} \BibitemShut
  {NoStop}%
\bibitem [{\citenamefont
  {L{\"u}tf{\"u}o{\u{g}}lu}(2025{\natexlab{b}})}]{Lutfuoglu:2025qkt}%
  \BibitemOpen
  \bibfield  {author} {\bibinfo {author} {\bibfnamefont {B.~C.}\ \bibnamefont
  {L{\"u}tf{\"u}o{\u{g}}lu}},\ }\bibfield  {title} {\bibinfo {title}
  {{Long-lived quasinormal modes and echoes in the Einstein-Gauss-Bonnet-Proca
  theory}},\ }\href@noop {} {\  (\bibinfo {year} {2025}{\natexlab{b}})},\
  \Eprint {https://arxiv.org/abs/2508.19194} {arXiv:2508.19194 [gr-qc]}
  \BibitemShut {NoStop}%
\bibitem [{\citenamefont {Konoplya}\ and\ \citenamefont
  {Zhidenko}(2005)}]{Konoplya:2004wg}%
  \BibitemOpen
  \bibfield  {author} {\bibinfo {author} {\bibfnamefont {R.~A.}\ \bibnamefont
  {Konoplya}}\ and\ \bibinfo {author} {\bibfnamefont {A.~V.}\ \bibnamefont
  {Zhidenko}},\ }\bibfield  {title} {\bibinfo {title} {{Decay of massive scalar
  field in a Schwarzschild background}},\ }\href
  {https://doi.org/10.1016/j.physletb.2005.01.078} {\bibfield  {journal}
  {\bibinfo  {journal} {Phys. Lett. B}\ }\textbf {\bibinfo {volume} {609}},\
  \bibinfo {pages} {377} (\bibinfo {year} {2005})},\ \Eprint
  {https://arxiv.org/abs/gr-qc/0411059} {arXiv:gr-qc/0411059} \BibitemShut
  {NoStop}%
\bibitem [{\citenamefont {Konoplya}\ and\ \citenamefont
  {Zhidenko}(2013)}]{Konoplya:2013rxa}%
  \BibitemOpen
  \bibfield  {author} {\bibinfo {author} {\bibfnamefont {R.~A.}\ \bibnamefont
  {Konoplya}}\ and\ \bibinfo {author} {\bibfnamefont {A.}~\bibnamefont
  {Zhidenko}},\ }\bibfield  {title} {\bibinfo {title} {{Massive charged scalar
  field in the Kerr-Newman background I: quasinormal modes, late-time tails and
  stability}},\ }\href {https://doi.org/10.1103/PhysRevD.88.024054} {\bibfield
  {journal} {\bibinfo  {journal} {Phys. Rev. D}\ }\textbf {\bibinfo {volume}
  {88}},\ \bibinfo {pages} {024054} (\bibinfo {year} {2013})},\ \Eprint
  {https://arxiv.org/abs/1307.1812} {arXiv:1307.1812 [gr-qc]} \BibitemShut
  {NoStop}%
\bibitem [{\citenamefont {Konoplya}\ and\ \citenamefont
  {Zhidenko}(2011)}]{Konoplya:2011qq}%
  \BibitemOpen
  \bibfield  {author} {\bibinfo {author} {\bibfnamefont {R.~A.}\ \bibnamefont
  {Konoplya}}\ and\ \bibinfo {author} {\bibfnamefont {A.}~\bibnamefont
  {Zhidenko}},\ }\bibfield  {title} {\bibinfo {title} {{Quasinormal modes of
  black holes: From astrophysics to string theory}},\ }\href
  {https://doi.org/10.1103/RevModPhys.83.793} {\bibfield  {journal} {\bibinfo
  {journal} {Rev. Mod. Phys.}\ }\textbf {\bibinfo {volume} {83}},\ \bibinfo
  {pages} {793} (\bibinfo {year} {2011})},\ \Eprint
  {https://arxiv.org/abs/1102.4014} {arXiv:1102.4014 [gr-qc]} \BibitemShut
  {NoStop}%
\bibitem [{\citenamefont {Vishveshwara}(1970)}]{Vishveshwara:1970zz}%
  \BibitemOpen
  \bibfield  {author} {\bibinfo {author} {\bibfnamefont {C.~V.}\ \bibnamefont
  {Vishveshwara}},\ }\bibfield  {title} {\bibinfo {title} {{Scattering of
  Gravitational Radiation by a Schwarzschild Black-hole}},\ }\href
  {https://doi.org/10.1038/227936a0} {\bibfield  {journal} {\bibinfo  {journal}
  {Nature}\ }\textbf {\bibinfo {volume} {227}},\ \bibinfo {pages} {936}
  (\bibinfo {year} {1970})}\BibitemShut {NoStop}%
\bibitem [{\citenamefont {Berti}\ \emph {et~al.}(2009)\citenamefont {Berti},
  \citenamefont {Cardoso},\ and\ \citenamefont {Starinets}}]{Berti:2009kk}%
  \BibitemOpen
  \bibfield  {author} {\bibinfo {author} {\bibfnamefont {E.}~\bibnamefont
  {Berti}}, \bibinfo {author} {\bibfnamefont {V.}~\bibnamefont {Cardoso}},\
  and\ \bibinfo {author} {\bibfnamefont {A.~O.}\ \bibnamefont {Starinets}},\
  }\bibfield  {title} {\bibinfo {title} {{Quasinormal modes of black holes and
  black branes}},\ }\href {https://doi.org/10.1088/0264-9381/26/16/163001}
  {\bibfield  {journal} {\bibinfo  {journal} {Class. Quant. Grav.}\ }\textbf
  {\bibinfo {volume} {26}},\ \bibinfo {pages} {163001} (\bibinfo {year}
  {2009})},\ \Eprint {https://arxiv.org/abs/0905.2975} {arXiv:0905.2975
  [gr-qc]} \BibitemShut {NoStop}%
\bibitem [{\citenamefont {Berti}\ \emph {et~al.}(2025)\citenamefont {Berti}
  \emph {et~al.}}]{Berti:2025hly}%
  \BibitemOpen
  \bibfield  {author} {\bibinfo {author} {\bibfnamefont {E.}~\bibnamefont
  {Berti}} \emph {et~al.},\ }\bibfield  {title} {\bibinfo {title} {{Black hole
  spectroscopy: from theory to experiment}},\ }\href@noop {} {\  (\bibinfo
  {year} {2025})},\ \Eprint {https://arxiv.org/abs/2505.23895}
  {arXiv:2505.23895 [gr-qc]} \BibitemShut {NoStop}%
\bibitem [{\citenamefont {Zerilli}(1970{\natexlab{b}})}]{Zerilli:1970wzz}%
  \BibitemOpen
  \bibfield  {author} {\bibinfo {author} {\bibfnamefont {F.~J.}\ \bibnamefont
  {Zerilli}},\ }\bibfield  {title} {\bibinfo {title} {{Gravitational field of a
  particle falling in a schwarzschild geometry analyzed in tensor harmonics}},\
  }\href {https://doi.org/10.1103/PhysRevD.2.2141} {\bibfield  {journal}
  {\bibinfo  {journal} {Phys. Rev. D}\ }\textbf {\bibinfo {volume} {2}},\
  \bibinfo {pages} {2141} (\bibinfo {year} {1970}{\natexlab{b}})}\BibitemShut
  {NoStop}%
\bibitem [{\citenamefont {Thompson}\ \emph {et~al.}(2017)\citenamefont
  {Thompson}, \citenamefont {Whiting},\ and\ \citenamefont
  {Chen}}]{Thompson:2016fxe}%
  \BibitemOpen
  \bibfield  {author} {\bibinfo {author} {\bibfnamefont {J.~E.}\ \bibnamefont
  {Thompson}}, \bibinfo {author} {\bibfnamefont {B.~F.}\ \bibnamefont
  {Whiting}},\ and\ \bibinfo {author} {\bibfnamefont {H.}~\bibnamefont
  {Chen}},\ }\bibfield  {title} {\bibinfo {title} {{Gauge Invariant
  Perturbations of the Schwarzschild Spacetime}},\ }\href
  {https://doi.org/10.1088/1361-6382/aa7f5b} {\bibfield  {journal} {\bibinfo
  {journal} {Class. Quant. Grav.}\ }\textbf {\bibinfo {volume} {34}},\ \bibinfo
  {pages} {174001} (\bibinfo {year} {2017})},\ \Eprint
  {https://arxiv.org/abs/1611.06214} {arXiv:1611.06214 [gr-qc]} \BibitemShut
  {NoStop}%
\bibitem [{\citenamefont {Jing}\ \emph
  {et~al.}(2022{\natexlab{b}})\citenamefont {Jing}, \citenamefont {Chen},
  \citenamefont {Sun}, \citenamefont {He}, \citenamefont {Wang},\ and\
  \citenamefont {Wang}}]{Jing:2021ahx}%
  \BibitemOpen
  \bibfield  {author} {\bibinfo {author} {\bibfnamefont {J.}~\bibnamefont
  {Jing}}, \bibinfo {author} {\bibfnamefont {S.}~\bibnamefont {Chen}}, \bibinfo
  {author} {\bibfnamefont {M.}~\bibnamefont {Sun}}, \bibinfo {author}
  {\bibfnamefont {X.}~\bibnamefont {He}}, \bibinfo {author} {\bibfnamefont
  {M.}~\bibnamefont {Wang}},\ and\ \bibinfo {author} {\bibfnamefont
  {J.}~\bibnamefont {Wang}},\ }\bibfield  {title} {\bibinfo {title}
  {{Self-consistent effective-one-body theory for spinless binaries based on
  post-Minkowskian approximation I: Hamiltonian and decoupled equation for
  $\psi _4^{\rm{B}}$}},\ }\href {https://doi.org/10.1007/s11433-022-1885-6}
  {\bibfield  {journal} {\bibinfo  {journal} {Sci. China Phys. Mech. Astron.}\
  }\textbf {\bibinfo {volume} {65}},\ \bibinfo {pages} {260411} (\bibinfo
  {year} {2022}{\natexlab{b}})},\ \Eprint {https://arxiv.org/abs/2112.09838}
  {arXiv:2112.09838 [gr-qc]} \BibitemShut {NoStop}%
\bibitem [{\citenamefont {Lenzi}\ and\ \citenamefont
  {Sopuerta}(2021)}]{Lenzi:2021wpc}%
  \BibitemOpen
  \bibfield  {author} {\bibinfo {author} {\bibfnamefont {M.}~\bibnamefont
  {Lenzi}}\ and\ \bibinfo {author} {\bibfnamefont {C.~F.}\ \bibnamefont
  {Sopuerta}},\ }\bibfield  {title} {\bibinfo {title} {{Master functions and
  equations for perturbations of vacuum spherically symmetric spacetimes}},\
  }\href {https://doi.org/10.1103/PhysRevD.104.084053} {\bibfield  {journal}
  {\bibinfo  {journal} {Phys. Rev. D}\ }\textbf {\bibinfo {volume} {104}},\
  \bibinfo {pages} {084053} (\bibinfo {year} {2021})},\ \Eprint
  {https://arxiv.org/abs/2108.08668} {arXiv:2108.08668 [gr-qc]} \BibitemShut
  {NoStop}%
\bibitem [{\citenamefont {Liu}\ \emph {et~al.}(2023{\natexlab{b}})\citenamefont
  {Liu}, \citenamefont {Fang}, \citenamefont {Jing},\ and\ \citenamefont
  {Wang}}]{Liu2023}%
  \BibitemOpen
  \bibfield  {author} {\bibinfo {author} {\bibfnamefont {W.}~\bibnamefont
  {Liu}}, \bibinfo {author} {\bibfnamefont {X.}~\bibnamefont {Fang}}, \bibinfo
  {author} {\bibfnamefont {J.}~\bibnamefont {Jing}},\ and\ \bibinfo {author}
  {\bibfnamefont {A.}~\bibnamefont {Wang}},\ }\bibfield  {title} {\bibinfo
  {title} {{Gauge invariant perturbations of general spherically symmetric
  spacetimes}},\ }\href {https://doi.org/10.1007/s11433-022-1956-4} {\bibfield
  {journal} {\bibinfo  {journal} {Sci. China Phys. Mech. Astron.}\ }\textbf
  {\bibinfo {volume} {66}},\ \bibinfo {pages} {210411} (\bibinfo {year}
  {2023}{\natexlab{b}})},\ \Eprint {https://arxiv.org/abs/2201.01259}
  {arXiv:2201.01259 [gr-qc]} \BibitemShut {NoStop}%
\bibitem [{\citenamefont {Tan}\ \emph {et~al.}(2025{\natexlab{a}})\citenamefont
  {Tan}, \citenamefont {Long}, \citenamefont {Deng},\ and\ \citenamefont
  {Jing}}]{Tan:2024aym}%
  \BibitemOpen
  \bibfield  {author} {\bibinfo {author} {\bibfnamefont {Q.}~\bibnamefont
  {Tan}}, \bibinfo {author} {\bibfnamefont {S.}~\bibnamefont {Long}}, \bibinfo
  {author} {\bibfnamefont {W.}~\bibnamefont {Deng}},\ and\ \bibinfo {author}
  {\bibfnamefont {J.}~\bibnamefont {Jing}},\ }\bibfield  {title} {\bibinfo
  {title} {{Graviscalar quasinormal modes and asymptotic tails of a thick
  brane}},\ }\href {https://doi.org/10.1016/j.physletb.2025.139667} {\bibfield
  {journal} {\bibinfo  {journal} {Phys. Lett. B}\ }\textbf {\bibinfo {volume}
  {868}},\ \bibinfo {pages} {139667} (\bibinfo {year} {2025}{\natexlab{a}})},\
  \Eprint {https://arxiv.org/abs/2409.06947} {arXiv:2409.06947 [gr-qc]}
  \BibitemShut {NoStop}%
\bibitem [{\citenamefont {Tan}\ \emph {et~al.}(2025{\natexlab{b}})\citenamefont
  {Tan}, \citenamefont {Long}, \citenamefont {Deng},\ and\ \citenamefont
  {Jing}}]{Tan:2024qij}%
  \BibitemOpen
  \bibfield  {author} {\bibinfo {author} {\bibfnamefont {Q.}~\bibnamefont
  {Tan}}, \bibinfo {author} {\bibfnamefont {S.}~\bibnamefont {Long}}, \bibinfo
  {author} {\bibfnamefont {W.}~\bibnamefont {Deng}},\ and\ \bibinfo {author}
  {\bibfnamefont {J.}~\bibnamefont {Jing}},\ }\bibfield  {title} {\bibinfo
  {title} {{Quasinormal modes and echoes of a double braneworld}},\ }\href
  {https://doi.org/10.1007/JHEP02(2025)055} {\bibfield  {journal} {\bibinfo
  {journal} {JHEP}\ }\textbf {\bibinfo {volume} {02}},\ \bibinfo {pages}
  {055}},\ \Eprint {https://arxiv.org/abs/2410.06945} {arXiv:2410.06945
  [gr-qc]} \BibitemShut {NoStop}%
\bibitem [{\citenamefont {Wu}\ \emph {et~al.}(2024{\natexlab{a}})\citenamefont
  {Wu}, \citenamefont {Teng}, \citenamefont {Li}, \citenamefont {Li},
  \citenamefont {Liu},\ and\ \citenamefont {Wang}}]{Wu:2023spa}%
  \BibitemOpen
  \bibfield  {author} {\bibinfo {author} {\bibfnamefont {S.~M.}\ \bibnamefont
  {Wu}}, \bibinfo {author} {\bibfnamefont {X.~W.}\ \bibnamefont {Teng}},
  \bibinfo {author} {\bibfnamefont {J.~X.}\ \bibnamefont {Li}}, \bibinfo
  {author} {\bibfnamefont {S.~H.}\ \bibnamefont {Li}}, \bibinfo {author}
  {\bibfnamefont {T.~H.}\ \bibnamefont {Liu}},\ and\ \bibinfo {author}
  {\bibfnamefont {J.}~\bibnamefont {Wang}},\ }\bibfield  {title} {\bibinfo
  {title} {{Genuinely accessible and inaccessible entanglement in Schwarzschild
  black hole}},\ }\href {https://doi.org/10.1016/j.physletb.2023.138334}
  {\bibfield  {journal} {\bibinfo  {journal} {Phys. Lett. B}\ }\textbf
  {\bibinfo {volume} {848}},\ \bibinfo {pages} {138334} (\bibinfo {year}
  {2024}{\natexlab{a}})},\ \Eprint {https://arxiv.org/abs/2311.12362}
  {arXiv:2311.12362 [gr-qc]} \BibitemShut {NoStop}%
\bibitem [{\citenamefont {Liu}\ and\ \citenamefont {Zeng}(2025)}]{Liu:2025reu}%
  \BibitemOpen
  \bibfield  {author} {\bibinfo {author} {\bibfnamefont {H.-F.}\ \bibnamefont
  {Liu}}\ and\ \bibinfo {author} {\bibfnamefont {D.-f.}\ \bibnamefont {Zeng}},\
  }\bibfield  {title} {\bibinfo {title} {{Master Functions of
  Reissner-Nordstrom Black Hole Perturbations and Their Darboux
  Transformation}},\ }\href@noop {} {\  (\bibinfo {year} {2025})},\ \Eprint
  {https://arxiv.org/abs/2505.04407} {arXiv:2505.04407 [gr-qc]} \BibitemShut
  {NoStop}%
\bibitem [{\citenamefont {Teukolsky}(1973)}]{Teukolsky:1973ha}%
  \BibitemOpen
  \bibfield  {author} {\bibinfo {author} {\bibfnamefont {S.~A.}\ \bibnamefont
  {Teukolsky}},\ }\bibfield  {title} {\bibinfo {title} {{Perturbations of a
  rotating black hole. 1. Fundamental equations for gravitational
  electromagnetic and neutrino field perturbations}},\ }\href
  {https://doi.org/10.1086/152444} {\bibfield  {journal} {\bibinfo  {journal}
  {Astrophys. J.}\ }\textbf {\bibinfo {volume} {185}},\ \bibinfo {pages} {635}
  (\bibinfo {year} {1973})}\BibitemShut {NoStop}%
\bibitem [{\citenamefont {Teukolsky}\ and\ \citenamefont
  {Press}(1974)}]{Teukolsky:1974yv}%
  \BibitemOpen
  \bibfield  {author} {\bibinfo {author} {\bibfnamefont {S.~A.}\ \bibnamefont
  {Teukolsky}}\ and\ \bibinfo {author} {\bibfnamefont {W.~H.}\ \bibnamefont
  {Press}},\ }\bibfield  {title} {\bibinfo {title} {{Perturbations of a
  rotating black hole. III - Interaction of the hole with gravitational and
  electromagnetic radiation}},\ }\href {https://doi.org/10.1086/153180}
  {\bibfield  {journal} {\bibinfo  {journal} {Astrophys. J.}\ }\textbf
  {\bibinfo {volume} {193}},\ \bibinfo {pages} {443} (\bibinfo {year}
  {1974})}\BibitemShut {NoStop}%
\bibitem [{\citenamefont {Jing}\ \emph
  {et~al.}(2023{\natexlab{b}})\citenamefont {Jing}, \citenamefont {Deng},
  \citenamefont {Long},\ and\ \citenamefont {Wang}}]{Jing:2023vzq}%
  \BibitemOpen
  \bibfield  {author} {\bibinfo {author} {\bibfnamefont {J.}~\bibnamefont
  {Jing}}, \bibinfo {author} {\bibfnamefont {W.}~\bibnamefont {Deng}}, \bibinfo
  {author} {\bibfnamefont {S.}~\bibnamefont {Long}},\ and\ \bibinfo {author}
  {\bibfnamefont {J.}~\bibnamefont {Wang}},\ }\bibfield  {title} {\bibinfo
  {title} {{Self-consistent effective-one-body theory for spinning binaries
  based on post-Minkowskian approximation}},\ }\href
  {https://doi.org/10.1007/s11433-023-2084-1} {\bibfield  {journal} {\bibinfo
  {journal} {Sci. China Phys. Mech. Astron.}\ }\textbf {\bibinfo {volume}
  {66}},\ \bibinfo {pages} {270411} (\bibinfo {year} {2023}{\natexlab{b}})},\
  \Eprint {https://arxiv.org/abs/2305.03225} {arXiv:2305.03225 [gr-qc]}
  \BibitemShut {NoStop}%
\bibitem [{\citenamefont {Deng}\ \emph {et~al.}(2024)\citenamefont {Deng},
  \citenamefont {Long},\ and\ \citenamefont {Jing}}]{Deng:2024ayh}%
  \BibitemOpen
  \bibfield  {author} {\bibinfo {author} {\bibfnamefont {W.}~\bibnamefont
  {Deng}}, \bibinfo {author} {\bibfnamefont {S.}~\bibnamefont {Long}},\ and\
  \bibinfo {author} {\bibfnamefont {J.}~\bibnamefont {Jing}},\ }\bibfield
  {title} {\bibinfo {title} {{Energy flux and waveform of gravitational wave
  generated by coalescing slow-spinning binary system in effective one-body
  theory}},\ }\href@noop {} {\  (\bibinfo {year} {2024})},\ \Eprint
  {https://arxiv.org/abs/2405.16423} {arXiv:2405.16423 [gr-qc]} \BibitemShut
  {NoStop}%
\bibitem [{\citenamefont {Pani}\ \emph {et~al.}(2011)\citenamefont {Pani},
  \citenamefont {Macedo}, \citenamefont {Crispino},\ and\ \citenamefont
  {Cardoso}}]{Pani2011}%
  \BibitemOpen
  \bibfield  {author} {\bibinfo {author} {\bibfnamefont {P.}~\bibnamefont
  {Pani}}, \bibinfo {author} {\bibfnamefont {C.~F.~B.}\ \bibnamefont {Macedo}},
  \bibinfo {author} {\bibfnamefont {L.~C.~B.}\ \bibnamefont {Crispino}},\ and\
  \bibinfo {author} {\bibfnamefont {V.}~\bibnamefont {Cardoso}},\ }\bibfield
  {title} {\bibinfo {title} {{Slowly rotating black holes in alternative
  theories of gravity}},\ }\href {https://doi.org/10.1103/PhysRevD.84.087501}
  {\bibfield  {journal} {\bibinfo  {journal} {Phys. Rev. D}\ }\textbf {\bibinfo
  {volume} {84}},\ \bibinfo {pages} {087501} (\bibinfo {year} {2011})},\
  \Eprint {https://arxiv.org/abs/1109.3996} {arXiv:1109.3996 [gr-qc]}
  \BibitemShut {NoStop}%
\bibitem [{\citenamefont {Pani}\ \emph
  {et~al.}(2012{\natexlab{a}})\citenamefont {Pani}, \citenamefont {Cardoso},
  \citenamefont {Gualtieri}, \citenamefont {Berti},\ and\ \citenamefont
  {Ishibashi}}]{Pani2012}%
  \BibitemOpen
  \bibfield  {author} {\bibinfo {author} {\bibfnamefont {P.}~\bibnamefont
  {Pani}}, \bibinfo {author} {\bibfnamefont {V.}~\bibnamefont {Cardoso}},
  \bibinfo {author} {\bibfnamefont {L.}~\bibnamefont {Gualtieri}}, \bibinfo
  {author} {\bibfnamefont {E.}~\bibnamefont {Berti}},\ and\ \bibinfo {author}
  {\bibfnamefont {A.}~\bibnamefont {Ishibashi}},\ }\bibfield  {title} {\bibinfo
  {title} {{Perturbations of slowly rotating black holes: massive vector fields
  in the Kerr metric}},\ }\href {https://doi.org/10.1103/PhysRevD.86.104017}
  {\bibfield  {journal} {\bibinfo  {journal} {Phys. Rev. D}\ }\textbf {\bibinfo
  {volume} {86}},\ \bibinfo {pages} {104017} (\bibinfo {year}
  {2012}{\natexlab{a}})},\ \Eprint {https://arxiv.org/abs/1209.0773}
  {arXiv:1209.0773 [gr-qc]} \BibitemShut {NoStop}%
\bibitem [{\citenamefont {Pani}\ \emph
  {et~al.}(2012{\natexlab{b}})\citenamefont {Pani}, \citenamefont {Cardoso},
  \citenamefont {Gualtieri}, \citenamefont {Berti},\ and\ \citenamefont
  {Ishibashi}}]{Pani2012prl}%
  \BibitemOpen
  \bibfield  {author} {\bibinfo {author} {\bibfnamefont {P.}~\bibnamefont
  {Pani}}, \bibinfo {author} {\bibfnamefont {V.}~\bibnamefont {Cardoso}},
  \bibinfo {author} {\bibfnamefont {L.}~\bibnamefont {Gualtieri}}, \bibinfo
  {author} {\bibfnamefont {E.}~\bibnamefont {Berti}},\ and\ \bibinfo {author}
  {\bibfnamefont {A.}~\bibnamefont {Ishibashi}},\ }\bibfield  {title} {\bibinfo
  {title} {{Black hole bombs and photon mass bounds}},\ }\href
  {https://doi.org/10.1103/PhysRevLett.109.131102} {\bibfield  {journal}
  {\bibinfo  {journal} {Phys. Rev. Lett.}\ }\textbf {\bibinfo {volume} {109}},\
  \bibinfo {pages} {131102} (\bibinfo {year} {2012}{\natexlab{b}})},\ \Eprint
  {https://arxiv.org/abs/1209.0465} {arXiv:1209.0465 [gr-qc]} \BibitemShut
  {NoStop}%
\bibitem [{\citenamefont {Pani}(2013)}]{Pani2013IJMPA}%
  \BibitemOpen
  \bibfield  {author} {\bibinfo {author} {\bibfnamefont {P.}~\bibnamefont
  {Pani}},\ }\bibfield  {title} {\bibinfo {title} {{Advanced Methods in
  Black-Hole Perturbation Theory}},\ }\href
  {https://doi.org/10.1142/S0217751X13400186} {\bibfield  {journal} {\bibinfo
  {journal} {Int. J. Mod. Phys. A}\ }\textbf {\bibinfo {volume} {28}},\
  \bibinfo {pages} {1340018} (\bibinfo {year} {2013})},\ \Eprint
  {https://arxiv.org/abs/1305.6759} {arXiv:1305.6759 [gr-qc]} \BibitemShut
  {NoStop}%
\bibitem [{\citenamefont {Pani}\ \emph
  {et~al.}(2013{\natexlab{a}})\citenamefont {Pani}, \citenamefont {Berti},\
  and\ \citenamefont {Gualtieri}}]{Pani2013prd}%
  \BibitemOpen
  \bibfield  {author} {\bibinfo {author} {\bibfnamefont {P.}~\bibnamefont
  {Pani}}, \bibinfo {author} {\bibfnamefont {E.}~\bibnamefont {Berti}},\ and\
  \bibinfo {author} {\bibfnamefont {L.}~\bibnamefont {Gualtieri}},\ }\bibfield
  {title} {\bibinfo {title} {{Scalar, Electromagnetic and Gravitational
  Perturbations of Kerr-Newman Black Holes in the Slow-Rotation Limit}},\
  }\href {https://doi.org/10.1103/PhysRevD.88.064048} {\bibfield  {journal}
  {\bibinfo  {journal} {Phys. Rev. D}\ }\textbf {\bibinfo {volume} {88}},\
  \bibinfo {pages} {064048} (\bibinfo {year} {2013}{\natexlab{a}})},\ \Eprint
  {https://arxiv.org/abs/1307.7315} {arXiv:1307.7315 [gr-qc]} \BibitemShut
  {NoStop}%
\bibitem [{\citenamefont {Pani}\ \emph
  {et~al.}(2013{\natexlab{b}})\citenamefont {Pani}, \citenamefont {Berti},\
  and\ \citenamefont {Gualtieri}}]{Pani2013prl}%
  \BibitemOpen
  \bibfield  {author} {\bibinfo {author} {\bibfnamefont {P.}~\bibnamefont
  {Pani}}, \bibinfo {author} {\bibfnamefont {E.}~\bibnamefont {Berti}},\ and\
  \bibinfo {author} {\bibfnamefont {L.}~\bibnamefont {Gualtieri}},\ }\bibfield
  {title} {\bibinfo {title} {{Gravitoelectromagnetic Perturbations of
  Kerr-Newman Black Holes: Stability and Isospectrality in the Slow-Rotation
  Limit}},\ }\href {https://doi.org/10.1103/PhysRevLett.110.241103} {\bibfield
  {journal} {\bibinfo  {journal} {Phys. Rev. Lett.}\ }\textbf {\bibinfo
  {volume} {110}},\ \bibinfo {pages} {241103} (\bibinfo {year}
  {2013}{\natexlab{b}})},\ \Eprint {https://arxiv.org/abs/1304.1160}
  {arXiv:1304.1160 [gr-qc]} \BibitemShut {NoStop}%
\bibitem [{\citenamefont {Kojima}(1992)}]{Kojima}%
  \BibitemOpen
  \bibfield  {author} {\bibinfo {author} {\bibfnamefont {Y.}~\bibnamefont
  {Kojima}},\ }\bibfield  {title} {\bibinfo {title} {{Equations governing the
  nonradial oscillations of a slowly rotating relativistic star}},\ }\href
  {https://doi.org/10.1103/PhysRevD.46.4289} {\bibfield  {journal} {\bibinfo
  {journal} {Phys. Rev. D}\ }\textbf {\bibinfo {volume} {46}},\ \bibinfo
  {pages} {4289} (\bibinfo {year} {1992})}\BibitemShut {NoStop}%
\bibitem [{\citenamefont {Tattersall}(2018)}]{Tattersall2018}%
  \BibitemOpen
  \bibfield  {author} {\bibinfo {author} {\bibfnamefont {O.~J.}\ \bibnamefont
  {Tattersall}},\ }\bibfield  {title} {\bibinfo {title}
  {{Kerr\textendash{}(anti\textendash{})de Sitter black holes: Perturbations
  and quasinormal modes in the slow rotation limit}},\ }\href
  {https://doi.org/10.1103/PhysRevD.98.104013} {\bibfield  {journal} {\bibinfo
  {journal} {Phys. Rev. D}\ }\textbf {\bibinfo {volume} {98}},\ \bibinfo
  {pages} {104013} (\bibinfo {year} {2018})},\ \Eprint
  {https://arxiv.org/abs/1808.10758} {arXiv:1808.10758 [gr-qc]} \BibitemShut
  {NoStop}%
\bibitem [{\citenamefont {Bl{\'a}zquez-Salcedo}\ and\ \citenamefont
  {Khoo}(2023)}]{Blazquez-Salcedo:2022eik}%
  \BibitemOpen
  \bibfield  {author} {\bibinfo {author} {\bibfnamefont {J.~L.}\ \bibnamefont
  {Bl{\'a}zquez-Salcedo}}\ and\ \bibinfo {author} {\bibfnamefont {F.~S.}\
  \bibnamefont {Khoo}},\ }\bibfield  {title} {\bibinfo {title} {{Quasinormal
  modes of slowly rotating Kerr-Newman black holes using the double series
  method}},\ }\href {https://doi.org/10.1103/PhysRevD.107.084031} {\bibfield
  {journal} {\bibinfo  {journal} {Phys. Rev. D}\ }\textbf {\bibinfo {volume}
  {107}},\ \bibinfo {pages} {084031} (\bibinfo {year} {2023})},\ \Eprint
  {https://arxiv.org/abs/2212.00054} {arXiv:2212.00054 [gr-qc]} \BibitemShut
  {NoStop}%
\bibitem [{\citenamefont {Brito}\ and\ \citenamefont
  {Pacilio}(2018)}]{Brito:2018hjh}%
  \BibitemOpen
  \bibfield  {author} {\bibinfo {author} {\bibfnamefont {R.}~\bibnamefont
  {Brito}}\ and\ \bibinfo {author} {\bibfnamefont {C.}~\bibnamefont
  {Pacilio}},\ }\bibfield  {title} {\bibinfo {title} {{Quasinormal modes of
  weakly charged Einstein-Maxwell-dilaton black holes}},\ }\href
  {https://doi.org/10.1103/PhysRevD.98.104042} {\bibfield  {journal} {\bibinfo
  {journal} {Phys. Rev. D}\ }\textbf {\bibinfo {volume} {98}},\ \bibinfo
  {pages} {104042} (\bibinfo {year} {2018})},\ \Eprint
  {https://arxiv.org/abs/1807.09081} {arXiv:1807.09081 [gr-qc]} \BibitemShut
  {NoStop}%
\bibitem [{\citenamefont {Wagle}\ \emph {et~al.}(2022)\citenamefont {Wagle},
  \citenamefont {Yunes},\ and\ \citenamefont {Silva}}]{Wagle:2021tam}%
  \BibitemOpen
  \bibfield  {author} {\bibinfo {author} {\bibfnamefont {P.}~\bibnamefont
  {Wagle}}, \bibinfo {author} {\bibfnamefont {N.}~\bibnamefont {Yunes}},\ and\
  \bibinfo {author} {\bibfnamefont {H.~O.}\ \bibnamefont {Silva}},\ }\bibfield
  {title} {\bibinfo {title} {{Quasinormal modes of slowly-rotating black holes
  in dynamical Chern-Simons gravity}},\ }\href
  {https://doi.org/10.1103/PhysRevD.105.124003} {\bibfield  {journal} {\bibinfo
   {journal} {Phys. Rev. D}\ }\textbf {\bibinfo {volume} {105}},\ \bibinfo
  {pages} {124003} (\bibinfo {year} {2022})},\ \Eprint
  {https://arxiv.org/abs/2103.09913} {arXiv:2103.09913 [gr-qc]} \BibitemShut
  {NoStop}%
\bibitem [{\citenamefont {Feng}\ and\ \citenamefont
  {Peng}(2024)}]{Feng:2024ygo}%
  \BibitemOpen
  \bibfield  {author} {\bibinfo {author} {\bibfnamefont {X.-H.}\ \bibnamefont
  {Feng}}\ and\ \bibinfo {author} {\bibfnamefont {J.}~\bibnamefont {Peng}},\
  }\bibfield  {title} {\bibinfo {title} {{Axial gravitational perturbations of
  slowly rotating compact objects in general relativity and beyond}},\ }\href
  {https://doi.org/10.1103/PhysRevD.110.064078} {\bibfield  {journal} {\bibinfo
   {journal} {Phys. Rev. D}\ }\textbf {\bibinfo {volume} {110}},\ \bibinfo
  {pages} {064078} (\bibinfo {year} {2024})},\ \Eprint
  {https://arxiv.org/abs/2404.16437} {arXiv:2404.16437 [gr-qc]} \BibitemShut
  {NoStop}%
\bibitem [{\citenamefont {Wu}\ \emph {et~al.}(2025{\natexlab{a}})\citenamefont
  {Wu}, \citenamefont {Teng}, \citenamefont {Li}, \citenamefont {Wang},\ and\
  \citenamefont {Lu}}]{Wu:2025euf}%
  \BibitemOpen
  \bibfield  {author} {\bibinfo {author} {\bibfnamefont {S.-M.}\ \bibnamefont
  {Wu}}, \bibinfo {author} {\bibfnamefont {X.-W.}\ \bibnamefont {Teng}},
  \bibinfo {author} {\bibfnamefont {W.-M.}\ \bibnamefont {Li}}, \bibinfo
  {author} {\bibfnamefont {Y.-X.}\ \bibnamefont {Wang}},\ and\ \bibinfo
  {author} {\bibfnamefont {J.}~\bibnamefont {Lu}},\ }\bibfield  {title}
  {\bibinfo {title} {{Nonseparability of multipartite systems in dilaton
  black~hole}},\ }\href {https://doi.org/10.1088/1475-7516/2025/09/030}
  {\bibfield  {journal} {\bibinfo  {journal} {JCAP}\ }\textbf {\bibinfo
  {volume} {09}},\ \bibinfo {pages} {030}},\ \Eprint
  {https://arxiv.org/abs/2503.17923} {arXiv:2503.17923 [gr-qc]} \BibitemShut
  {NoStop}%
\bibitem [{\citenamefont {Azreg-A\"\i{}nou}(2014)}]{Azreg-Ainou:2014pra}%
  \BibitemOpen
  \bibfield  {author} {\bibinfo {author} {\bibfnamefont {M.}~\bibnamefont
  {Azreg-A\"\i{}nou}},\ }\bibfield  {title} {\bibinfo {title} {{Generating
  rotating regular black hole solutions without complexification}},\ }\href
  {https://doi.org/10.1103/PhysRevD.90.064041} {\bibfield  {journal} {\bibinfo
  {journal} {Phys. Rev. D}\ }\textbf {\bibinfo {volume} {90}},\ \bibinfo
  {pages} {064041} (\bibinfo {year} {2014})},\ \Eprint
  {https://arxiv.org/abs/1405.2569} {arXiv:1405.2569 [gr-qc]} \BibitemShut
  {NoStop}%
\bibitem [{\citenamefont {Newman}\ and\ \citenamefont
  {Janis}(1965)}]{Newman:1965tw}%
  \BibitemOpen
  \bibfield  {author} {\bibinfo {author} {\bibfnamefont {E.~T.}\ \bibnamefont
  {Newman}}\ and\ \bibinfo {author} {\bibfnamefont {A.~I.}\ \bibnamefont
  {Janis}},\ }\bibfield  {title} {\bibinfo {title} {{Note on the Kerr spinning
  particle metric}},\ }\href {https://doi.org/10.1063/1.1704350} {\bibfield
  {journal} {\bibinfo  {journal} {J. Math. Phys.}\ }\textbf {\bibinfo {volume}
  {6}},\ \bibinfo {pages} {915} (\bibinfo {year} {1965})}\BibitemShut {NoStop}%
\bibitem [{\citenamefont {Ding}\ \emph {et~al.}(2020)\citenamefont {Ding},
  \citenamefont {Liu}, \citenamefont {Casana},\ and\ \citenamefont
  {Cavalcante}}]{Ding2020}%
  \BibitemOpen
  \bibfield  {author} {\bibinfo {author} {\bibfnamefont {C.}~\bibnamefont
  {Ding}}, \bibinfo {author} {\bibfnamefont {C.}~\bibnamefont {Liu}}, \bibinfo
  {author} {\bibfnamefont {R.}~\bibnamefont {Casana}},\ and\ \bibinfo {author}
  {\bibfnamefont {A.}~\bibnamefont {Cavalcante}},\ }\bibfield  {title}
  {\bibinfo {title} {{Exact Kerr-like solution and its shadow in a gravity
  model with spontaneous Lorentz symmetry breaking}},\ }\href
  {https://doi.org/10.1140/epjc/s10052-020-7743-y} {\bibfield  {journal}
  {\bibinfo  {journal} {Eur. Phys. J. C}\ }\textbf {\bibinfo {volume} {80}},\
  \bibinfo {pages} {178} (\bibinfo {year} {2020})},\ \Eprint
  {https://arxiv.org/abs/1910.02674} {arXiv:1910.02674 [gr-qc]} \BibitemShut
  {NoStop}%
\bibitem [{\citenamefont {Leaver}(1985)}]{Leaver:1985ax}%
  \BibitemOpen
  \bibfield  {author} {\bibinfo {author} {\bibfnamefont {E.~W.}\ \bibnamefont
  {Leaver}},\ }\bibfield  {title} {\bibinfo {title} {{An Analytic
  representation for the quasi normal modes of Kerr black holes}},\ }\href
  {https://doi.org/10.1098/rspa.1985.0119} {\bibfield  {journal} {\bibinfo
  {journal} {Proc. Roy. Soc. Lond. A}\ }\textbf {\bibinfo {volume} {402}},\
  \bibinfo {pages} {285} (\bibinfo {year} {1985})}\BibitemShut {NoStop}%
\bibitem [{\citenamefont {Leaver}(1990)}]{Leaver:1990zz}%
  \BibitemOpen
  \bibfield  {author} {\bibinfo {author} {\bibfnamefont {E.~W.}\ \bibnamefont
  {Leaver}},\ }\bibfield  {title} {\bibinfo {title} {{Quasinormal modes of
  Reissner-Nordstrom black holes}},\ }\href
  {https://doi.org/10.1103/PhysRevD.41.2986} {\bibfield  {journal} {\bibinfo
  {journal} {Phys. Rev. D}\ }\textbf {\bibinfo {volume} {41}},\ \bibinfo
  {pages} {2986} (\bibinfo {year} {1990})}\BibitemShut {NoStop}%
\bibitem [{\citenamefont {Percival}\ and\ \citenamefont
  {Dolan}(2020)}]{Percival:2020skc}%
  \BibitemOpen
  \bibfield  {author} {\bibinfo {author} {\bibfnamefont {J.}~\bibnamefont
  {Percival}}\ and\ \bibinfo {author} {\bibfnamefont {S.~R.}\ \bibnamefont
  {Dolan}},\ }\bibfield  {title} {\bibinfo {title} {{Quasinormal modes of
  massive vector fields on the Kerr spacetime}},\ }\href
  {https://doi.org/10.1103/PhysRevD.102.104055} {\bibfield  {journal} {\bibinfo
   {journal} {Phys. Rev. D}\ }\textbf {\bibinfo {volume} {102}},\ \bibinfo
  {pages} {104055} (\bibinfo {year} {2020})},\ \Eprint
  {https://arxiv.org/abs/2008.10621} {arXiv:2008.10621 [gr-qc]} \BibitemShut
  {NoStop}%
\bibitem [{\citenamefont {Guo}\ \emph {et~al.}(2023{\natexlab{b}})\citenamefont
  {Guo}, \citenamefont {Tan},\ and\ \citenamefont {Liu}}]{Guo:2022rms}%
  \BibitemOpen
  \bibfield  {author} {\bibinfo {author} {\bibfnamefont {W.-D.}\ \bibnamefont
  {Guo}}, \bibinfo {author} {\bibfnamefont {Q.}~\bibnamefont {Tan}},\ and\
  \bibinfo {author} {\bibfnamefont {Y.-X.}\ \bibnamefont {Liu}},\ }\bibfield
  {title} {\bibinfo {title} {{Gravitoelectromagnetic coupled perturbations and
  quasinormal modes of a charged black hole with scalar hair}},\ }\href
  {https://doi.org/10.1103/PhysRevD.107.124046} {\bibfield  {journal} {\bibinfo
   {journal} {Phys. Rev. D}\ }\textbf {\bibinfo {volume} {107}},\ \bibinfo
  {pages} {124046} (\bibinfo {year} {2023}{\natexlab{b}})},\ \Eprint
  {https://arxiv.org/abs/2212.08784} {arXiv:2212.08784 [gr-qc]} \BibitemShut
  {NoStop}%
\bibitem [{\citenamefont {Lin}\ and\ \citenamefont {Qian}(2017)}]{Lin:2016sch}%
  \BibitemOpen
  \bibfield  {author} {\bibinfo {author} {\bibfnamefont {K.}~\bibnamefont
  {Lin}}\ and\ \bibinfo {author} {\bibfnamefont {W.-L.}\ \bibnamefont {Qian}},\
  }\bibfield  {title} {\bibinfo {title} {{A Matrix Method for Quasinormal
  Modes: Schwarzschild Black Holes in Asymptotically Flat and (Anti-) de Sitter
  Spacetimes}},\ }\href {https://doi.org/10.1088/1361-6382/aa6643} {\bibfield
  {journal} {\bibinfo  {journal} {Class. Quant. Grav.}\ }\textbf {\bibinfo
  {volume} {34}},\ \bibinfo {pages} {095004} (\bibinfo {year} {2017})},\
  \Eprint {https://arxiv.org/abs/1610.08135} {arXiv:1610.08135 [gr-qc]}
  \BibitemShut {NoStop}%
\bibitem [{\citenamefont {Lin}\ \emph {et~al.}(2017)\citenamefont {Lin},
  \citenamefont {Qian}, \citenamefont {Pavan},\ and\ \citenamefont
  {Abdalla}}]{Lin:2017oag}%
  \BibitemOpen
  \bibfield  {author} {\bibinfo {author} {\bibfnamefont {K.}~\bibnamefont
  {Lin}}, \bibinfo {author} {\bibfnamefont {W.-L.}\ \bibnamefont {Qian}},
  \bibinfo {author} {\bibfnamefont {A.~B.}\ \bibnamefont {Pavan}},\ and\
  \bibinfo {author} {\bibfnamefont {E.}~\bibnamefont {Abdalla}},\ }\bibfield
  {title} {\bibinfo {title} {{A matrix method for quasinormal modes: Kerr and
  Kerr{\textendash}Sen black holes}},\ }\href
  {https://doi.org/10.1142/S0217732317501346} {\bibfield  {journal} {\bibinfo
  {journal} {Mod. Phys. Lett. A}\ }\textbf {\bibinfo {volume} {32}},\ \bibinfo
  {pages} {1750134} (\bibinfo {year} {2017})},\ \Eprint
  {https://arxiv.org/abs/1703.06439} {arXiv:1703.06439 [gr-qc]} \BibitemShut
  {NoStop}%
\bibitem [{\citenamefont {Lin}\ and\ \citenamefont {Qian}(2019)}]{Lin:2019mmf}%
  \BibitemOpen
  \bibfield  {author} {\bibinfo {author} {\bibfnamefont {K.}~\bibnamefont
  {Lin}}\ and\ \bibinfo {author} {\bibfnamefont {W.-L.}\ \bibnamefont {Qian}},\
  }\bibfield  {title} {\bibinfo {title} {{On matrix method for black hole
  quasinormal modes}},\ }\href {https://doi.org/10.1088/1674-1137/43/3/035105}
  {\bibfield  {journal} {\bibinfo  {journal} {Chin. Phys. C}\ }\textbf
  {\bibinfo {volume} {43}},\ \bibinfo {pages} {035105} (\bibinfo {year}
  {2019})},\ \Eprint {https://arxiv.org/abs/1902.08352} {arXiv:1902.08352
  [gr-qc]} \BibitemShut {NoStop}%
\bibitem [{\citenamefont {Lin}\ and\ \citenamefont {Qian}(2023)}]{Lin:2022ynv}%
  \BibitemOpen
  \bibfield  {author} {\bibinfo {author} {\bibfnamefont {K.}~\bibnamefont
  {Lin}}\ and\ \bibinfo {author} {\bibfnamefont {W.-L.}\ \bibnamefont {Qian}},\
  }\bibfield  {title} {\bibinfo {title} {{High-order matrix method with
  delimited expansion domain}},\ }\href
  {https://doi.org/10.1088/1361-6382/acc50f} {\bibfield  {journal} {\bibinfo
  {journal} {Class. Quant. Grav.}\ }\textbf {\bibinfo {volume} {40}},\ \bibinfo
  {pages} {085019} (\bibinfo {year} {2023})},\ \Eprint
  {https://arxiv.org/abs/2209.11612} {arXiv:2209.11612 [gr-qc]} \BibitemShut
  {NoStop}%
\bibitem [{\citenamefont {Shen}\ \emph {et~al.}(2022)\citenamefont {Shen},
  \citenamefont {Qian}, \citenamefont {Lin}, \citenamefont {Shao},\ and\
  \citenamefont {Pan}}]{Shen:2022xdp}%
  \BibitemOpen
  \bibfield  {author} {\bibinfo {author} {\bibfnamefont {S.-F.}\ \bibnamefont
  {Shen}}, \bibinfo {author} {\bibfnamefont {W.-L.}\ \bibnamefont {Qian}},
  \bibinfo {author} {\bibfnamefont {K.}~\bibnamefont {Lin}}, \bibinfo {author}
  {\bibfnamefont {C.-G.}\ \bibnamefont {Shao}},\ and\ \bibinfo {author}
  {\bibfnamefont {Y.}~\bibnamefont {Pan}},\ }\bibfield  {title} {\bibinfo
  {title} {{Matrix method for perturbed black hole metric with
  discontinuity}},\ }\href {https://doi.org/10.1088/1361-6382/ac95f1}
  {\bibfield  {journal} {\bibinfo  {journal} {Class. Quant. Grav.}\ }\textbf
  {\bibinfo {volume} {39}},\ \bibinfo {pages} {225004} (\bibinfo {year}
  {2022})},\ \Eprint {https://arxiv.org/abs/2203.14320} {arXiv:2203.14320
  [gr-qc]} \BibitemShut {NoStop}%
\bibitem [{\citenamefont {Lei}\ \emph {et~al.}(2021)\citenamefont {Lei},
  \citenamefont {Wang},\ and\ \citenamefont {Jing}}]{Lei2021}%
  \BibitemOpen
  \bibfield  {author} {\bibinfo {author} {\bibfnamefont {Y.}~\bibnamefont
  {Lei}}, \bibinfo {author} {\bibfnamefont {M.}~\bibnamefont {Wang}},\ and\
  \bibinfo {author} {\bibfnamefont {J.}~\bibnamefont {Jing}},\ }\bibfield
  {title} {\bibinfo {title} {{Maxwell perturbations in a cavity with Robin
  boundary conditions: two branches of modes with spectrum bifurcation on
  Schwarzschild black holes}},\ }\href
  {https://doi.org/10.1140/epjc/s10052-021-09942-8} {\bibfield  {journal}
  {\bibinfo  {journal} {Eur. Phys. J. C}\ }\textbf {\bibinfo {volume} {81}},\
  \bibinfo {pages} {1129} (\bibinfo {year} {2021})},\ \Eprint
  {https://arxiv.org/abs/2108.04146} {arXiv:2108.04146 [gr-qc]} \BibitemShut
  {NoStop}%
\bibitem [{\citenamefont {Liu}\ \emph {et~al.}(2024{\natexlab{c}})\citenamefont
  {Liu}, \citenamefont {Fang}, \citenamefont {Jing},\ and\ \citenamefont
  {Wang}}]{Liu:2024oeq}%
  \BibitemOpen
  \bibfield  {author} {\bibinfo {author} {\bibfnamefont {W.}~\bibnamefont
  {Liu}}, \bibinfo {author} {\bibfnamefont {X.}~\bibnamefont {Fang}}, \bibinfo
  {author} {\bibfnamefont {J.}~\bibnamefont {Jing}},\ and\ \bibinfo {author}
  {\bibfnamefont {J.}~\bibnamefont {Wang}},\ }\bibfield  {title} {\bibinfo
  {title} {{Lorentz violation induces isospectrality breaking in
  Einstein-bumblebee gravity theory}},\ }\href
  {https://doi.org/10.1007/s11433-024-2405-y} {\bibfield  {journal} {\bibinfo
  {journal} {Sci. China Phys. Mech. Astron.}\ }\textbf {\bibinfo {volume}
  {67}},\ \bibinfo {pages} {280413} (\bibinfo {year} {2024}{\natexlab{c}})},\
  \Eprint {https://arxiv.org/abs/2402.09686} {arXiv:2402.09686 [gr-qc]}
  \BibitemShut {NoStop}%
\bibitem [{\citenamefont {Liu}\ \emph {et~al.}(2023{\natexlab{d}})\citenamefont
  {Liu}, \citenamefont {Fang}, \citenamefont {Jing},\ and\ \citenamefont
  {Wang}}]{Liu:2023uft}%
  \BibitemOpen
  \bibfield  {author} {\bibinfo {author} {\bibfnamefont {W.}~\bibnamefont
  {Liu}}, \bibinfo {author} {\bibfnamefont {X.}~\bibnamefont {Fang}}, \bibinfo
  {author} {\bibfnamefont {J.}~\bibnamefont {Jing}},\ and\ \bibinfo {author}
  {\bibfnamefont {J.}~\bibnamefont {Wang}},\ }\bibfield  {title} {\bibinfo
  {title} {{Gravito-electromagnetic perturbations of MOG black holes with a
  cosmological constant: quasinormal modes and ringdown waveforms}},\ }\href
  {https://doi.org/10.1088/1475-7516/2023/11/057} {\bibfield  {journal}
  {\bibinfo  {journal} {JCAP}\ }\textbf {\bibinfo {volume} {11}},\ \bibinfo
  {pages} {057}},\ \Eprint {https://arxiv.org/abs/2306.03599} {arXiv:2306.03599
  [gr-qc]} \BibitemShut {NoStop}%
\bibitem [{\citenamefont {Fuentes-Schuller}\ and\ \citenamefont
  {Mann}(2005)}]{Fuentes-Schuller:2004iaz}%
  \BibitemOpen
  \bibfield  {author} {\bibinfo {author} {\bibfnamefont {I.}~\bibnamefont
  {Fuentes-Schuller}}\ and\ \bibinfo {author} {\bibfnamefont {R.~B.}\
  \bibnamefont {Mann}},\ }\bibfield  {title} {\bibinfo {title} {{Alice falls
  into a black hole: Entanglement in non-inertial frames}},\ }\href
  {https://doi.org/10.1103/PhysRevLett.95.120404} {\bibfield  {journal}
  {\bibinfo  {journal} {Phys. Rev. Lett.}\ }\textbf {\bibinfo {volume} {95}},\
  \bibinfo {pages} {120404} (\bibinfo {year} {2005})},\ \Eprint
  {https://arxiv.org/abs/0410172} {arXiv:0410172 [quant-ph]} \BibitemShut
  {NoStop}%
\bibitem [{\citenamefont {Wu}\ \emph {et~al.}(2025{\natexlab{b}})\citenamefont
  {Wu}, \citenamefont {Wang}, \citenamefont {Shang},\ and\ \citenamefont
  {Liu}}]{Wu:2025ncd}%
  \BibitemOpen
  \bibfield  {author} {\bibinfo {author} {\bibfnamefont {S.-M.}\ \bibnamefont
  {Wu}}, \bibinfo {author} {\bibfnamefont {Y.-X.}\ \bibnamefont {Wang}},
  \bibinfo {author} {\bibfnamefont {S.-H.}\ \bibnamefont {Shang}},\ and\
  \bibinfo {author} {\bibfnamefont {W.}~\bibnamefont {Liu}},\ }\bibfield
  {title} {\bibinfo {title} {{Influence of dark matter on quantum entanglement
  and coherence in curved spacetime}},\ }\href@noop {} {\  (\bibinfo {year}
  {2025}{\natexlab{b}})},\ \Eprint {https://arxiv.org/abs/2507.16142}
  {arXiv:2507.16142 [gr-qc]} \BibitemShut {NoStop}%
\bibitem [{\citenamefont {Wu}\ \emph {et~al.}(2024{\natexlab{b}})\citenamefont
  {Wu}, \citenamefont {Wang}, \citenamefont {Wang}, \citenamefont {Li},
  \citenamefont {Huang},\ and\ \citenamefont {Zeng}}]{Wu:2023vis}%
  \BibitemOpen
  \bibfield  {author} {\bibinfo {author} {\bibfnamefont {S.-M.}\ \bibnamefont
  {Wu}}, \bibinfo {author} {\bibfnamefont {C.-X.}\ \bibnamefont {Wang}},
  \bibinfo {author} {\bibfnamefont {R.-D.}\ \bibnamefont {Wang}}, \bibinfo
  {author} {\bibfnamefont {J.-X.}\ \bibnamefont {Li}}, \bibinfo {author}
  {\bibfnamefont {X.-L.}\ \bibnamefont {Huang}},\ and\ \bibinfo {author}
  {\bibfnamefont {H.-S.}\ \bibnamefont {Zeng}},\ }\bibfield  {title} {\bibinfo
  {title} {{Curvature-enhanced multipartite coherence in the multiverse*}},\
  }\href {https://doi.org/10.1088/1674-1137/ad32bf} {\bibfield  {journal}
  {\bibinfo  {journal} {Chin. Phys. C}\ }\textbf {\bibinfo {volume} {48}},\
  \bibinfo {pages} {075107} (\bibinfo {year} {2024}{\natexlab{b}})},\ \Eprint
  {https://arxiv.org/abs/2307.00698} {arXiv:2307.00698 [gr-qc]} \BibitemShut
  {NoStop}%
\bibitem [{\citenamefont {Huang}\ \emph {et~al.}(2024)\citenamefont {Huang},
  \citenamefont {Wu},\ and\ \citenamefont {Wu}}]{Huang:2024vyc}%
  \BibitemOpen
  \bibfield  {author} {\bibinfo {author} {\bibfnamefont {X.}~\bibnamefont
  {Huang}}, \bibinfo {author} {\bibfnamefont {H.}~\bibnamefont {Wu}},\ and\
  \bibinfo {author} {\bibfnamefont {S.}~\bibnamefont {Wu}},\ }\bibfield
  {title} {\bibinfo {title} {{Generated genuine tripartite steering and its
  monogamy in the background of a Kerr-Newman black hole}},\ }\href
  {https://doi.org/10.1088/1674-1137/ad6e5f} {\bibfield  {journal} {\bibinfo
  {journal} {Chin. Phys. C}\ }\textbf {\bibinfo {volume} {48}},\ \bibinfo
  {pages} {115106} (\bibinfo {year} {2024})}\BibitemShut {NoStop}%
\bibitem [{\citenamefont {Wu}\ \emph {et~al.}(2024{\natexlab{c}})\citenamefont
  {Wu}, \citenamefont {Wang}, \citenamefont {Huang},\ and\ \citenamefont
  {Wang}}]{Wu:2024qhd}%
  \BibitemOpen
  \bibfield  {author} {\bibinfo {author} {\bibfnamefont {S.-M.}\ \bibnamefont
  {Wu}}, \bibinfo {author} {\bibfnamefont {R.-D.}\ \bibnamefont {Wang}},
  \bibinfo {author} {\bibfnamefont {X.-L.}\ \bibnamefont {Huang}},\ and\
  \bibinfo {author} {\bibfnamefont {Z.}~\bibnamefont {Wang}},\ }\bibfield
  {title} {\bibinfo {title} {{Does gravitational wave assist vacuum steering
  and Bell nonlocality?}},\ }\href {https://doi.org/10.1007/JHEP07(2024)155}
  {\bibfield  {journal} {\bibinfo  {journal} {JHEP}\ }\textbf {\bibinfo
  {volume} {07}},\ \bibinfo {pages} {155}},\ \Eprint
  {https://arxiv.org/abs/2405.07235} {arXiv:2405.07235 [gr-qc]} \BibitemShut
  {NoStop}%
\bibitem [{\citenamefont {Li}\ \emph {et~al.}(2025)\citenamefont {Li},
  \citenamefont {Shang},\ and\ \citenamefont {Wu}}]{Li:2025bzd}%
  \BibitemOpen
  \bibfield  {author} {\bibinfo {author} {\bibfnamefont {S.-H.}\ \bibnamefont
  {Li}}, \bibinfo {author} {\bibfnamefont {S.-H.}\ \bibnamefont {Shang}},\ and\
  \bibinfo {author} {\bibfnamefont {S.-M.}\ \bibnamefont {Wu}},\ }\bibfield
  {title} {\bibinfo {title} {{Does acceleration always degrade quantum
  entanglement for tetrapartite Unruh-DeWitt detectors?}},\ }\href
  {https://doi.org/10.1007/JHEP05(2025)214} {\bibfield  {journal} {\bibinfo
  {journal} {JHEP}\ }\textbf {\bibinfo {volume} {05}},\ \bibinfo {pages}
  {214}},\ \Eprint {https://arxiv.org/abs/2502.05881} {arXiv:2502.05881
  [gr-qc]} \BibitemShut {NoStop}%
\bibitem [{\citenamefont {Frolov}\ \emph {et~al.}(2018)\citenamefont {Frolov},
  \citenamefont {Krtou{\v{s}}}, \citenamefont {Kubiz{\v{n}}{\'a}k},\ and\
  \citenamefont {Santos}}]{Frolov:2018ezx}%
  \BibitemOpen
  \bibfield  {author} {\bibinfo {author} {\bibfnamefont {V.~P.}\ \bibnamefont
  {Frolov}}, \bibinfo {author} {\bibfnamefont {P.}~\bibnamefont
  {Krtou{\v{s}}}}, \bibinfo {author} {\bibfnamefont {D.}~\bibnamefont
  {Kubiz{\v{n}}{\'a}k}},\ and\ \bibinfo {author} {\bibfnamefont {J.~E.}\
  \bibnamefont {Santos}},\ }\bibfield  {title} {\bibinfo {title} {{Massive
  Vector Fields in Rotating Black-Hole Spacetimes: Separability and Quasinormal
  Modes}},\ }\href {https://doi.org/10.1103/PhysRevLett.120.231103} {\bibfield
  {journal} {\bibinfo  {journal} {Phys. Rev. Lett.}\ }\textbf {\bibinfo
  {volume} {120}},\ \bibinfo {pages} {231103} (\bibinfo {year} {2018})},\
  \Eprint {https://arxiv.org/abs/1804.00030} {arXiv:1804.00030 [hep-th]}
  \BibitemShut {NoStop}%
\bibitem [{\citenamefont {Wei}\ \emph {et~al.}(2022)\citenamefont {Wei},
  \citenamefont {Liu},\ and\ \citenamefont {Mann}}]{Wei:2022dzw}%
  \BibitemOpen
  \bibfield  {author} {\bibinfo {author} {\bibfnamefont {S.-W.}\ \bibnamefont
  {Wei}}, \bibinfo {author} {\bibfnamefont {Y.-X.}\ \bibnamefont {Liu}},\ and\
  \bibinfo {author} {\bibfnamefont {R.~B.}\ \bibnamefont {Mann}},\ }\bibfield
  {title} {\bibinfo {title} {{Black Hole Solutions as Topological Thermodynamic
  Defects}},\ }\href {https://doi.org/10.1103/PhysRevLett.129.191101}
  {\bibfield  {journal} {\bibinfo  {journal} {Phys. Rev. Lett.}\ }\textbf
  {\bibinfo {volume} {129}},\ \bibinfo {pages} {191101} (\bibinfo {year}
  {2022})},\ \Eprint {https://arxiv.org/abs/2208.01932} {arXiv:2208.01932
  [gr-qc]} \BibitemShut {NoStop}%
\bibitem [{\citenamefont {Wei}\ \emph {et~al.}(2024)\citenamefont {Wei},
  \citenamefont {Liu},\ and\ \citenamefont {Mann}}]{Wei:2024gfz}%
  \BibitemOpen
  \bibfield  {author} {\bibinfo {author} {\bibfnamefont {S.-W.}\ \bibnamefont
  {Wei}}, \bibinfo {author} {\bibfnamefont {Y.-X.}\ \bibnamefont {Liu}},\ and\
  \bibinfo {author} {\bibfnamefont {R.~B.}\ \bibnamefont {Mann}},\ }\bibfield
  {title} {\bibinfo {title} {{Universal topological classifications of black
  hole thermodynamics}},\ }\href {https://doi.org/10.1103/PhysRevD.110.L081501}
  {\bibfield  {journal} {\bibinfo  {journal} {Phys. Rev. D}\ }\textbf {\bibinfo
  {volume} {110}},\ \bibinfo {pages} {L081501} (\bibinfo {year} {2024})},\
  \Eprint {https://arxiv.org/abs/2409.09333} {arXiv:2409.09333 [gr-qc]}
  \BibitemShut {NoStop}%
\bibitem [{\citenamefont {Wu}\ \emph {et~al.}(2025{\natexlab{c}})\citenamefont
  {Wu}, \citenamefont {Liu}, \citenamefont {Wu},\ and\ \citenamefont
  {Mann}}]{Wu:2024asq}%
  \BibitemOpen
  \bibfield  {author} {\bibinfo {author} {\bibfnamefont {D.}~\bibnamefont
  {Wu}}, \bibinfo {author} {\bibfnamefont {W.}~\bibnamefont {Liu}}, \bibinfo
  {author} {\bibfnamefont {S.-Q.}\ \bibnamefont {Wu}},\ and\ \bibinfo {author}
  {\bibfnamefont {R.~B.}\ \bibnamefont {Mann}},\ }\bibfield  {title} {\bibinfo
  {title} {{Novel topological classes in black hole thermodynamics}},\ }\href
  {https://doi.org/10.1103/PhysRevD.111.L061501} {\bibfield  {journal}
  {\bibinfo  {journal} {Phys. Rev. D}\ }\textbf {\bibinfo {volume} {111}},\
  \bibinfo {pages} {L061501} (\bibinfo {year} {2025}{\natexlab{c}})},\ \Eprint
  {https://arxiv.org/abs/2411.10102} {arXiv:2411.10102 [hep-th]} \BibitemShut
  {NoStop}%
\bibitem [{\citenamefont {Liu}\ \emph {et~al.}(2025{\natexlab{e}})\citenamefont
  {Liu}, \citenamefont {Zhang}, \citenamefont {Wu},\ and\ \citenamefont
  {Wang}}]{Liu:2025iyl}%
  \BibitemOpen
  \bibfield  {author} {\bibinfo {author} {\bibfnamefont {W.}~\bibnamefont
  {Liu}}, \bibinfo {author} {\bibfnamefont {L.}~\bibnamefont {Zhang}}, \bibinfo
  {author} {\bibfnamefont {D.}~\bibnamefont {Wu}},\ and\ \bibinfo {author}
  {\bibfnamefont {J.}~\bibnamefont {Wang}},\ }\bibfield  {title} {\bibinfo
  {title} {{Thermodynamic topological classes of the rotating, accelerating
  black holes}},\ }\href {https://doi.org/10.1088/1361-6382/ade35b} {\bibfield
  {journal} {\bibinfo  {journal} {Class. Quant. Grav.}\ }\textbf {\bibinfo
  {volume} {42}},\ \bibinfo {pages} {125007} (\bibinfo {year}
  {2025}{\natexlab{e}})},\ \Eprint {https://arxiv.org/abs/2409.11666}
  {arXiv:2409.11666 [hep-th]} \BibitemShut {NoStop}%
\bibitem [{\citenamefont {Zhu}\ \emph {et~al.}(2025)\citenamefont {Zhu},
  \citenamefont {Liu},\ and\ \citenamefont {Wu}}]{Zhu:2024zcl}%
  \BibitemOpen
  \bibfield  {author} {\bibinfo {author} {\bibfnamefont {X.-D.}\ \bibnamefont
  {Zhu}}, \bibinfo {author} {\bibfnamefont {W.}~\bibnamefont {Liu}},\ and\
  \bibinfo {author} {\bibfnamefont {D.}~\bibnamefont {Wu}},\ }\bibfield
  {title} {\bibinfo {title} {{Universal thermodynamic topological classes of
  rotating black holes}},\ }\href
  {https://doi.org/10.1016/j.physletb.2024.139163} {\bibfield  {journal}
  {\bibinfo  {journal} {Phys. Lett. B}\ }\textbf {\bibinfo {volume} {860}},\
  \bibinfo {pages} {139163} (\bibinfo {year} {2025})},\ \Eprint
  {https://arxiv.org/abs/2409.12747} {arXiv:2409.12747 [hep-th]} \BibitemShut
  {NoStop}%
\bibitem [{\citenamefont {Wu}\ \emph {et~al.}(2024{\natexlab{d}})\citenamefont
  {Wu}, \citenamefont {Gu}, \citenamefont {Zhu}, \citenamefont {Jiang},\ and\
  \citenamefont {Yang}}]{Wu:2024rmv}%
  \BibitemOpen
  \bibfield  {author} {\bibinfo {author} {\bibfnamefont {D.}~\bibnamefont
  {Wu}}, \bibinfo {author} {\bibfnamefont {S.-Y.}\ \bibnamefont {Gu}}, \bibinfo
  {author} {\bibfnamefont {X.-D.}\ \bibnamefont {Zhu}}, \bibinfo {author}
  {\bibfnamefont {Q.-Q.}\ \bibnamefont {Jiang}},\ and\ \bibinfo {author}
  {\bibfnamefont {S.-Z.}\ \bibnamefont {Yang}},\ }\bibfield  {title} {\bibinfo
  {title} {{Topological classes of thermodynamics of the static multi-charge
  AdS black holes in gauged supergravities: novel temperature-dependent
  thermodynamic topological phase transition}},\ }\href
  {https://doi.org/10.1007/JHEP06(2024)213} {\bibfield  {journal} {\bibinfo
  {journal} {JHEP}\ }\textbf {\bibinfo {volume} {06}},\ \bibinfo {pages}
  {213}},\ \Eprint {https://arxiv.org/abs/2402.00106} {arXiv:2402.00106
  [hep-th]} \BibitemShut {NoStop}%
\bibitem [{\citenamefont {Wu}(2023{\natexlab{a}})}]{Wu:2023fcw}%
  \BibitemOpen
  \bibfield  {author} {\bibinfo {author} {\bibfnamefont {D.}~\bibnamefont
  {Wu}},\ }\bibfield  {title} {\bibinfo {title} {{Consistent thermodynamics and
  topological classes for the four-dimensional Lorentzian charged Taub-NUT
  spacetimes}},\ }\href {https://doi.org/10.1140/epjc/s10052-023-11782-7}
  {\bibfield  {journal} {\bibinfo  {journal} {Eur. Phys. J. C}\ }\textbf
  {\bibinfo {volume} {83}},\ \bibinfo {pages} {589} (\bibinfo {year}
  {2023}{\natexlab{a}})},\ \Eprint {https://arxiv.org/abs/2306.02324}
  {arXiv:2306.02324 [gr-qc]} \BibitemShut {NoStop}%
\bibitem [{\citenamefont {Wu}(2023{\natexlab{b}})}]{Wu:2023xpq}%
  \BibitemOpen
  \bibfield  {author} {\bibinfo {author} {\bibfnamefont {D.}~\bibnamefont
  {Wu}},\ }\bibfield  {title} {\bibinfo {title} {{Classifying topology of
  consistent thermodynamics of the four-dimensional neutral Lorentzian
  NUT-charged spacetimes}},\ }\href
  {https://doi.org/10.1140/epjc/s10052-023-11561-4} {\bibfield  {journal}
  {\bibinfo  {journal} {Eur. Phys. J. C}\ }\textbf {\bibinfo {volume} {83}},\
  \bibinfo {pages} {365} (\bibinfo {year} {2023}{\natexlab{b}})},\ \Eprint
  {https://arxiv.org/abs/2302.01100} {arXiv:2302.01100 [gr-qc]} \BibitemShut
  {NoStop}%
\bibitem [{\citenamefont {Wu}\ and\ \citenamefont {Wu}(2023)}]{Wu:2023sue}%
  \BibitemOpen
  \bibfield  {author} {\bibinfo {author} {\bibfnamefont {D.}~\bibnamefont
  {Wu}}\ and\ \bibinfo {author} {\bibfnamefont {S.-Q.}\ \bibnamefont {Wu}},\
  }\bibfield  {title} {\bibinfo {title} {{Topological classes of thermodynamics
  of rotating AdS black holes}},\ }\href
  {https://doi.org/10.1103/PhysRevD.107.084002} {\bibfield  {journal} {\bibinfo
   {journal} {Phys. Rev. D}\ }\textbf {\bibinfo {volume} {107}},\ \bibinfo
  {pages} {084002} (\bibinfo {year} {2023})},\ \Eprint
  {https://arxiv.org/abs/2301.03002} {arXiv:2301.03002 [hep-th]} \BibitemShut
  {NoStop}%
\bibitem [{\citenamefont {Wu}(2023{\natexlab{c}})}]{Wu:2022whe}%
  \BibitemOpen
  \bibfield  {author} {\bibinfo {author} {\bibfnamefont {D.}~\bibnamefont
  {Wu}},\ }\bibfield  {title} {\bibinfo {title} {{Topological classes of
  rotating black holes}},\ }\href {https://doi.org/10.1103/PhysRevD.107.024024}
  {\bibfield  {journal} {\bibinfo  {journal} {Phys. Rev. D}\ }\textbf {\bibinfo
  {volume} {107}},\ \bibinfo {pages} {024024} (\bibinfo {year}
  {2023}{\natexlab{c}})},\ \Eprint {https://arxiv.org/abs/2211.15151}
  {arXiv:2211.15151 [gr-qc]} \BibitemShut {NoStop}%
\bibitem [{\citenamefont {Liu}\ \emph {et~al.}(2024{\natexlab{d}})\citenamefont
  {Liu}, \citenamefont {Wu},\ and\ \citenamefont {Wang}}]{Liu:2024soc}%
  \BibitemOpen
  \bibfield  {author} {\bibinfo {author} {\bibfnamefont {W.}~\bibnamefont
  {Liu}}, \bibinfo {author} {\bibfnamefont {D.}~\bibnamefont {Wu}},\ and\
  \bibinfo {author} {\bibfnamefont {J.}~\bibnamefont {Wang}},\ }\bibfield
  {title} {\bibinfo {title} {{Light rings and shadows of static black holes in
  effective quantum gravity}},\ }\href
  {https://doi.org/10.1016/j.physletb.2024.139052} {\bibfield  {journal}
  {\bibinfo  {journal} {Phys. Lett. B}\ }\textbf {\bibinfo {volume} {858}},\
  \bibinfo {pages} {139052} (\bibinfo {year} {2024}{\natexlab{d}})},\ \Eprint
  {https://arxiv.org/abs/2408.05569} {arXiv:2408.05569 [gr-qc]} \BibitemShut
  {NoStop}%
\bibitem [{\citenamefont {Liu}\ \emph {et~al.}(2024{\natexlab{e}})\citenamefont
  {Liu}, \citenamefont {Wu},\ and\ \citenamefont {Wang}}]{Liu:2024iec}%
  \BibitemOpen
  \bibfield  {author} {\bibinfo {author} {\bibfnamefont {W.}~\bibnamefont
  {Liu}}, \bibinfo {author} {\bibfnamefont {D.}~\bibnamefont {Wu}},\ and\
  \bibinfo {author} {\bibfnamefont {J.}~\bibnamefont {Wang}},\ }\bibfield
  {title} {\bibinfo {title} {{Light rings and shadows of static black holes in
  effective quantum gravity II: A new solution without Cauchy horizons}},\
  }\href@noop {} {\  (\bibinfo {year} {2024}{\natexlab{e}})},\ \Eprint
  {https://arxiv.org/abs/2412.18083} {arXiv:2412.18083 [gr-qc]} \BibitemShut
  {NoStop}%
\bibitem [{\citenamefont {Cunha}\ \emph {et~al.}(2024)\citenamefont {Cunha},
  \citenamefont {Herdeiro},\ and\ \citenamefont {Novo}}]{Cunha:2024ajc}%
  \BibitemOpen
  \bibfield  {author} {\bibinfo {author} {\bibfnamefont {P.~V.~P.}\
  \bibnamefont {Cunha}}, \bibinfo {author} {\bibfnamefont {C.~A.~R.}\
  \bibnamefont {Herdeiro}},\ and\ \bibinfo {author} {\bibfnamefont {J.~P.~A.}\
  \bibnamefont {Novo}},\ }\bibfield  {title} {\bibinfo {title} {{Light rings on
  stationary axisymmetric spacetimes: Blind to the topology and able to
  coexist}},\ }\href {https://doi.org/10.1103/PhysRevD.109.064050} {\bibfield
  {journal} {\bibinfo  {journal} {Phys. Rev. D}\ }\textbf {\bibinfo {volume}
  {109}},\ \bibinfo {pages} {064050} (\bibinfo {year} {2024})},\ \Eprint
  {https://arxiv.org/abs/2401.05495} {arXiv:2401.05495 [gr-qc]} \BibitemShut
  {NoStop}%
\end{thebibliography}
%

\end{document}